\documentclass[
superscriptaddress,
twocolumn,
nofootinbib,
amsmath,amssymb,
aps,pra,
floatfix,
]{revtex4-2}
\usepackage{lipsum}
\usepackage{ulem}
\usepackage{xcolor}
\usepackage{dsfont}
\usepackage{relsize}

\newcommand{\initial}{\psi}

\newcommand{\ii}{\mathrm{i}}
\newcommand{\cF}{\mathcal{F}}

\newcommand{\be}{\begin{equation}} 	
\newcommand{\ee}{\end{equation}}
\newcommand{\e}[1]{e^{#1}}

\newcommand{\revadd}[1]{{#1}}

\newcommand{\addw}{\omega}

\usepackage{cancel}

\usepackage{array}
\newcolumntype{C}[1]{>{\centering\arraybackslash}p{#1}}

\usepackage[hidelinks]{hyperref}

\usepackage[draft]{changes}
\usepackage{comment}

\makeatletter 
    
\renewcommand\onecolumngrid{
\do@columngrid{one}{\@ne}
\def\set@footnotewidth{\onecolumngrid}
\def\footnoterule{\kern-6pt\hrule width 1.5in\kern6pt}%
}

\renewcommand\twocolumngrid{
        \def\footnoterule{
        \dimen@\skip\footins\divide\dimen@\thr@@
        \kern-\dimen@\hrule width.5in\kern\dimen@}
        \do@columngrid{mlt}{\tw@}
}

\makeatother

\usepackage{graphicx}

\usepackage{enumitem}

\usepackage{dcolumn}    
\usepackage{bm,bbm}
\usepackage{physics}
\usepackage{amsmath}
\usepackage{amsthm}
\usepackage{amsfonts}
\usepackage[hidelinks]{hyperref}
\usepackage{xcolor}
\usepackage{float}
\usepackage{soul}
\sethlcolor{yellow}

\newtheorem{result}{Result}

\newtheorem{proposition}{Proposition}

\usepackage{lipsum}

\usepackage{cancel}

\newcommand{\1}{\mathbbm{1}}

\newcommand{\f}{\mathcal{F}}

\newcommand{\eqq}[1]{Eq. \eqref{eq:#1}}

\newcommand{\figg}[1]{Fig. \ref{fig:#1}}

\renewcommand{\vec}[1]{\boldsymbol{#1}}

\renewcommand{\var}{{\rm Var}}
\newcommand{\vecc}[1]{|{{#1\rangle}\!\rangle}}

\usepackage{tikz}

\begin{document}

\preprint{APS/123-QED}

\title{From dynamical to steady-state many-body metrology: \\ Precision limits and their attainability with two-body interactions}
\author{Ricard Puig}
\email{ricard.puigivalls@epfl.ch}
\affiliation{Institute of Physics, Ecole Polytechnique F\'{e}d\'{e}rale de Lausanne (EPFL), CH-1015 Lausanne, Switzerland}
\affiliation{F\'isica Te\`orica: Informaci\'o i Fen\`omens Qu\`antics, Department de F\'isica, Universitat Aut\`onoma de Barcelona, 08193 Bellaterra (Barcelona), Spain}
\affiliation{D\'{e}partement de Physique Appliqu\'{e}e,  Universit\'{e} de Gen\`{e}ve,  1211 Gen\`{e}ve,  Switzerland}

\author{Pavel Sekatski}
\affiliation{D\'{e}partement de Physique Appliqu\'{e}e,  Universit\'{e} de Gen\`{e}ve,  1211 Gen\`{e}ve,  Switzerland}

\author{Paolo Andrea Erdman}
\affiliation{Freie Universit{\" a}t Berlin, Department of Mathematics and Computer Science, Arnimallee 6, 14195 Berlin, Germany}

\author{Paolo~Abiuso}
\affiliation{Institute for Quantum Optics and Quantum Information - IQOQI Vienna,
Austrian Academy of Sciences, Boltzmanngasse 3, A-1090 Vienna, Austria}

\author{John Calsamiglia}
\affiliation{F\'isica Te\`orica: Informaci\'o i Fen\`omens Qu\`antics, Department de F\'isica, Universitat Aut\`onoma de Barcelona, 08193 Bellaterra (Barcelona), Spain}

\author{Martí Perarnau-Llobet}
\email{marti.perarnau@uab.cat}
\affiliation{F\'isica Te\`orica: Informaci\'o i Fen\`omens Qu\`antics, Department de F\'isica, Universitat Aut\`onoma de Barcelona, 08193 Bellaterra (Barcelona), Spain}
\affiliation{D\'{e}partement de Physique Appliqu\'{e}e,  Universit\'{e} de Gen\`{e}ve,  1211 Gen\`{e}ve,  Switzerland}

\date{\today}

\begin{abstract}
We consider the estimation of an unknown parameter $\theta$ via a many-body probe. The probe is initially prepared in a product state and  many-body \revadd{time-independent} interactions enhance its \mbox{$\theta$-sensitivity} during the dynamics and/or in the steady state. We present  bounds on the Quantum Fisher Information, and corresponding optimal interacting Hamiltonians, for two paradigmatic scenarios for encoding~$\theta$: (i)~via unitary Hamiltonian dynamics (dynamical metrology), and (ii)~in  the Gibbs and diagonal ensembles (time-averaged dephased state), two ubiquitous steady states of many-body open dynamics. We then move to the specific problem of estimating the strength of a magnetic field via interacting spins and derive two-body interacting Hamiltonians that can approach the fundamental precision bounds. In this case, we additionally analyze the transient regime leading to the steady states and characterize tradeoffs between equilibration times and measurement precision. Overall, our results provide a comprehensive picture of the potential of many-body control in quantum sensing.
\end{abstract}
\maketitle

\section{Introduction}

Quantum metrology is concerned with the precise estimation of physical quantities that can range from the strength of a magnetic field to the temperature of ultra-cold gases,  beyond the capability of conventional classical sensors by exploiting quantum resources such as quantum coherence or entanglement~\cite{qmqi, optical_interferometry_review}. 
A central problem in quantum metrology is the estimation of a parameter $\theta$ of a Hamiltonian $H_\theta$. For time-independent $H_\theta$, this can typically be expressed as  
\begin{equation}\label{eq:paradigm}
    H_{\theta}=\theta H_S + H_C\, .
\end{equation}
Here, $H_S$ is the part of the Hamiltonian encoding the signal, whereas $H_C$ includes any other contribution. The latter can range from the action of an uncontrolled environment to a control term designed to enhance the measurement precision. Considering  a quantum state $\rho_0$  
evolving under $U=\exp(-i t H_\theta)$, where we set $\hbar=1$,  the Cramér-Rao bound tells us that the uncertainty $\Delta \theta$  is bounded as~\cite{boixogeneralised} 
\begin{align}
\label{eq:boundCR}
    (\Delta \theta)^2 \geq \frac{1}{\mu\,t^2 \|H_S\|^2}\, ,
\end{align}
where $\mu$ is the number of \revadd{identical repetitions of the experiment,} and  we 
defined the pseudonorm $\|H_S\| = \lambda_{\rm max}^{S} - \lambda_{\rm min}^{S}$ with $\lambda_{\rm max}^{S}$ ($\lambda_{\rm min}^{S}$) the maximal (minimal) eigenvalue of $H_S$. For example, in magnetometry with $N$ spins, $H_S=\frac{\addw}{2} \sum_{j=1}^N \sigma_z^{(j)}$ \revadd{(where $\addw$ dictates the energy scale),} we recover the standard (dynamical) Heisenberg limit $(\Delta \theta)^{-2} \leq \mu t^2 \revadd{\addw^2} N^2$. This can be contrasted to the standard shot-noise limit (SNL) $\mu N \revadd{\addw^2} t^2$ obtained for independent coherently evolving systems~\cite{qmqi, optical_interferometry_review}.

\noindent\begin{table*}
    \centering
    \begin{tabular}{|C{6.2cm}| C{3.8cm}| C{4cm} |C{3.8cm}|}
         \hline\rule{0pt}{1.1\normalbaselineskip} 
           & Dynamical  with an
           & Steady state:  & Steady state:   \\ 
          & initial ground state of $H_S$
          & Diagonal ensemble  &  Gibbs ensemble  \\[1.ex]
         \hline \rule{0pt}{1.2\normalbaselineskip} 
         General upper bound & $\mathcal{F} \leq  t^2  \|H_S\|^2$   \hspace{1.5mm} (Ref.~\cite{boixogeneralised})
         & $\mathcal{F} \leq \frac{3+\pi^2}{3} \frac{\|H_S\|^2}{E^2} $   \hspace{0.5mm} (Result \ref{res: dephasing upperbound})   & $\mathcal{F} \leq  \frac{1}{4}\beta^2 \|H_S\|^2 $ \hspace{0.5mm} (Ref. \cite{abiuso2024fundamental})  \\[1.ex] 
         \hline\rule{0pt}{1.2\normalbaselineskip} 
         Best-known protocol with unrestricted $H_C$
         & $\mathcal{F} \approx  \frac{1}{4}t^2  \|H_S\|^2$    \hspace{0.5mm} (Result \ref{res:pinchedlocal}) 
         & $\mathcal{F} = \frac{3}{2} \frac{\|H_S\|^2}{E^2}  $    \hspace{5.5mm} (Result~\ref{res: dephasing attainable})   & $\mathcal{F} =  \frac{1}{4}\beta^2 \|H_S\|^2 $  \hspace{0.5mm} (Ref.~\cite{abiuso2024fundamental}) \\ [1.ex]
         \hline
    \end{tabular}
    \caption{\textbf{Summary of the main results for arbitrary encoding $H_S$.}  All upper bounds depend on $||H_S|| = \lambda_{\rm max}^{S} - \lambda_{\rm min}^{S}$ with $\lambda_{\rm max}^{S}$ ($\lambda_{\rm min}^{S}$) the maximal (minimal) eigenvalue of $H_S$. Up to constants, the time $t$ is replaced by $\beta$ for Gibbs state metrology and by $E^{-1}$ for the diagonal ensemble with $E$ the minimal non-zero energy gap of $H_\theta$ (we choose units in which $\hbar=1$).  All upper bounds can be saturated, up to constants, by explicit Hamiltonian control $H_C$; see the details below.
    } 
    \label{tab:1}
\end{table*}

\noindent\begin{table*}
    \centering
    \begin{tabular}{|C{6.2cm}|C{3.8cm}|C{4cm}|C{3.8cm}|}
         \hline\rule{0pt}{1.1\normalbaselineskip} 
           & Dynamical  with
           & Steady state:  & Steady state:   \\
          & initial product state & Diagonal ensemble  &  Gibbs ensemble  \\[1.ex]
         \hline \rule{0pt}{1.2\normalbaselineskip} 
        Bound for $N$-spin probe ($\|H_S \| = \addw N$) & $\mathcal{F} \leq  t^2 \revadd{\addw^2} N^2$    
         & $\mathcal{F} \leq \frac{3+\pi^2}{3} \frac{\revadd{\addw^2} N^2}{E^2}  $     & $\mathcal{F} \leq  \frac{1}{4}\beta^2 \revadd{\addw^2} N^2$   \\ [1.ex]
         \hline\rule{0pt}{1.2\normalbaselineskip} 
         Best-known protocol with  two-body $H_C$ 
         & $\mathcal{F} \approx  \frac{1}{4}t^2 \revadd{\addw^2} N^2$  \hspace{0.5mm} (Result \ref{res:quantum-adv-twobody}) 
         & $\mathcal{F} \approx  \frac{1.34 \revadd{\addw^2} N^{3/2}}{E^2}  $ \hspace{0.5mm} (Result~\ref{res:quantum-adv-twobody-dephasing})    & $\mathcal{F} \approx  \frac{1}{4}\beta^2 \revadd{\addw^2} N^2 $ \hspace{0.5mm} (Result \ref{res:quantum-adv-twobody-gibbs})  \\ [1.ex]
         \hline
    \end{tabular}
    \caption{\textbf{Summary of the main results for sensing a magnetic field with an $N$-spin interacting probe.} \revadd{We assume that $\|H_S\| = \addw N$.} In this work, we show that a central spin model can approach the Heisenberg limit (up to a $1/4$ factor). We also show numerically that with a collective spin coupling as in Eq.~\eqref{eq:all-to-allintro} we can achieve a comparable performance. With a suitable choice of interaction strengths, this model can also saturate the QFI bound for the Gibbs ensemble and achieve an  $N^{3/2}$ scaling for the QFI of the diagonal ensemble.
    } 
    \label{tab:2}
\end{table*}

The most common approach  to overcome the SNL consists of preparing $\rho_0$ as a carefully entangled state, which requires a prior preparation stage, while assuming that $H_C=0$~\cite{qmqi, optical_interferometry_review}. Yet, recent years have witnessed an increased interest in understanding the potential of many-body interactions, i.e. taking $H_C \neq 0$ and $\rho_0$ as a product state, for overcoming the SNL and achieving enhanced metrology~\cite{montenegro2024quantum, macieszczak2016dynamical, Braun2018,Hotter2024Combining,baak2024self,zhou2020quantum}. Within the study of quantum metrology with many-body systems, we distinguish two main scenarios: dynamical and static (or steady state). 

The dynamical scenario corresponds to the situation described above in which information on $\theta$ is encoded into the many-body probe through unitary evolution under~$H_\theta$, and both time~$t$ and the number of particles $N$ are regarded as resources. By properly engineering $H_\theta$ close to a dynamical phase transition, it is possible for the measurement sensitivity $\f$ to scale beyond the SNL~\cite{macieszczak2016dynamical, Tsang2013, Chu2021Dynamic}. This has been shown for the Lipkin-Meshkov-Glick model, obtaining up to $\mathcal{F} \sim t^2 \revadd{\addw^2} N^{1.75}$~\cite{Guan2021Identifying}  as well as more general long-range Ising models~\cite{Shi2024Universal}; see also results for stark localized systems~\cite{He2023Stark,manshouri2024quantum}, fermionic lattices~\cite{Sahoo2024Localization} and quantum-optics systems~\cite{Garbe2020Critical, Chu2021Dynamic, Garbe2022Critical, Ilias2022Criticality, Gietka2022,gorecki2024interplay,Alushi2024Optimality}. Using the Lieb-Robinson bound, it has recently been shown that long-range interactions are required to beat the SNL when starting from a product state~\cite{Shi2024Universal, Chu2023Strong}.

In the static scenario, information on $\theta$ is instead encoded in the steady state of the system, which may be a ground, thermal, or nonequilibrium steady state. \revadd{These states are  naturally robust to environmental noise. } This \revadd{approach} can be connected to the dynamical \revadd{one} by assuming the presence of an environment\footnote{Formally speaking, both the environment's Hamiltonian and the system-environment interaction can be included as a fixed term in $H_C$. In this case, only partial control is available, but we stress that the bound~\eqref{eq:boundCR} remains valid.  }  
and taking the limit $t\rightarrow \infty$ of the dissipative evolution. \revadd{Obviously, in this case bound \eqref{eq:boundCR} becomes uninformative as, in principle, the measurement uncertainty can be arbitrarily small in the limit $t\rightarrow \infty$. However, when restricting  to steady states, the measurement precision is typically finite and has been extensively analyzed for various open systems~\cite{montenegro2024quantum}. }
As in the dynamical scenario, by exploiting phase transitions in the steady state, one can  obtain a measurement sensitivity that scales beyond linear \revadd{in N}~\cite{Zanardi2008, Banchi2014, Rams2018Limits}. It is in principle possible to obtain scalings beyond~$N^2$, but this comes at the expense of a diverging preparation time so that the general bound~\eqref{eq:boundCR} is always respected for any (dissipative) dynamics~\cite{Rams2018Limits, Gietka2021adiabaticcritical}. Within the static approach, numerous works have shown enhancements in the sensitivity of many-body probes close to a critical point, either of ground~\cite{Invernizzi2008, Sarkar2022Free, Salvia2023Critical,montenegro2024quantum,Mukhopadhyay2024Modular}, thermal~\cite{Zanardi2007Mixed, Invernizzi2008, Gammelmark2011, Mehboudi2016, Mehboudi2019,Abiuso_2024Optimal,Yu2024Criticality,Ostermann2024Temperature} or nonequilibrium steady states~\cite{FernandezLorenzo2017,fernandez2018heisenberg, Marzolino2017, Raghunandan2018, DiCandia2023, Ilias2024Criticality}.

Despite this relevant progress, a general understanding of the limits and potential of many-body quantum metrology is far from complete, both in the dynamic and static  scenarios. Relevant questions include the following. 
\begin{itemize} 
\item Can we approach the fundamental dynamical bound in Eq.~\eqref{eq:boundCR}, starting from product states, via physically relevant, time independent, $H_C$ (e.g., limited to two-body interactions)? 
\item Can we derive bounds similar to Eq.~\eqref{eq:boundCR}, but applicable instead to steady-state metrology? 
\end{itemize}
The main goal of this work is to investigate such questions for closed and open systems. For that, we derive bounds   for the 
quantum Fisher information (QFI)~\cite{paris2008quantum},  which sets an ultimate limit on the measurement precision, and investigates their saturability.  We consider time independent $H_\theta$ and follow a twofold approach. 

On the one hand, we characterize the QFI given a generic signal $H_S$ and arbitrary Hamiltonian control $H_C$. For dynamical metrology, we take as an initial state an eigenstate of $H_S$; this covers relevant cases as initial product states when $H_S$ is a sum of local terms and the ground state of $H_S$. Given an $H_S$, we then derive an $H_C$ that approaches the bound in Eq.~\eqref{eq:boundCR} up to a constant $1/4$, which we believe to be optimal. For open systems, we derive a new bound on the QFI applicable to the long time-averaged state or diagonal ensemble~\cite{Rigol2008, Kollar2008, Cassidy2011Generalized, Eisert2015}. Together with the recent upper bound for the QFI of the Gibbs ensemble~\cite{abiuso2024fundamental}, these bounds can be seen as the counterpart of that in Eq.~\eqref{eq:boundCR} for relevant cases of steady state metrology ($t\rightarrow \infty$). We also develop optimal $H_C$ that can saturate these bounds. Our results for generic $H_S$ are summarized in Table~\ref{tab:1}.

On the other hand,  we consider the particular case of magnetometry via a many-body spin probe, i.e.,
\begin{equation}
H_S = \addw S_z\, ,
\end{equation}
with $S_j=\frac{1}{2}\sum_{i=1}^N \sigma_{j}^{(i)}$ and $j=x, y, z$. 
We then look for the control Hamiltonian $H_C$ featuring only two-body interactions that can approach the upper bounds on the QFI (for dynamic and static metrology).  In the case of closed systems, we analytically show that a central spin model can reach the scaling $\revadd{\addw^2}N^2t^2$,  but with a worse prefactor of $1/4$ when compared to the fundamental bound in Eq.~\eqref{eq:boundCR}. Beyond the central spin model, we focus on spin-squeezing control Hamiltonians with two-body couplings of the form 
\begin{equation}
\label{eq:all-to-allintro}
    H_C =  a S_x^2 + b S_y^2 +c S_z^2\, .
\end{equation}
This model involving collective spin variables is very natural in atomic ensembles used in magnetometry and interferometry~\cite{baamara2021squeezingnonlinearspinobservables, Chalopin_2018, 
 Gross_2010,Riedel_2010}. By appropriate choice of the $(a,b,c)$ coefficients, we numerically show that a Heisenberg scaling can be obtained in dynamical metrology (as for the central spin model). We also show that the control in Eq.~\eqref{eq:all-to-allintro} enables the saturation of the QFI bound for Gibbs states. For the diagonal ensemble, we find a scaling $N^{3/2}$ of the QFI for an appropriate choice of $(a,b,c)$. Our main results for magnetometry are summarized in Table~\ref{tab:2}. 

Finally, in the same spin-probe scenario we  study the transient regime to steady states. That is, we introduce appropriate noise models (corresponding to dephasing and thermalization) and evaluate the value of the QFI at all intermediate times leading to equilibration. We again find strong improvements attainable via many-body interactions, and we analyze the different tradeoffs between the scaling of equilibration time and the QFI.

The paper is structured as follows. In Section~\ref{Sec:Background}, we introduce the main quantities of interest. In Section~\ref{sec:dynamical_metro}, we focus on closed systems and discuss various choices of~$H_C$ while contrasting them to the bound in Eq.~\eqref{eq:boundCR}. In Section~\ref{sec:steadystate}, we focus on steady-state metrology of the diagonal and Gibbs ensemble, by deriving both bounds on the QFI and strategies to approach them. In Section~\ref{sec:transient}, we build connections between both previous scenarios, by analyzing the finite-time performance of~$H_C$ that becomes optimal in the steady state regime. Finally, we conclude in Section~\ref{sec:conclusions}.

\section{framework}\label{Sec:Background}

To estimate an unknown scalar parameter $\theta$, we consider an estimator $\hat{\theta}$. This estimator is a random variable dependent on the measurement results $\vec{x}$ sampled from $p(\vec{x}|\theta)$. The precision of an estimator can be quantified by its mean squared error, which is the expected square deviation of the estimator from the true value of the parameter.

Remarkably, the mean squared error of any locally unbiased estimator can be lower bounded with the Cram\' er-Rao bound~\cite{first_quantum_fisher}
\begin{equation}\label{eq: CRB}
    (\Delta \theta)^2 \geq \frac{1}{\mu \,\mathcal{I}}\, .
\end{equation}
Here $\mu$ is the number of \revadd{identical repetitions of the experiment} and $\mathcal{I}= \mathds{E}\left[\partial_\theta \log p(\vec{x}|\theta)\right]$ is the classical Fisher information (CFI), which depends only on the outcome distribution $p(\vec{x}|\theta)$ and its derivative $\partial_\theta p(\vec{x}|\theta)$ at $\theta$. In addition, this inequality can be saturated asymptotically~\cite{assymptotic}. The CFI is thus a natural, estimator-independent, figure of merit to benchmark sensing strategies.

In a quantum framework, the probability distribution $p(\vec{x}|\theta)$ arises by measuring a quantum state $\rho_\theta$, which encodes information on $\theta$.  
Assuming that the optimal measurement is performed, we can substitute the CFI with its quantum version: the QFI $\f$. The QFI can be expressed in terms of the probe state evolved under the parameter-dependent dynamical process $\rho_\theta$ and its derivative with respect to the unknown parameter $\dot{\rho}_\theta =\partial_\theta\rho_\theta$. In its most general form, the QFI is 
\begin{align}
\mathcal{F} &= \Tr[ \mathds{J}^{-1}_{ \rho_\theta}(\dot \rho_\theta) \dot \rho_\theta]\, ,
\end{align}
where we have defined $\mathds{J}^{-1}_{ \rho_\theta}(\dot \rho_\theta)$, to be the solution of the Lyapunov equation
\begin{equation}
    \dot \rho_\theta = \frac{1}{2} \left[\mathds{J}^{-1}_{ \rho_\theta}(\dot \rho_\theta) \rho_\theta + \rho_\theta \mathds{J}^{-1}_{ \rho_\theta}(\dot \rho_\theta)\right]\, ,
\end{equation}
i.e., it is given by the inverse of the Bures multiplication superoperator $\mathds{J}_{\rho_\theta}[A]= \frac{1}{2}(A \rho_\theta + \rho_\theta A)$ applied on $\dot \rho_\theta$.  Note that $\mathds{J}^{-1}_{ \rho_\theta}(\dot \rho_\theta)$ is equivalent to the symmetric logarithmic derivative~\cite{paris2008quantum}. Under some constraints, the form of the QFI is simplified. In particular, when we have a pure state, $\rho_\theta = \ketbra{\psi_\theta}$, the QFI is given by
 \begin{equation}
        \f = 4 \left(\braket{\Dot{\psi}_\theta} - \left| \braket{\Dot{\psi}_\theta}{{\psi}_\theta} \right|^2\right)\, ,
        \label{eq:fishd}
\end{equation}
where $\ket{\Dot{\psi}_\theta} = \partial_\theta \ket{\psi_\theta}$.

\begin{figure}[ht!]
    \centering
    \begin{tikzpicture}
        \pgftext{\includegraphics[width=0.55\linewidth]{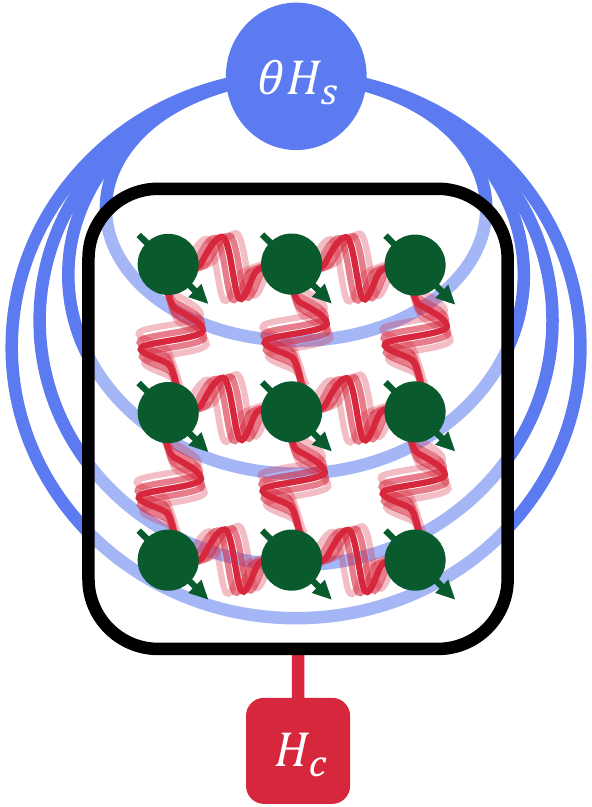}}
        \node at (-1.5,-2.4) {\large Sensor $\rho_\theta$};
    \end{tikzpicture}
    \caption{In this figure we illustrate the process of encoding the signal $\theta$ (via $H_S$, in blue) into the sensor (state, in green), while also adding an extra controlled interaction $H_C$ (in red) to the system to enhance the amount of information we can extract from the signal.}
    \label{fig:possible1_differentsystems}
\end{figure}

Information on $\theta$ is encoded on $\rho_\theta$ via some dynamical process driven by Hamiltonian $H_\theta$ given in Eq.~\eqref{eq:paradigm}. Importantly, in this work we assume that the $H_\theta$ can be split into two contributions (see Fig.~\ref{fig:possible1_differentsystems}): 
\begin{enumerate}[label=(\alph*)]
    \item $H_S$: this corresponds to the signal, i.e., the part of $H_\theta$ encoding the unknown parameter $\theta$.  We assume $H_S$ to be known and fixed, and not subject to control.
    \item $H_C$: this accounts for any other contribution that is $\theta$~independent. We assume (partial) control on $H_C$ to enhance sensitivity~$\f_\theta$. We focus on time independent~$H_C$. Note that, in principle $H_C$ can also contain noncontrollable environmental interactions. 
\end{enumerate}
At this point, a comment is in order. In several places, we seek the optimal $H_C^*$ maximizing $\f_\theta$ for the available control. As expected, $H_C^*$ can implicitly depend on~$\theta$. However, it has to be clear that, given a fixed $\theta$, $H_C^*$ is also fixed and independent of small variations $\delta \theta$ on which the QFI, being a pointwise quantity, directly depends (the conventional method in the local estimation scenario~\cite{abiuso2024fundamental,li2018frequentist}). Furthermore, we stress that because we are focusing on local estimation, the multiplicative dependence of $\theta H_S$ is not restrictive\revadd{~\cite{abiuso2024fundamental}}.

We consider three main scenarios in which information on $\theta$ is encoded upon $\rho_\theta$ via $H_\theta$. We discuss them in what follows. 

\noindent\textbf{Scenario 1: dynamical metrology.} In this scenario we consider the time-evolved state under $H_\theta$ in Eq. \eqref{eq:paradigm}. Consider $\initial$ the initial state of the probe, or sensor, evolving in time as: 
\begin{equation}
\label{eq:DynamicalEncoding}
    \rho_\theta(t) = e^{-\ii tH_\theta}\initial e^{\ii tH_\theta}\, . 
\end{equation}
If we know the time $t$ (with enough precision), then we can use this to infer the value of $\theta$. In particular, if $\rho_\theta(t) = \ketbra{\psi_\theta(t)}$, we can see from Eq. \eqref{eq:fishd} that the QFI scales quadratically with time. Furthermore, in Ref.~\cite{Pang_2014} it is shown that we can write the QFI as 
\begin{equation}\label{eq:fid_psit}
    \f = 4t^2 {\rm Var}(H_{\rm eff})_{\ket{\psi_{\theta}(t)}}\, ,
\end{equation}
where
\begin{align}\label{eq:Heff}
    H_\mathrm{eff} = \int_0^1 e^{-\ii t s H_\theta}H_S e^{\ii   t s H_\theta} d s\, .
\end{align}

In general (even for mixed states), we can bound the QFI via the variance of $H_{\rm eff}$~\cite{boixogeneralised}. This means that for any $\rho_\theta(t)$ the QFI is bounded as 
\begin{align}\label{eq:boundFdyn1}
    \f \leq 4t^2{\rm Var}(H_{\rm eff})_{\rho_{\theta}(t)}\, .
\end{align}
Furthermore, we can trivially show that $|\lambda_{\max,\min}^{\rm eff}|\leq |\lambda_{\max,\min}^{S}|$:
\begin{align}
    \lambda_{\max}^{\rm eff}&=
\max_{\ket{\psi}}\bra{\psi}H_\mathrm{eff}\ket{\psi}=\nonumber\\
&=\max_{\ket{\psi}}\int_0^1 ds \bra{\psi}\e{-\ii t s H_\theta}
H_S \e{\ii t s H_\theta}\ket{\psi}\nonumber\\
&\leq \int_0^1 ds \lambda_{\max}^{S}\, ,
\end{align}
and equivalently for $\lambda_{\min}^{\rm eff}$,  where $\lambda_{\max,\min}^{\bullet}$ 
refers to the maximum or minimum eigenvalue of $H_{\bullet}$. 
Then we just need to apply the fact that $\max_{\rho}{\rm Var}(H)_{\rho}\leq (\lambda_{\max}-\lambda_{\min})^2/4=  \|H_S\|^2/4$. Finally we can now simplify the upper-bound in Eq.~\eqref{eq:boundFdyn1} as 
\begin{align}\label{eq:boundFdyn}
    \f \leq t^2 \left(\lambda_{\max}^{S}-\lambda_{\min}^{S}\right)^2\, ,
\end{align}
and apply this to the bound in Eq.~\eqref{eq: CRB} to obtain
\begin{align}
    (\Delta \theta)^2 \geq \frac{1}{ \mu\,  t^2 \|H_S\|^2 }\, , 
\end{align}
where we recover the bound given in Eq.~\eqref{eq:boundCR} and in Ref.~\cite{boixogeneralised}.

\vspace{2mm}
\noindent\textbf{Scenario 2: steady-state or static metrology.} In many practical scenarios, environmental noise, and lack of time precision, make it desirable to consider alternative sensing scenarios~\cite{montenegro2024quantum}. A well-studied scenario is steady state metrology: measurements are performed in the steady-state $\rho_\theta$ of an open quantum system. Conceptually, this can be seen as a particular scenario of dynamical metrology, in which $t\rightarrow \infty$. In this case, bound~\eqref{eq:boundFdyn} becomes useless, and in this paper, we search for alternative time-independent upper bounds on $\f$. We consider two physically relevant steady states: the time-averaged state (or diagonal ensemble) and the Gibbs state.

\vspace{1mm}
\noindent\textit{Time-averaged state or diagonal ensemble.} When the time evolution is coherent but our time-keeping device is not precise enough, our system will be effectively described by its time average. 
In general, the expression of the time average state will be
\begin{align}
\label{eq:timeAverageState}
    \rho_\theta(T) = \frac{1}{T}\int_{0}^T  e^{-\ii t H_\theta}\initial e^{\ii t H_\theta}d t \, . 
\end{align}
A steady state naturally arises in the long-time limit $T\rightarrow \infty$, in which we can write the time-averaged state as
\begin{align}\label{eq:dephasin}
    \rho_\theta = \mathcal{D}_{H_\theta}(\initial)\, ,
\end{align}
where we have defined the dephasing map or ``pinching'' as 
\begin{align}\label{eq:pinching_map}
    \mathcal{D}_{H_\theta}(\cdot) = \sum_{k}\Pi_k\cdot \Pi_k\, , 
\end{align}
with $\Pi_k$ being the projector into the $k$th eigenspace of~$H_\theta$.

It is worth noting that the long-time averaged state commonly appears in the dynamics of closed many-body systems, as the steady state describing the equilibration of local observables~\cite{Rigol2008, Kollar2008, Cassidy2011Generalized, Eisert2015}. In this context, it is also known as the diagonal ensemble.

\vspace{1mm}
\noindent\textit{Gibbs state metrology.}
We also consider the ubiquitous scenario where the probe is weakly coupled to a thermal bath at temperature $T$. 
In this case, the  steady state is given by the Gibbs state: 
 \begin{align}\label{eq:thermal_state}
     \rho_\theta = \frac{e^{-\beta H_\theta}}{\Tr[e^{-\beta H_\theta}]}
 \end{align}
 with $\beta=(k_{\rm B}T)^{-1}$ and $k_{\rm B}$ the Boltzmann constant. 
 The thermal state depends only on $H_\theta$ and is independent of the initial state.

 For both the diagonal and Gibbs ensembles, we will look for optimal Hamiltonian controls $H_C$ as well as upper bounds on $\mathcal{F}$   (see Tables \ref{tab:1} and \ref{tab:2}). 

\vspace{2mm}
\noindent{\bf Scenario 3: transient regime.} Finally, we also study how the different regimes connect, see Section~\ref{sec:transient} below. For that, we  consider different noise models via the Lindblad master equation
\begin{align}\label{eq:limbladian_general}
     \dot{\rho}_{\theta}(t) =& -\ii [H_\theta,\rho_{\theta}(t)] \\
     &+ \sum_i \gamma_i \left( L_{\theta, i} \rho_{\theta}(t) L_{\theta, i}^\dagger -\frac{1}{2}\{\rho_{\theta}(t),L_{\theta, i}^\dagger L_{\theta, i}\}\right)\, .\nonumber
 \end{align}
 We characterize how the QFI behaves in the transitions between the different regimes presented above assuming that the signal $H_S$ is fixed and $H_C$ can be externally controlled (note that this might induce changes in  $L_{i,\theta}$).

\section{Dynamical Metrology} 
\label{sec:dynamical_metro}

We devote this section to studying dynamical metrology under a coherent evolution by $H_\theta$ and how to enhance it via $H_C$. In this scenario we study two settings. First, we consider a general encoding, where both the signal $H_S$ and the controllable term $H_C$ are arbitrary Hermitian operators. Second, we focus on the interacting spins probe, where we assume the signal $H_S$ to be a sum of local terms, particularly a sum of spins, and we restrict $H_C$ to be a sum of two-body terms.

\subsection{General encoding}
Here we study the effect of $H_C$ on the task of estimating $\theta$ via some noiseless dynamics. We start by giving a simple and general expression of the QFI for a pure state $\ket{\psi_{\theta}(t)}$ that has evolved for some time $t$ under the Hamiltonian $H_\theta$.

\begin{result}[Dynamical quantum Fisher information]\label{res:pinched}
   Let $H_\theta$ be the Hamiltonian of the form of Eq.~\eqref{eq:paradigm}, $H_\theta =  \theta H_S + H_C$, where $\theta$ is the unknown parameter, $H_S$ is the signal, and $H_C$ is a controlled term. Let the final state after the evolution be $\ket{\psi_\theta(t)} = e^{-\ii t H_\theta}\ket{\psi}$ for some initial pure state $\ket{\psi}$. The QFI of this state is given by the variance
    \begin{equation}
    \f = 4 t^2 \var( H_P )_{\ket{\psi}} + \order{t\revadd{\|H_S\|^2/\Delta_g}} \ 
\label{eq:Fisher_Pinched}
\end{equation}
of the pinched signal Hamiltonian
\begin{equation}\label{eq:refpinched}
H_P = \mathcal{D}_{H_\theta}(H_S)= \sum_k \Pi_k H_S \Pi_{\revadd{k}}\, .
\end{equation}
Here, $\mathcal{D}_{H_\theta}$ is the self-adjoint pinching (dephasing) map in the eigenbasis of $H_\theta$ introduced in Eq. \eqref{eq:pinching_map}, and the $\Pi_k$ are the projectors on the eigenspaces of $H_\theta = \sum_{k} E_k \Pi_k$ associated with different eigenvalues. 

\revadd{In Eq.~\eqref{eq:Fisher_Pinched}, we define $\Delta_g =\min_{i,j} |E_i-E_j|$ as the minimum non-zero energy difference between energy levels of $H_\theta$. } 
\end{result}

\begin{proof} 
We provide an outline of the proof below; for further details, see Appendix~\ref{app: proof 1}.
Following Ref.~\cite{Pang_2014}, we use the integral representation of the derivative of an exponential to write the derivative of the evolved state appearing in Eq.~\eqref{eq:fishd} as
\begin{equation}
    \partial_\theta \ket{\psi_\theta{(t)}} = -\ii t H_{\rm eff} \ket{\psi_\theta{(t)}}\, ,
\end{equation}
where generator $H_{\rm eff}$ was introduced in Eq.~\eqref{eq:Heff}
\begin{equation}
    H_\mathrm{eff} = \int_0^1 e^{-\ii t s H_\theta}H_S e^{\ii   t s H_\theta} d s\, .
\end{equation}
Next, via Eq.~\eqref{eq:fishd}, we can write the QFI as in Eq.~\eqref{eq:fid_psit}
\begin{equation}\label{eq:QFIHeff}
    \f = 4t^2\var (H_\mathrm{eff})_{\ket{\psi_\theta(t)}}\, .
\end{equation}

If we express the full many-body Hamiltonian in its spectral decomposition, $H_\theta = \sum_k \Pi_k E_k$, it can be seen that in the limit of large times, all coherences in Eq.~\eqref{eq:Heff}   with different energies $E_k\neq E_{k'}$ dephase and become vanishingly small. Consequently, to leading order, the effective Hamiltonian simplifies to a pinched version of the signal Hamiltonian: 
\begin{equation}\label{eq:A_as_sum}
     H_{\mathrm{eff}} =  \sum_k \Pi_k H_S \Pi_k +\order{\frac{\revadd{\|H_S\|^2}}{\revadd{\Delta_g}t}}\, 
\end{equation}
Here we defined $\Delta_g = \min_{i,j}|E_i-E_j|$ as the minimum non-zero energy difference between energy levels of $H_\theta$. Then Eq.~\eqref{eq:fid_psit} becomes
\begin{align}\label{eq:final_eq_of_scketch_pinched}
    \f =& 4t^2 \var ( H_P)_{\ket{\psi_\theta(t)}} + \order{t\revadd{\|H_S\|{^2}/\Delta_g}}\, .
\end{align}
Here, to bound the correction term, we used the Proposition~\ref{prop:gersh-energy} in Appendix~\ref{app:circleth}, based on the Gershgorin circle theorem \cite{circle}.
Finally, since $[H_P,H_\theta] = 0$, the variance remains time independent, allowing us to rewrite Eq.~\eqref{eq:final_eq_of_scketch_pinched} as Eq.~\eqref{eq:Fisher_Pinched}, thereby completing the proof.
\end{proof}\bigskip

\revadd{We already discussed that in general, this approach cannot surpass the Heisenberg scaling of the QFI, $\f\propto \addw ^2 N^2 t^2$. In particular, we illustrate this point in Eq.~\eqref{eq:boundFdyn}. }

Even though the maximum achievable resolution is not changed via the control term $H_C$, this result highlights how these controlled interactions can enhance (or worsen if one is not careful) the sensitivity for a particular initial state. \revadd{In particular, we point out that, without loss of generality, the control term $H_C$ can modify $\Delta_g$ as desired, and thus by increasing the strength of $H_C$, we can make the residual term go to zero.}

\begin{result}\label{res:pinchedlocal}
    Let $H_\theta$ be the Hamiltonian of the form $H_\theta =\theta H_S + H_C $, where $\theta$ is the unknown parameter, $H_S$ is the signal, and $H_C$ a control Hamiltonian on which we have full control. Let sensor $\ket{\psi}$ be in the ground state of signal $H_S$, $\ket{\psi} = \ket{\Phi_\downarrow}$. Let it evolve coherently with $H_\theta$ for a time $t$. The final state after the evolution is $\ket{\Phi_{\downarrow,\theta}(t)} = e^{-\ii t H_\theta} \ket{\Phi_\downarrow}$. Then for a good choice of $H_C$ the following QFI is attainable
    \begin{equation}
    \f = t^2\frac{\|H_S\|^2}{4} + \order{t\revadd{\|H_S\|^2/\Delta_g}}
    \label{eq:finch_res2}
\end{equation}
with $\|H_S\| = \lambda_{\rm max}^{S} - \lambda_{\rm min}^{S}$ and $\lambda_{\rm max}^{S}$ ($\lambda_{\rm min}^{S}$) the maximal (minimal) eigenvalue of $H_S$.
\end{result}

\begin{proof} 
We can devise a particular strategy that attains this bound. We write the signal Hamiltonian as 
\begin{equation}
    H_S=\lambda_{\max}^{S} \ketbra{\Phi_\uparrow}{\Phi_\uparrow}+ \lambda_{\min}^{S}\ketbra{\Phi_\downarrow}{\Phi_\downarrow}+ {H}_\perp\, ,
\end{equation}
where we have defined $\ket{\Phi_\uparrow}$ to be the highest energy state,  
and $\ket{\Phi_\downarrow}$ the ground state. Finally, $ H_\perp$ is orthogonal to the subspace spanned by $\{\ket{\Phi_\downarrow}, \ket{\Phi_\uparrow}\}$.

Now, choose the control Hamiltonian  $H_C$ such that the total Hamiltonian has spectral decomposition (or any other Hamiltonian with the same eigenspaces)
\begin{equation}
    H_\theta=\revadd{\frac{\Delta_g}{2}(}\ketbra{\Phi_{\scriptscriptstyle \nearrow}} - \ketbra{\Phi_{\scriptscriptstyle \nwarrow}}\revadd{)} + H_\perp\,
\end{equation}
with $\ket{\Phi_{\scriptscriptstyle \nearrow}}=\cos\tfrac{\pi}{8}\ket{\Phi_\uparrow}+\sin\tfrac{\pi}{8}\ket{\Phi_\downarrow}$ and 
$\ket{\Phi_{\scriptscriptstyle \nwarrow}}=\sin\tfrac{\pi}{8}\ket{\Phi_\uparrow}-\cos\tfrac{\pi}{8}\ket{\Phi_\downarrow}$. It is important to notice that $H_\theta$ is still dependent on $\theta$. However, this dependence is hidden by the control term $H_C$, as discussed in Section \ref{Sec:Background}. 

Now it is straightforward  to show that
\begin{equation}\label{eq:effham}
    H_P=\sum_i P_i H_S P_i= h_P + H'_\perp \, 
\end{equation}
with $h_P$ and $H_\perp'$ living in orthogonal subspaces (i.e. $h_P H_\perp '=0$), and 
\begin{align}
    h_P=\frac{\Delta_+}{2}\1 + \frac{\Delta_-}{2\sqrt{2}}\widehat\sigma_{{}_\nearrow} \, ,
\label{eq:hpPauli}
\end{align}
where $\widehat\sigma_{\scriptscriptstyle \nearrow}=\frac{1}{\sqrt{2}}(\widehat\sigma_x+\widehat\sigma_z)$, with
 the Pauli matrices  written in the $\{\ket{\Phi_{\uparrow}},\ket{\Phi_{\downarrow}}\}$ basis, and $\Delta_\pm=\lambda_{\max}^{S}\pm\lambda_{\min}^{S}$. Thus, the variance for the state $\ket{\Phi_\downarrow}$ is readily obtained:
\begin{equation}
\var(H_P)_{\downarrow}=\var(h_P)_{\downarrow}=
    \frac{\Delta_-^2}{8}\var(\widehat\sigma_{\scriptscriptstyle\nearrow})=\frac{\Delta_-^2}{16}\, .
\end{equation}
The last equality easily follows from the properties of  Pauli matrices:
$\langle\widehat\sigma_{\vec{n}} ^2\rangle_\downarrow=\langle\1\rangle_{\downarrow}=1$ and $\langle\widehat\sigma_{\vec{n}}\rangle_{\downarrow}=
\vec{n}\cdot\hat{z}=n_z$
and $n_z=\frac{1}{\sqrt{2}}$ for $\sigma_{\scriptscriptstyle\nearrow}$.
From \eqref{eq:Fisher_Pinched} we  get the QFI
\begin{equation}
\label{eq:optfisher}
    \f = t^2 \frac{\left(\lambda_{\max}^{S}-\lambda_{\min}^{S}\right)^2}{4} +\order{t\revadd{\|H_S\|^2/}\revadd{\Delta_g}}\, .
\end{equation} 
\end{proof}
Result~\ref{res:pinchedlocal} shows that for a good, time-independent, choice of control $H_C$, it is possible to reach the maximum scaling of the QFI discussed in Eq.~\eqref{eq:boundFdyn}. Note that there is still a factor of $4$ with the maximum QFI achievable. Notably, this is achieved using the ground state of the signal~$H_S$, whose QFI would vanish in the absence of~$H_C$.

\revadd{The time-evolved state takes the form 
\begin{align}\label{eq:ghztype2q}
    e^{-\ii t H_\theta}\ket{\Phi_\downarrow} = \sin\frac{\pi}{8}\ket{\Phi_{\scriptscriptstyle \nearrow}} - e^{-\ii {\Delta_g}t}\cos\frac{\pi}{8}\ket{\Phi_{\scriptscriptstyle \nwarrow}}\,.
\end{align}

It is interesting to note that  this state at hand fluctuates between a highly entangled state (a GHZ-like for $\Delta_g t=\pi+2\pi n,\,n\in\mathbb{N}$) and a non entangled state (for example for $\Delta_g t= 2\pi n,\,n\in\mathbb{N}$). Nevertheless, the QFI is proportional to $\|H_S\|^2$ for all $t$, regardless of the time-evolved state being entangled or not. This highlights a crucial difference between the standard noninteracting metrology protocols (where the QFI only grows as $\|H_S\|^2$ when the state is entangled) and the considered many-body setup. In our scenario,  the parameter is encoded into the state in a nontrivial fashion due to the presence of interactions, so that even separable states can show a  sensitivity beyond shot noise -- at least for a period of time. }

\subsection{Interacting spins probe}\label{sec:spin_network}
So far, our results have not explicitly referenced the structure of the signal or control Hamiltonians. In this section, we focus on a physically relevant setting where the sensor consists of multiple constituents and the signal Hamiltonian acts independently on each of them. For concreteness, we consider the paradigmatic case of magnetometry, where the sensor is composed of $N$ spins $\tfrac{1}{2}$ paricles, with a signal Hamiltonian given by  $H_S = \addw S_z = \frac{\addw}{2}\sum_{i=1}^N\sigma_z^{(i)}$. 

 In this context, the number of spins, $N$, naturally emerges as a key resource against which the estimation precision is benchmarked. The seminal works on quantum metrology (see, e.g., Ref.~\cite{original_heisenberg} and the references therein) demonstrated that preparing a suitably entangled probe state, allowing it to evolve under the external field (encoding the unknown parameter via the unitary \(U = e^{-\ii t \theta \addw H_S}\)), and performing an appropriate measurement one can estimate the parameter with a mean squared error scaling as $N^{-2}$ --saturating the upper bound in Table~\ref{tab:1} with $\mathcal{F}=t^2 \revadd{\addw^2} N^2$). This surpasses the classical shot-noise limit of $N^{-1}$, the best precision attainable with separable probe states.
 
However, preparing and maintaining the highly entangled probe state requires further resources both in preparation time and in the ability to implement controlled multiqubit interactions. 
Our results above suggest an alternative quantum metrology strategy where the required entanglement is not preprepared but instead emerges dynamically due to the presence of spin interactions ($H_C$), which act alongside or concurrently with the signal Hamiltonian (see also Ref.~\cite{Hayes2018Making}).

We can immediately use Result~\ref{res:pinchedlocal} to show that such concurrent metrology strategies can reach the optimal scaling, specifically $\mathcal{F}=\tfrac{1}{4}t^2 \revadd{\addw^2} N^2$, with an initial product state $\ket{\Phi_\downarrow}=\ket{1}^{\otimes N}$ (where $\sigma_z\ket{1} = -\ket{1}$) by the application of suitable control Hamiltonian. In addition, as shown in Appendix~\ref{app:proof_measurement_general}, local 
measurements suffice to attain this bound (see also Ref.~\cite{Zhou_2020} for more general proof of the attainability of the QFI by LOCC measurements, i.e. measurements that only involve local measurements and classical communication). This means that all the generation and detection of quantum correlation can be delegated to the control Hamiltonian.

These promising results raise the practical question of how to implement the required control Hamiltonian using physically realizable resources. To address this, we now focus on systems governed by two-body interactions, which are more readily available in laboratory settings. 

\begin{result}\label{res:quantum-adv-twobody}
    Let $H_\theta$ be the Hamiltonian of the form $H_\theta = \theta H_S+ H_C $, with $H_S = \frac{\addw}{2}\sum_{i=1}^N \sigma_z^{(i)}$ a local Hamiltonian, and $H_C$ the control term with only two-body interactions.
    That is,
   \begin{equation}\label{eq:spin net}
       H_C  =\sum_{k,l}\sum_{i,j} \alpha^{(i,j)}_{k,l} \sigma^{(i)}_k\otimes\sigma^{(j)}_l\, .
    \end{equation}
    Let the initial state be a product state. Then, by a proper choice of $H_C$, we show that the QFI behaves as 
    \begin{equation}
        \f =  \nu t^2 \revadd{\addw^2} N^2 +\revadd{\order{t \addw^2 N^2/\beta}}\, ,
    \end{equation}
    where $\nu\in(0,1]$ depends on the setting. We show this behavior for a central spin probe ($\nu = 1/4$) and for a spin-squeezing model as given in Eq.~\eqref{eq:all-to-allintro} ($\nu \in (0.14,0.2)$). Finally, $\beta$ is a parameter of the control Hamiltonian that dictates its strength.
\end{result}

This result illustrates the possibility of reaching the maximum scaling of the QFI  via two-body interactions. 

\subsubsection{Central-spin model}\label{sec:central_spin}
First, we focus on a central-spin model. This consists of a two-body system in which a single particle interacts with all the (otherwise noninteracting) remaining $N-1$ particles. 
Central-spin models have a long history of interest, as they can be theoretically solved in various configurations~\cite{gaudin1976diagonalisation,breuer2004non-markovian,hutton2004mediated,bortz2007exact} and can be used to model systems of particular physical relevance, such as the
interaction of a single electron with surrounding nuclear spins~\cite{schliemann2003electron} (as in, e.g., quantum dots), as well as diamond nitrogen vacancies~\cite{gurudev2007quantum}.
These models constitute a promising playground for the study of quantum
Hamiltonian control~\cite{arenz2014control}, and
have been proposed as archetypes of quantum memories~\cite{denning2019collective} and quantum batteries~\cite{liu2021entanglement} enhanced via collective effects.
More recently, classical central-spin models have been shown to feature collective advantages in temperature estimation~\cite{Abiuso_2024Optimal}, as well as in thermal protocols~\cite{rolandi2023collective}.

For our purposes we thus consider the $\theta$-encoded Hamiltonian to be $H_S = \addw S_z = \frac{\addw}{2}\sum_{i=1}^{N}\sigma_z^{(i)}$. Indeed, the signal we study is the typical magnetic field component in the $Z$ direction. Consider now the control Hamiltonian
\begin{align}\label{eq: central spin}
   H_C=\alpha \ketbra{0}{0}\otimes\mathbf{1}^{(N-1)}+\beta \ketbra{1}{1}\otimes S_x^{(N-1)}\;,
\end{align}
where $\ketbra{0}$ and $\ketbra{1}$ act on the central spin while $\mathbf{1}^{(N-1)}$ and $S_x^{(N-1)}\equiv \frac{1}{2}\sum_{i=2}^{N}\sigma_x^{(i)}$ act on the remaining ones.
It is then straightforward to diagonalize $H_\theta=\theta H_S + H_C$ for $\theta=0$ (notice that $\theta\neq 0$ can always be reabsorbed by adding local $\sigma_z^{(i)}$ terms in the control $H_C$). Furthermore, we choose the probe's preparation to be
\begin{align}\label{eq:initial_central}
    \ket{\psi} = \ket{+}\otimes\ket{0}^{\otimes( N-1)}\, ,
\end{align}
where $\ket{+}$ is the eigenstate of $\sigma_x$ with positive eigenvalue, and $\ket{0}$ is the equivalent eigenstate of $\sigma_z$. The time-evolved state for is of the form 
\begin{align}
    e^{-\ii tH_{\theta}}\ket{\psi} = \frac{1}{\sqrt{2}}\left[ \ket{0}^{\otimes N} + e^{-\ii t \alpha }\ket{1} \left( e^{-\ii t \beta \sigma_x}\ket{0} \right)^{\otimes N-1}\right],
\end{align}
that is,  conditional on the state of the central spin being at $\ket{1}$, the remaining spins undergo a coherent rotation around the x axis.
 As in Eq.~\eqref{eq:ghztype2q}, the state transitions between a GHZ-like entangled state and a product state. Yet, again, as we will shortly  see, the QFI growth remains proportional to $\|H_S\|^2$ at all times, regardless of the entanglement present in the state at that time. Once more, this  showcases the  subtle relation between the QFI and entanglement in our setup. The necessary quantum correlations for a quantum enhanced precision manifest here in the effective  Hamiltonian \eqref{eq:Heff}, which, can also be computed exactly for all times:
\begin{align}\label{eq:Heffstar}
    H_\mathrm{eff} &=\frac{\omega}{2} \sigma_z\otimes\mathbf{1}+ \omega\int_0^1 \left(\ketbra{0}{0}\otimes S_z +
    \ketbra{1}{1}\otimes S_\phi\right)d s\nonumber\\
    &=\frac{\omega}{2}\sigma_z\otimes\mathbf{1}+ \omega \ketbra{0}{0}\otimes S_z +
    \frac{\omega}{\beta t}\ketbra{1}{1}\otimes \bar S 
\end{align}
with $S_\phi=\cos(\beta s t) S_z-\sin(\beta s t) S_y$ and
    $\bar S=\sin(\beta t) S_z+(1-\cos(\beta t))S_y $.
As expected, we find that the effective Hamiltonian converges to the pinched Hamiltonian $H_P$ given by 
\begin{align}\label{eq:HPstar}
H_P=\omega \ketbra{0}{0}\otimes S_z+ \frac{\omega}{2}\sigma_z\otimes \mathbf{1}^{(N-1)}\;.
 \end{align}
 which indeed exhibits quantum correlations.
 Eq.~\eqref{eq:Heffstar} shows that the correction term  decreases inversely proportional to time, at a rate that can be arbitrarily increased by tuning the strength of the control Hamiltonian:
    \begin{align}
        H_\mathrm{eff} = H_P + \order{\frac{\omega N}{\beta t}}\, .
    \end{align}
Here we have used that  $\|\bar S\| = \order{N}$. 

We can finally use Eq.~\eqref{eq:QFIHeff} to compute the quantum Fisher information, which quantifies the precision in estimating $\theta$, to find that
\begin{equation}
    \f = t^2{\addw^2}\frac{(N+1)^2}{4} + \order{\frac{t{\addw^2}N^2}{\beta}}\, ,
\end{equation}
 again attaining the optimal quadratic scaling in $N$ and $t$.

\subsubsection{Spin-squeezing Hamiltonian}\label{sec:all_to_all_seciton}

Now we focus on the case of a control Hamiltonian $H_C$ of the form~\eqref{eq:all-to-allintro}. Assume a signal of the form of $\addw S_z$, the total Hamiltonian ruling the dynamics of the system is now
\begin{equation}\label{eq: spin squeez}
    H_\theta(a,b,c) = \theta S_z + a S_x^2 + b S_y^2 + c S_z^2 \,,
\end{equation}
which is a particular case of Eq. \eqref{eq:spin net} with collective (all-to-all) coupling $\alpha_{x,x}^{(i,j)}=\frac{a}{4},\alpha_{y,y}^{(i,j)}=\frac{b}{4}$ and $\alpha_{z,z}^{(i,j)}=\frac{c}{4}$.
For this interaction, and with a good choice of parameters $(a,b,c)$ and (product) initial state, we show that the QFI scales quadratically with the number of qubits $N$. Furthermore, we can give analytical guarantees for a superlinear scaling with the number of particles $N$. 

First, we show that we can achieve $\f\sim t^2\revadd{\addw^2}N^2$. We do this with the parameters $(a^*,b^*, c^*) = (a, 1/\sqrt{N}, 0)$, i.e. with the following Hamiltonian
\begin{equation}
    H_{\theta}(a^*,b^*, c^*)= \theta \addw S_z + a S_x^2 + \frac{1}{\sqrt{N}} S_y ^2\, 
\end{equation}
where $a$ needs to be a large constant such that we are in the regime of $\theta\addw/a\approx 0$.  Furthermore, we need a particular initial (product) state of the form:
\begin{equation}
    \ket{\vartheta^*, \phi^*}^{\otimes N} = \ket{\frac{\pi}{4}, \frac{\pi}{2}}^{\otimes N}\, .
\end{equation}

\begin{figure*}[ht!]
    \centering
    \includegraphics[width = 0.99\textwidth]{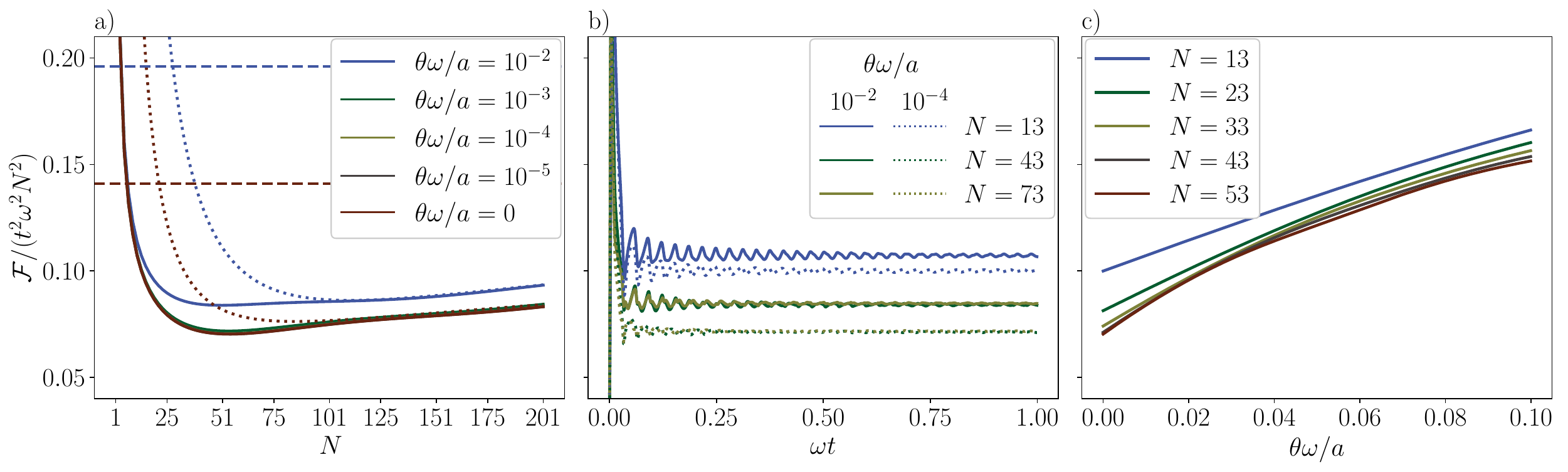}
    \caption{Representation of the different relations of $\f/(t^2 \revadd{\addw^2} N^{2})$ with respect to important parameters of the problem. We use a value of $a = 100$. We only plot $N$-odd values of the number of particles (the value is 0 otherwise). \textbf{a)} Relation between $\f/(t^2 \revadd{\addw^2}N^{2})$ and the number of particles used (recall that $N$ is always odd) for different ratios $\theta\addw/a$. The simulation is performed in a time $\addw t = 1000$. It is clear in the figure that we reach maximum scaling $\f \sim t^2 \revadd{\addw^2} N^2$. Two regression functions (dotted lines) of the form $f_{\rm reg}(N) = k_1 + k_2/\sqrt{N} + k_3/N$ are shown with their asymptotic convergence (dashed lines); we show $\theta\addw/a=0$ in brown, and $\theta\addw/a=10^{-2}$ in blue. The regressions are $f_{\theta\addw/a=0}(N) = 0.141 - 1.193/\sqrt{N} + 5.498 /N, \ f_{\theta\addw/a=10^{-2}}(N)=0.196 - 2.311/\sqrt{N} + 12.159 /N$. We also plot the limits of the regression functions with dashed lines. Both of them converge to the maximum scaling in the limit of a large number of particles. \textbf{b)} Relation of $\f/(t^2 \revadd{\addw^2} N^{2})$ with $\addw t$ for different values of $\theta\addw/a$ and $N$. It can be seen that both parameters do not affect how rapidly we reach the approximate time-independent QFI. \textbf{c)} Relation between $\f/(t^2\revadd{\addw^2} N^2)$ and $\theta\addw/a$ for the different numbers of particles $N$. The simulation is performed in a time $\addw t = 1000$.}
    \label{fig:ml_rescaling}
\end{figure*}

Under these constraints and for an odd number of particles $N$ we can numerically verify as seen in Fig. \ref{fig:ml_rescaling}a) that the QFI scales as $\f\sim t^2\revadd{\addw^2} N^2$, or in other words, we find the maximal scaling in $N$ that the signal allows. Furthermore, in this figure, we study the robustness of the protocol when moving away from the aforementioned regimes, i.e. different ratios of $\theta\addw/a$, and the possible change with the number of particles $N$. Finally, we analyze the convergence to the steady-state regime of the QFI, as proven in Result \ref{res:pinched}.

For a simpler choice of parameters, we can prove a superlinear scaling of the QFI with the number of particles. In particular, $\f \sim t^2 \revadd{\addw^2} N^{3/2}$. To achieve this we need $b=0,\,c = 0$, i.e.,
\begin{equation}
    H_{\theta}(a,0, 0)= \theta \addw S_z + a S_x^2\, .
\end{equation}
Then we choose an initial state of the form of product state of eigenstates of $\sigma_y$, i.e. $\ket{\psi} = \ket{+y}^{\otimes N}$. Finally, we also require the number of particles to be odd. Then we can analytically prove that in the large time regime\revadd{, for $\theta = 0$} the QFI has the form
\begin{equation}
        \f = t^2 \revadd{\addw^2} \frac{N^{3/2}}{\sqrt{2\pi}}+\order{t\revadd{\omega^2 N^2/a}}\ .
        \label{eq:fish2b}
\end{equation}
The proof can be found in Appendix \ref{app:oneaxistwisting}. Furthermore, in \figg{evolution2body} we show how sensible these results are to different ratios of $\theta\addw/a$ and different numbers of qubits $N$, and how fast the convergence to a time-independent regime proven in Result \ref{res:pinched} is. Furthermore, the state resulting from the dynamical evolution at $\theta = 0$ oscillates between a product and a spin-squeezed state, i.e. $e^{-\ii t a S_x^2}\ket{+y}^{\otimes N}$. Similarly to the cases before, we see that the QFI surpasses the SNL even when the state is not entangled.

\begin{figure*}[ht!]
		\centering
		\includegraphics[width=0.99\linewidth]{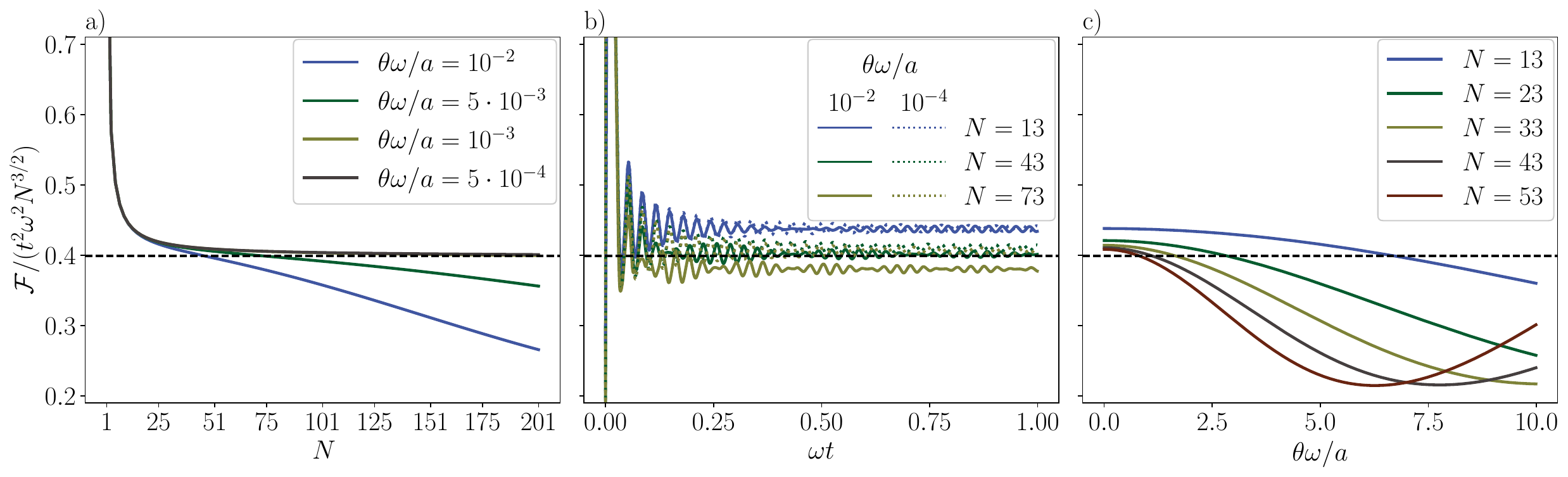} 
     \caption{
     Representation of the different relations of the normalized QFI ($\f/(t^2 \revadd{\addw^2} N^{3/2})$) with parameters of the problem. The dashed line represents the QFI from \eqq{fish2b}. Furthermore, we use a value of $a = 100$.  \textbf{a)} Relation between the $\f/(t^2 \revadd{\addw^2} N^{3/2})$ and the number of particles used (recall that $N$ is always odd) for different ratios $\theta\addw/a$. The simulation is performed in a time $\addw t = 1000$. We see that for a smaller number of particles, we do not require $\theta\addw/a$ to be small. This changes when the number of particles increases. \textbf{b)} Relation of $\f/(t^2\revadd{\addw^2} N^{3/2})$ with $\addw t$ for two different values of ${\theta}\addw/{a}$ and $N$. Changing $\theta\addw/a$ does not affect how fast we reach the time-independent QFI. However, there is a small effect due to $N$.  \textbf{c)} Relation between $\f/(t^2 \revadd{\addw^2} N^{3/2})$ and $\theta\addw /a$ for different number of particles. The simulation is performed in a time $\addw t = 1000$.}
    \label{fig:evolution2body}
\end{figure*}

Finally, we show that the optimal measurement (i.e. when the CFI and the QFI are equal), is just $S_x$. Intuitively, this is the observable that maximizes the CFI as it is the one that will potentially measure the most change from $S_z$ without being affected by the one-axis twisting caused by $S_x^2$. A proof can be found in Appendix \ref{app:measurement}.

\section{Steady-state metrology}
\label{sec:steadystate}

So far, we have considered the evolution of the probe to be coherent and focused on the $t^2$ scaling of the QFI in such a dynamical scenario. Nevertheless, in the limit of large~$t$, both environmental noise and timekeeping imprecision are expected to eventually break the unitarity of evolution. \revadd{In such a case, a particularly relevant class of states is steady states, namely,  the asymptotic state reached in the limit $t\rightarrow \infty$ of the time evolution. These states are particularly appealing due to their natural robustness to noise and their   potential as quantum sensors has been characterized for different models~\cite{montenegro2024quantum}. Our purpose is to derive upper limits on their QFI (given arbitrary control) and analyze its reachability\footnote{\revadd{Note that despite the Heisenberg limit diverges in the limit $t\rightarrow \infty$, the QFI of steady states typically remains finite. This is in contrast to unitarily time-evolved states whose QFI can in principle grow arbitrarily with $t$ (saturating this QFI in practice would  however require increasingly high dynamical control and measurements precision). }}. Specifically, } we consider the following two canonical and physically-motivated classes of steady states: those induced by \textit{A) dephasing} and \textit{B) thermalization}.

\subsection{Dephasing metrology: the case of the diagonal ensemble }
\label{subsec:dephasing}

As treated in Section~\ref{sec:dynamical_metro}, the long-term evolution governed by Hamiltonian $H_\theta$ in Eq.~\eqref{eq:paradigm} leads to Fisher information that is governed by the pinching (dephasing) of signal $H_S$ in the eigenbasis of $H_\theta$. That is, the dynamical QFI then corresponds to the variance of the pinched signal Hamiltonian on the initial state, Eq.~\eqref{eq:Fisher_Pinched}. However, to resolve such dynamical susceptibility in the final state one needs to have an increasingly precise timekeeping device. In contrast, if the timekeeping precision is finite, in the large-$t$ limit the system will be effectively described by the steady state of the dephasing map (pinching) applied on the initial state itself as shown in Eq.~\eqref{eq:dephasin}. In sharp contrast to the dynamical case this state has no time dependence. Nevertheless, since the eigenbasis of $H_\theta$ depends on the parameter so does the steady-state $\rho_\theta$.

We now assume that the Hamiltonian $H_\theta$ has a non-degenerate spectrum around $\theta$, with a minimal energy gap $E$. When the energy gap is closing, the dephasing map in Eq.~\eqref{eq:dephasin} becomes discontinuous and 
the QFI of $\rho_\theta$  diverges. For a nondegenerate $H_\theta$ the dephasing map is thus of the form
\begin{equation}\label{eq: rho dephasing}
    \rho_\theta = \mathcal{D}_{H_\theta}(\initial)=  \sum_k \ketbra{\varphi_k^\theta} \initial \ketbra{\varphi_k^\theta},
\end{equation}
where $\{\ket{\varphi_k^\theta}\}$ is the eigenbasis of $H_\theta$. Here, first-order perturbation theory yields 
\begin{equation}\label{eq: eig perturation}
    \ket{\dot \varphi_k} = \sum_{j\neq k} \ket{\varphi_j} \frac{\bra{\varphi_j} H_S \ket{\varphi_k}}{E_k-E_j},
\end{equation}
where we dropped the $\theta$ index and introduced $E_k = \bra{\varphi_k} H_\theta \ket{\varphi_k}$. This can be conveniently rewritten as 
\begin{equation}\label{eq: S of Hs}
\ket{\dot \varphi_k} = -\ii S \ket{\varphi_k}   \quad \text{with} \quad S:= \ii \sum_{j\neq k} \frac{\ketbra{\varphi_j} H_S \ketbra{\varphi_k}}{E_k-E_j}
\end{equation}
where $S$ is Hermitian $S^\dag =S$ and ``off diagonal" $\bra{\varphi_k} S \ket{\varphi_k}=0$. That is, $S$ is the generator of the local rotation of the eigenvectors $\ket{\varphi_k}$ of $H_\theta$.

To bound the QFI of the dephasing channel in Eq.~\eqref{eq: rho dephasing} we proceed in two steps. First, we derive a general bound in terms of operator $S$, valid whenever $\ket{\dot \varphi_k} = -\ii S \ket{\varphi_k}$. Second, we compute the bound in terms of $H_S$ for the specific form of $S$ in Eq.~\eqref{eq: S of Hs}.\\

\paragraph{QFI of general dephasing maps}-- 
Observe that the output of the dephasing channel $\rho_{\theta}=\mathcal{D}_{H_\theta}(\initial)$ in Eq.~\eqref{eq: rho dephasing}  with $\ket{\dot \varphi_k} = -\ii S \ket{\varphi_k}$   reads
\begin{align}
\dot \rho_\theta &=  - \ii  [S, \mathcal{D}_{H_\theta}(\initial)]+\ii\,  \mathcal{D}_{H_\theta}([S,\initial])
= \dot \rho_{\rm ext} + \dot \rho_{\rm int}\, .
\end{align}
Clearly, $\rho_\theta$ and  $\dot \rho_{\rm int}= \ii \mathcal{D}_{H_\theta}([S, \psi])$ are diagonal in the dephasing basis, while $\dot \rho_{\rm ext}= -\ii [S, \mathcal{D}_{H_\theta}(\initial)]$ is off-diagonal. This decomposition can be used to simplify the QFI as
\begin{align}\label{eq: QFI def 1}
    \mathcal{F} &=\Tr[(\dot{\rho}_{\rm ext}+\dot{\rho}_{\rm int})\mathbb{J}^{-1}_{\rho_\theta}[\dot{\rho}_{\rm ext}+\dot{\rho}_{\rm int}]]\, .
\end{align}
In fact, $\mathbb{J}^{-1}_{\rho_\theta}$  is linear, and satisfies by definition $\dot \rho_{\rm int(ext)}=\frac{1}{2}\{\mathbb{J}^{-1}_{\rho_\theta}[\dot{\rho}_{\rm int(ext)}],\rho_\theta \}$. From this identity it is clear that $\mathbb{J}^{-1}_{\rho_\theta}[\dot{\rho}_{\rm int(ext)}]$ is diagonal (off-diagonal) just like $\dot \rho_{\rm int(ext)}$ is. Hence, the cross terms in Eq.~\eqref{eq: QFI def 1} vanish, and the QFI decomposes as the sum of two contributions
\begin{align}
    \mathcal{F} =\underbrace{
    \Tr[\dot{\rho}_{\rm ext}\mathbb{J}^{-1}_{\rho_\theta}[\dot{\rho}_{\rm ext}]]}_{=\mathcal{F}_{\rm ext}}
    + 
    \underbrace{\Tr[\dot{\rho}_{\rm int}\mathbb{J}^{-1}_{\rho_\theta}[\dot{\rho}_{\rm int}]]}_{=\mathcal{F}_{\rm int}}\;.
\end{align}

The two terms admit a simple intuitive interpretation, as summarized by the following result.

\begin{result}\label{res: dephasing general}
The quantum Fisher information of the dephasing channel  $\rho_\theta = \mathcal{D}_{H_\theta}(\initial)=\sum_{k} \ketbra{\varphi_k} \initial \ketbra{\varphi_k}$ with $\ket{\dot \varphi_k} = -\ii S \ket{\varphi_k}$ is given by
\begin{equation}\label{eq: UB1 dephasing}
    \f = \f_{\rm ext} +\f_{\rm int}\, .
\end{equation}
Here, $\f_\text{\rm ext}$ is the QFI due to the rotation $\dot \rho_{\rm ext} = -\ii [S,\rho_\theta]$, generated by the Hermitian operator $S$ on the dephased state $\rho_\theta=\mathcal{D}_{H_\theta}(\initial)$, and $\f_{\rm int}$ is the Fisher information of the distribution $p_k = \bra{\varphi_k} \initial \ket{\varphi_k}$.
\end{result}

\paragraph{Upper bounds for dephasing metrology with Hamiltonian control}-- Let us now take $S$ to be of the particular form given in the Eq.~\eqref{eq: S of Hs} with a bounded energy gap $|E_k-E_j|\geq E\,\, \forall \,\, k,j$. We want to express the right-hand side of Eq.~\eqref{eq: UB1 dephasing} in terms of the Hamiltonian $H_S$ and the gap $E$. The expression we obtain is given by the following result.

\begin{result} \label{res: dephasing upperbound}
For any input state $\initial$ let  $\rho_\theta=\mathcal{D}_{H_\theta}(\initial)$ be the state resulting from the application of the dephasing Eq.~\eqref{eq: rho dephasing} with a nondegenerate Hamiltonian $H_\theta=H_C +\theta H_S$ with all energies gapped by at least $E$. The QFI of the state $\rho_\theta$ is upper bounded by
\begin{align}
     \f \leq \frac{(3+\pi^2)}{3} \frac{\|H_S\|^2}{E^2}\;,
     \label{eq:dephasing_bound}
\end{align}
where $\|H_S\| = \lambda_{\rm max}^{S} - \lambda_{\rm min}^{S}$, as in Section~\ref{sec:dynamical_metro}.
\end{result}

The proof of this result can be found in Appendix~\ref{app:dephasing} and relies on separately maximizing $\f_{\rm ext}$ and $\f_{\rm int}$. Thanks to the interpretation of both $\f_{\rm ext}$ and $\f_{\rm int}$ corresponding to QFI induced by (external and internal) rotations induced by the generator $S$, the proof relies on bounding the operator norm of $S$ using the Proposition.~\ref{prop:gersh-energy} in Appendix~\ref{app:circleth} based on Gershgorin circle theorem (see Appendix~\ref{app:circleth} for details). \\

\subsubsection{Saturable scaling for dephasing metrology with Hamiltonian control}
The bound in Eq.~\eqref{eq:dephasing_bound} cannot be saturated in general. However, if Hamiltonian control is unrestricted it is generally possible to reach the same scaling $\cF \sim \|H_S\|^2/E^2$ up to a constant prefactor. In particular, let us assume that one can prepare an initial product state $\ket{\psi}$ that is an eigenstate of $H_S$ exactly the middle of the spectrum. This is motivated by local $H_S$, where (for even N) $\ket{\psi}$ can be prepared by initializing half of the particles ``up'' and half ``down''. In Appendix~\ref{app: dephasing example} we then show the following result. 

\begin{result}\label{res: dephasing attainable}
For any signal Hamiltonian $H_S$ with extremal eigenvalues $\lambda_{\rm min}^S$ and $\lambda_{\rm max}^S$ and the initial state $\ket{\psi}$ satisfying $H_S \ket{\psi} =\frac{\lambda_{\rm min}^S+\lambda_{\rm max}^S}{2}\ket{\psi}$, there exist a control Hamiltonian $H_C$ such that the QFI of the dephased state $\rho_\theta=\mathcal{D}_{H_\theta}(\psi)$ in Eq.~\eqref{eq: rho dephasing} is given by 
\begin{equation}\label{eq: QFI dephasing go part}
\cF = \frac{3}{2} \frac{\|H_S \|^2}{E^2}\, ,
\end{equation}
where $E$ is the minimal energy gap of the (nondegenerate) Hamiltonian $H_\theta= H_C+ \theta H_S$, and $\|H_S\| = \lambda_{\rm max}^{S} - \lambda_{\rm min}^{S}$.
\end{result}

We now sketch a proof of the result. Without loss of generality, we can consider the situation around $\theta=0$. Let us denote by $\ket{\Phi_\downarrow}$ and $\ket{\Phi_\uparrow}$ the extremal eigenstates of  $H_S$ corresponding to eigenvalues $\lambda_\text{min}^S$ and  $\lambda_\text{max}^S$ respectively.  Since we have full freedom to choose the control Hamiltonian $H_C$, we take it to be block diagonal with respect to the qutrit subspace $\text{span}\{\ket{\Phi_\uparrow},\ket{\psi},\ket{\Phi_\downarrow}\}$.
For such a choice, the dephasing dynamics for any $\theta$ leaves the initial state $\ket{\psi}$ inside this subspace. Thus, in this setting, the freedom to choose the control Hamiltonian with a bounded energy gap boils down to letting $H_C$ run through all $3\times3$ Hermitian matrices with non-degenerate eigenvalues separated by at least $E$. With the help of the Result~\ref{res: dephasing general}, in Appendix~\ref{res: dephasing general} we optimize $\cF_\text{int}$ for all real $H_C$ and find that the choice 
\begin{equation}
H_C =\frac{E}{\sqrt{2}}\left(
\begin{array}{ccc}
 0 & 1 & 0 \\
 1 & 0 & 1 \\
 0 & 1 & 0 \\
\end{array}
\right)
\end{equation}
in the basis $\{\ket{\Phi_\uparrow},\ket{\psi},\ket{\Phi_\downarrow}\}$
has the energy spectrum $(E,0,-E)$, and for the dephased state $\rho_\theta = \mathcal{D}_{H_\theta}(\psi)$ gives $\cF=\cF_\text{int}+\cF_\text{ext}$ in Eq.~\eqref{eq: QFI dephasing go part}. \\

\subsubsection{Interacting spin probe}

Similar to Section~\ref{sec:dynamical_metro}, we now apply our general considerations to the estimation of the magnetic field strength ($H_S =  \frac{\addw}{2}\sum_{i=1}^N\sigma_z^{(i)}$) via a interacting spin probe with two-body control as given in Eq.~\eqref{eq:all-to-allintro}. We consider as initial product state
\begin{equation}
    \ket{\psi} = \ket{-y}^{\otimes N},
    \label{eq:psi0_decoh}
\end{equation}
where $\ket{-y}$ is an eigenstate of $\sigma_y$. To compute the asymptotic QFI, $\mathcal{F}$,  
in the long-time limit, we project the initial state $\ket{\psi}$ onto the eigenbasis of $H_\theta$, and remove coherence between different eigenspaces. Note that, in this process, we do not remove coherence between eigenstates belonging to the same eigenspace, and we do not remove coherence between eigenstates whose energy difference is proportional to the infinitesimal increment of $\theta$ necessary to compute the QFI. 

\begin{figure}[!tb]
	\centering
	\includegraphics[width=0.99\columnwidth]{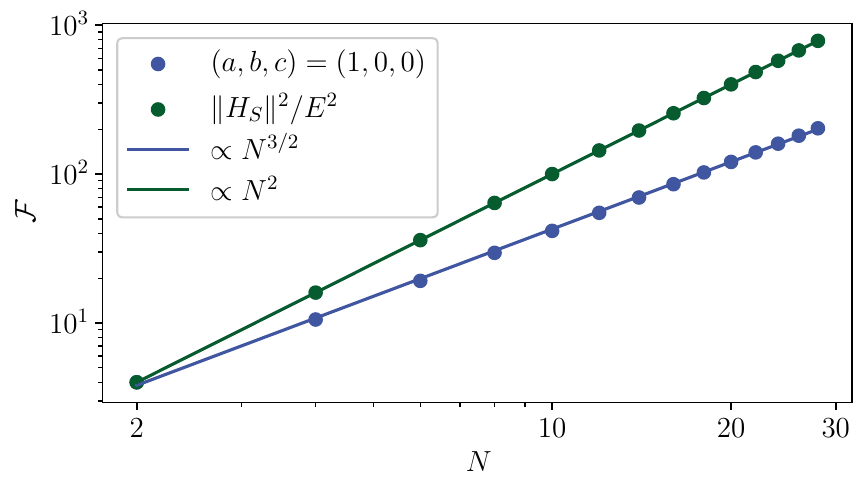}
	\caption{QFI, as a function of $N$, for $\theta=0$. The blue dots correspond to $\mathcal{F}$ computed for $(a,b,c)=(1,0,0)$, while the green dots represent $\|H_S\|^2/E^2$. The solid lines are references to visualize the $N^{3/2}$ and $N^2$ scaling. 
    }
	\label{fig:decoherence_jx}
\end{figure}

\begin{result}\label{res:quantum-adv-twobody-dephasing}
    Let $H_\theta$ be the Hamiltonian of the form $H_\theta = \theta H_S+ H_C $, with $H_S = \frac{\addw}{2}\sum_{i=1}^N \sigma^{(i)}_z$ a local Hamiltonian, and $H_C$ the control term with only two-body interactions as given in Eq.~\eqref{eq:all-to-allintro}. Let the initial state be a product state as given in Eq.~\eqref{eq:psi0_decoh}, and consider the corresponding long-time averaged state as given by $\mathcal{D}_{H_\theta}(\psi)$. 
    Then, by a proper choice of $H_C$,  the  QFI evaluated at $\theta=0$ of $\mathcal{D}_{H_{\theta = 0}}(\psi)$ behaves as 
    \begin{equation}
        \f \approx   A \frac{\revadd{\addw^2} N^{3/2}}{E^2} +\order{N}\, ,
    \end{equation}
    where $A\approx 1.34$ at least up to $N\leq 30$ (see numerical results in Fig.~\ref{fig:decoherence_jx}).  
\end{result}

Indeed, in Fig.~\ref{fig:decoherence_jx} we plot $\mathcal{F}$, 
as a function of $N$, as blue dots in the $(a,b,c)=(1,0,0)$ case, and we compute the QFI around $\theta=0$. We consider even values of $N$ since odd values yield a lower QFI. Interestingly, from the numerics we find an $N^{3/2}$ scaling, confirming a super-linear scaling of the QFI, enabled by interactions, also in this scenario with decoherence.

As a comparison, in Fig.~\ref{fig:decoherence_jx} we further plot as green dots the ratio $\|H_S\|^2/E^2$ that appears in the upper bound in Eq.~\eqref{eq:dephasing_bound}. While this upper bound is saturated for $N=2$, we see that it scales as $N^2$; thus, it is not saturated with this choice of the interaction for larger values of $N$. 

Finally, we computed the ratio between the QFI for arbitrary choices of $(a,b,c)$, and the upper bound $\|H_S\|^2/E^2$, and we found numerical evidence that $(a,b,c)=(1,0,0)$ maximizes this ratio [note, however, that other choices of $(a,b,c)$ could yield a higher QFI]. It remains an interesting open question to explore alternative interacting spin geometries that can approach the $N^2$  scaling for the long time-averaged state.  

\subsection{Gibbs ensemble metrology}
\label{subsec:equilibrium_metro}

While dephasing might be interpreted as an effective noise induced by the finiteness of time resolution, non-unitary dynamics can, more generally, affect metrological probes that are not perfectly isolated from environmental degrees of freedom. A rather general example of noisy evolutions are those leading to thermal equilibrium, i.e. where the steady state of the probe is given by the Gibbs state
\begin{equation}
\label{eq:Gibbs_state}
    \rho_\theta = \frac{e^{-\beta H_\theta}}{\Tr[ e^{-\beta H_\theta}]}\, 
\end{equation}
for some temperature $T=(k_{\rm B}\beta)^{-1}$.

The expression for the Fisher information in such case takes the form of a generalized variance (see e.g.,Refs.~\cite{zanardi2007bures,abiuso2024fundamental,scandi2023quantum}) 
\begin{align}
    \label{eq:QFI_thermal}
    \mathcal{F} =\beta^2\left(\Tr[H_S\mathcal{J}_{\rho_\theta}[H_S] ] - \Tr[\rho_\theta H_S]^2\right)\;,
\end{align}
where $\mathcal{J}_\rho$ is a superoperator that acts, in the basis of eigenvectors of $\rho\equiv\sum_i p_i\ketbra{i}$, as\footnote{More precisely, \eqref{eq:therm_calJ_def} is valid for $p_i\neq p_j$, while for $p_i=p_j$ it holds $\mathcal{J}_{\rho}[\ketbra{i}{j}]=p_i \ketbra{i}{j}$, as the limit $p_i\rightarrow p_j$ would suggest. See~\cite{scandi2023quantum,abiuso2024fundamental} for details.}
\begin{align}
\label{eq:therm_calJ_def}
    \mathcal{J}_{\rho}[\ketbra{i}{j}]=2\frac{(p_i-p_j)^2}{(\ln p_i-\ln p_j)^2(p_i+p_j)}\ketbra{i}{j}\;.
\end{align}
In the semiclassical case in which $[\rho_\theta,H_S]=0$ (equivalently, $[H_C,H_S]=0$) the expression in Eq.~\eqref{eq:QFI_thermal} reduces to the standard variance of $H_S$ computed on $\rho_\theta$, while it is smaller otherwise~\cite{abiuso2024fundamental}.

While previous works have considered metrological probes at thermal equilibrium, the fundamental limits of this approach and the corresponding optimal control (that is, $\max_{H_C}\f_\theta$ for given $H_S$) have recently been obtained in~\cite{abiuso2024fundamental}. For completeness, we now briefly mention the main results obtained there.
\begin{result}[Optimal metrology at thermal equilibrium~\cite{abiuso2024fundamental}]
\label{res:MaximalQFIGibbs}
When assuming no restriction on the possible form of the control values $H_C$ and finite temperature $0<\beta<\infty$, the maximum (saturable) value of the Fisher information at thermal equilibrium in Eq.~\eqref{eq:QFI_thermal} is given by
    \begin{align}
    \f\leq \beta^2 \frac{\| H_S \|^2}{4}\;.
    \label{eq:max_thermal_FI}
\end{align}
where $\|H_S\| = \lambda_{\max}^{S}-\lambda_{\min}^{S}$.
\end{result}

Note that this bound, as the other ones depending on~$\|H_S\|^2$, entails a Heisenberg-like scaling $\propto N^2$ of the Fisher information in terms of the number of particles, and diverges in the limit of zero temperature $\beta\rightarrow\infty$. This is because the state in Eq.~\eqref{eq:thermal_state} becomes the ground state in such limit, which can in principle have infinitely sharp transitions between microstates whenever there are crossovers of energy between the ground and first-excited  states. Indeed, a meaningful bound can still be obtained in this limit,  under the additional restriction of the system being gapped~\cite{abiuso2024fundamental}
\begin{align}
     \f_{\rm{low-T}} \leq \frac{\|H_S\|^2}{\Delta_g^2}+\mathcal{O}(e^{-\beta\Delta_g})\;,
\label{eq:max_ground_QFI_gap}
\end{align}
with $\Delta_g$ is the spectral gap. 

\subsubsection{Interacting spin probe}

As in the previous sections, we now focus on the estimation of the strength of a magnetic field  ($H_S =  \frac{1}{2}\sum_{i=1}^N\sigma_z^{(i)}$) via an interacting spin probe with two-body control as given in Eq.~\eqref{eq:all-to-allintro}, which is now prepared in a Gibbs state as in Eq.~\eqref{eq:Gibbs_state}. Our main result is to show that the upper bound in Eq.~\eqref{eq:max_thermal_FI} can be easily saturated in this case.
\begin{result}\label{res:quantum-adv-twobody-gibbs}
    Let $H_\theta$ be the Hamiltonian of the form $H_\theta = \theta H_S+ H_C $, with $H_S = \frac{\addw}{2}\sum_{i=1}^N \sigma^{(i)}_j$ a local Hamiltonian, and $H_C$ the control term with only two-body interactions as given in~\eqref{eq:all-to-allintro}. 
    Then, by taking $c \gg 1$ and $a=b=0$, the thermal  QFI evaluated at $\theta=0$ reads 
    \begin{equation}
        \f \approx     \frac{\beta^2 \revadd{\addw^2} N^2}{4}\,,
        \label{eq:maxthermal}
    \end{equation}
   thus saturating the upper bound  \eqref{eq:max_thermal_FI}. 
\end{result}

To show this result, first note that $H_S$ and $H_C$ commute. In this case the QFI in~\eqref{eq:QFI_thermal} simplifies to the variance of $H_S$:
\begin{align}
    \label{eq:QFI_thermal2}
    \mathcal{F} =\beta^2\left(\Tr[H_S^2 \rho_\theta  ] - \Tr[\rho_\theta H_S]^2\right)\,.
\end{align}
This variance is maximized for state $\rho_\theta = \frac{1}{2}\left(\ketbra{\lambda_{\max }^S}+\ketbra{\lambda_{\min }^S}\right)$ where $\ket{\lambda_{\min }^{S}}=\ket{1,...,1}$ and $\ket{\lambda_{\max }^{S}}=\ket{0,...,0}$. Now, we wish to encode this state in a Gibbs state as in \eqref{eq:Gibbs_state}. For $\theta=0$, the corresponding Gibbs state is simply 
\begin{equation}
\label{eq:Gibbs_state_theta0}
    \rho_0 = \frac{e^{-\beta H_C}}{\Tr[ e^{-\beta H_C}]}\, .
\end{equation}
By taking $c \gg 1$  (with $a=b=0$), the corresponding Gibbs state becomes the desired state $\rho_\theta \approx \frac{1}{2}\left(\ketbra{1,...,1}+\ketbra{0,...,0}\right)$ up to exponentially small corrections of order $\mathcal{O}(\exp(-\beta c))$. A direct application of Eq.~\eqref{eq:QFI_thermal2} then gives the desired result \eqref{eq:maxthermal}.

\section{Transient regime: From dynamical to steady state 
}
\label{sec:transient}

The main goal of this last section is to characterize the behavior of the QFI in the transient regime, before reaching the steady states described in the previous sections (diagonal and Gibbs ensembles). Furthermore, we also analyze other types of physically relevant noise, e.g. local and global decoherence. 

The dynamics of the probe are assumed Markovian and described by the Lindblad master equation introduced in Eq.~\eqref{eq:limbladian_general}. Following the previous sections, we take a probe consisting of $N$ spin-$\frac{1}{2}$ systems with a symmetric Hamiltonian 
where signal $H_S$ is fixed, $H_C$ can be controlled and the dissipators $L_{\theta,i}$ can depend on $H_\theta = \theta H_S + H_C$. The total  Hamiltonian reads
 \begin{equation}
     H_\theta(a,b,c) = \theta \addw S_z + aS_x^2+ bS_y^2+ cS_z^2\,,
    \label{eq:h_decoh}
\end{equation}
where the signal term is $S_z$ and the control term is of the form of Eq.~\eqref{eq:all-to-allintro} ($H_C =  aS_x^2+ bS_y^2+ cS_z^2$) and features collective two-body interactions with tunable parameters $(a,b,c)$. For the noise model, specified by the dissipators $\sqrt{\gamma_i } L_{\theta, i}$, we consider four physically motivated scenarios in the sections below: dephasing, thermalization, global noise, and local noise.

\subsection{Dephasing}
\label{subsec:dephasingtransient}

In this subsection, we study the QFI in the presence of decoherence between the eigenstates. We consider the Hamiltonian in Eq.~\eqref{eq:h_decoh} and the initial state 
\begin{equation}
    \ket{\psi} = \ket{-y}^{\otimes N},
    \label{eq:psi0_decoh2}
\end{equation}
where $\ket{-y}$ is an eigenstate of $\sigma_y\ket{-y} = -\ket{-y}$. 

We now consider a finite-time model to analyze the time necessary to reach the asymptotic values leading to Result~\ref{res:quantum-adv-twobody-dephasing}. The model is based on our ignorance about the exact time $t$ for which the evolution takes place.  In particular, we assume that the initial state $\ket{\psi}$ in Eq.~(\ref{eq:psi0_decoh2}) evolves according to $H_\theta$ in Eq.~(\ref{eq:h_decoh}) for an unknown time $t$ that satisfies a probability distribution $P(t)$. For simplicity, we assume a flat distribution $P(t)=1/T$, so that the quantum state is described by the time-averaged state, i.e.,
\begin{equation}
    \rho_\theta(T) = \frac{1}{T}\int_{0}^T  e^{-\ii t H_\theta}\ketbra{\psi} e^{\ii t H_\theta}d t\,.
    \label{eq:decoh_state}
\end{equation}
In the limit $T\rightarrow \infty$, we obtain a fully dephased state.

\begin{figure}[!tb]
	\centering
	\includegraphics[width=0.99\columnwidth]{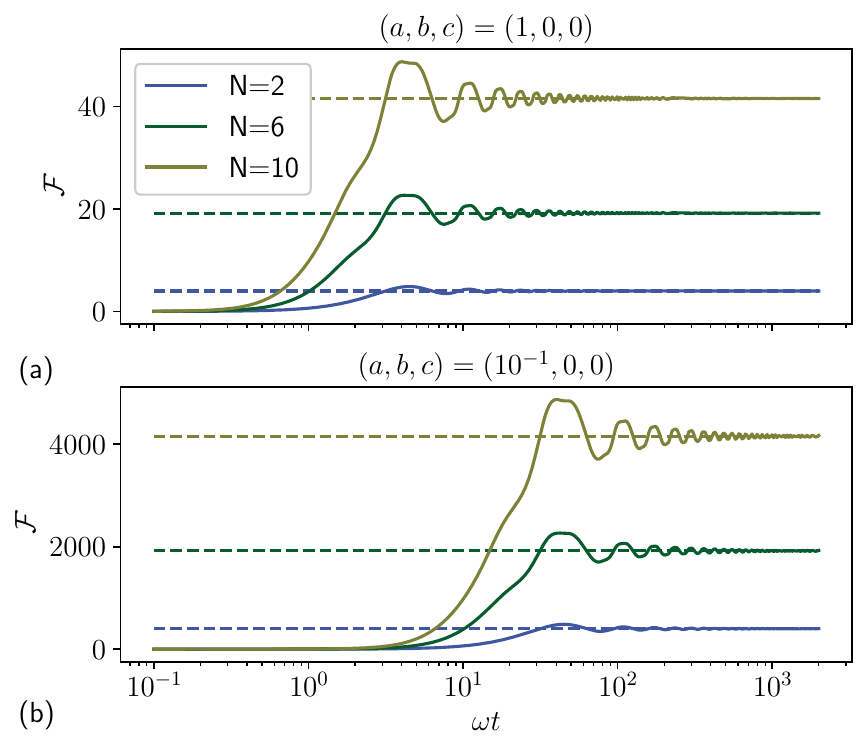}
	\caption{QFI, as a function of time, for the three values of $N$ shown in the legend and $\theta=0$. The solid lines correspond to the finite-time QFI of $\rho_\theta(T)$ in Eq.~(\ref{eq:decoh_state}). The dashed lines represent the asymptotic values of the $\mathcal{F}$. The results are computed with $(a,b,c)=(1,0,0)$ in \textbf{a)}, and with $(a,b,c)=(10^{-1},0,0)$ in \textbf{b)} .}
	\label{fig:decoherence_time}
\end{figure}
In Fig.~\ref{fig:decoherence_time} we plot as solid lines the QFI, as a function of time, for the three values of $N$ shown in the legend. The dashed lines correspond to the asymptotic values of $\mathcal{F}$. Fig.~\ref{fig:decoherence_time}a) corresponds to $(a,b,c)=(1,0,0)$, as in Fig.~\ref{fig:decoherence_jx}, and Fig.~\ref{fig:decoherence_time}b) corresponds to $(a,b,c)=(10^{-1},0,0)$.
As expected, the finite-time QFI tends to the asymptotic value after a transient time characterized by an oscillating behavior of $\mathcal{F}$. This time appears to be independent of $N$. 

Interestingly, we notice the following fact. If we reduce the interaction by a factor $\lambda$, i.e. we consider $(a,b,c) \to (a,b,c)/\lambda$, the gap between the spectrum scales as $1/\lambda$, leading to the asymptotic value of the $\mathcal{F}$ scaling as $\lambda^2$. This is consistent with the first-order perturbation theory result in  Eq.~\eqref{eq: eig perturation}, and confirmed by the comparison between Fig.~\ref{fig:decoherence_time}a) and Fig.~\ref{fig:decoherence_time}b), where $\lambda=10$, and the asymptotic value of $\mathcal{F}$ increases by $10^2$. This higher value of the QFI is reached after a longer transient time. However, this time only scales with $\lambda$ (as confirmed by Fig.~\ref{fig:decoherence_time}). Therefore, by rescaling the interaction by $\lambda$, the asymptotic value of the QFI scales as $\mathcal{F}\propto \lambda^2$, whereas the transient time only increases as $\lambda$. In other words, the QFI rate, $\mathcal{F}/t\addw$, increases linearly with~$\lambda$. As we will see, this is in stark contrast with the thermal case, studied in Section~\ref{subsec:thermaltransient}, where an enhancement of the QFI comes at the expense of an exponential increase in the equilibration time.

\subsection{Thermalization}
\label{subsec:thermaltransient}

\begin{figure}[!tb]
	\centering
	\includegraphics[width=0.99\columnwidth]{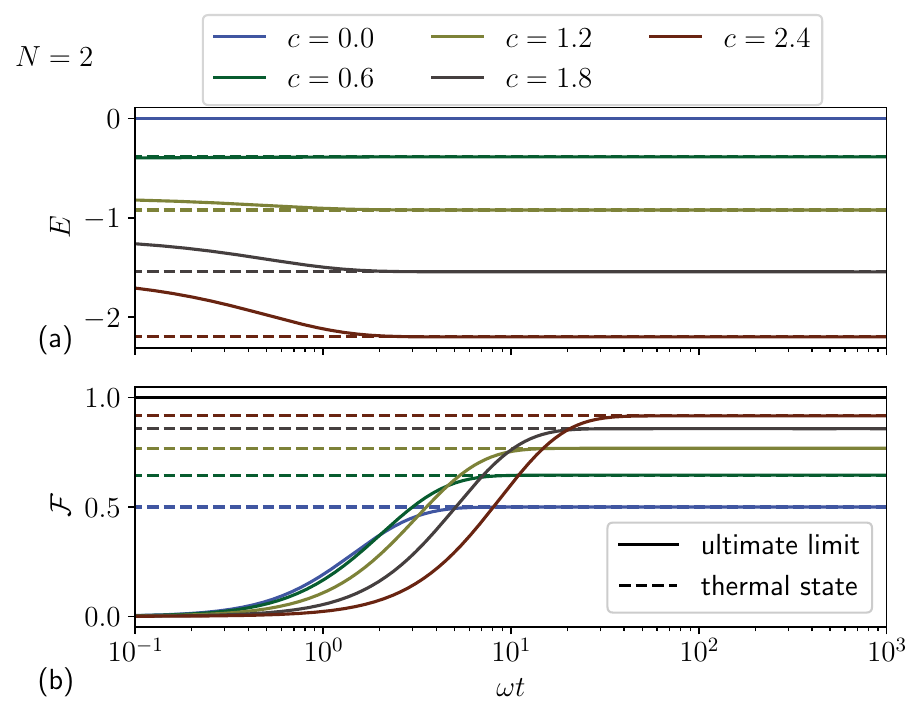}
 	\includegraphics[width=0.99\columnwidth]{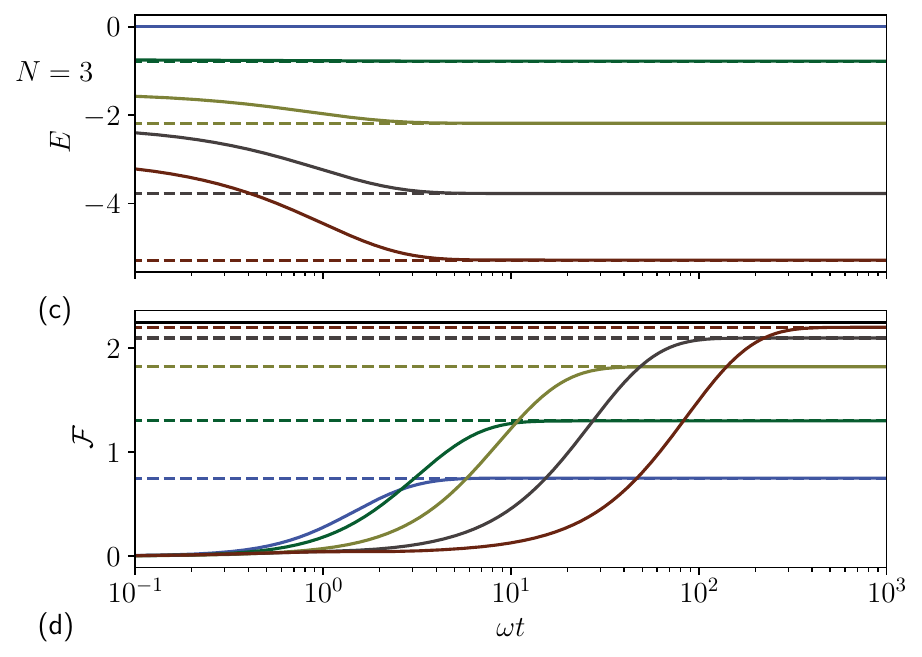}
  	\includegraphics[width=0.99\columnwidth]{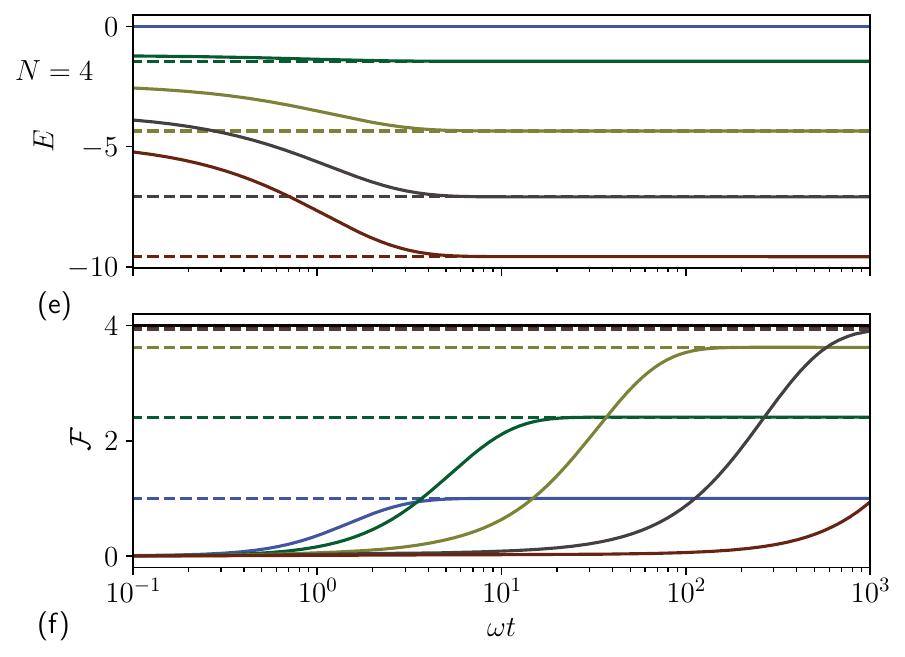}
	\caption{Energy and Fisher information respectively as a function of time. The number of qubits are $N = 2$ for \textbf{a),b)}; $N=3$ for \textbf{c), d)}; $N = 4$ for \textbf{e), f)}. Parameters: $\beta=1$, $\gamma=1$, $\theta=0$, initial state $p_n=1/(N+1)$. }
	\label{fig:small_n}
\end{figure}

Here we study how the Fisher information changes, as a function of time, during thermalization. We consider the same Hamiltonian as in the previous section but set $a=b=0$ and $c\neq 0$ so that time evolution can be effectively described by classical stochastic dynamics. Indeed, recall that by taking $c\gg 1$ we can achieve the maximal possible thermal QFI as discussed in Result~\ref{res:quantum-adv-twobody-gibbs}.  
We assume that each qubit is locally coupled to a thermal bath, which drives the global state towards a thermal state (detailed balance). Denoting with $p_n$ the probability of having $n$ excitations in the system, we consider the rate equation
\begin{multline}
    \dot{p}_n = -p_n\left[ \Gamma_{n\to n+1} + \Gamma_{n\to n-1} \right] \\
    + p_{n-1}\Gamma_{n-1\to n} + p_{n+1}\Gamma_{n+1\to n}\,,
\end{multline}
where $\Gamma_{n\to m}$ represents the transition rate from $n$ to $m$ excitations in the sytem. They are given by
\begin{equation}
\begin{aligned}
    \Gamma_{n\to n+1} &= \gamma (N-n) f\left[ U(n+1)-U(n) \right], \\
    \Gamma_{n\to n-1} &= \gamma n f\left[ U(n-1)-U(n) \right]\,, 
\end{aligned}
\end{equation}
where $\gamma$ is the thermalization rate, $U(n)$ is the energy of a state with $n$ excitations, and 
\begin{equation}
    f(x) = \left[ 1+e^{ \beta x }  \right]^{-1}\,,
\end{equation}
where $\beta$ is the inverse temperature of the system.

\begin{figure}[!tb]
	\centering
	\includegraphics[width=0.99\columnwidth]{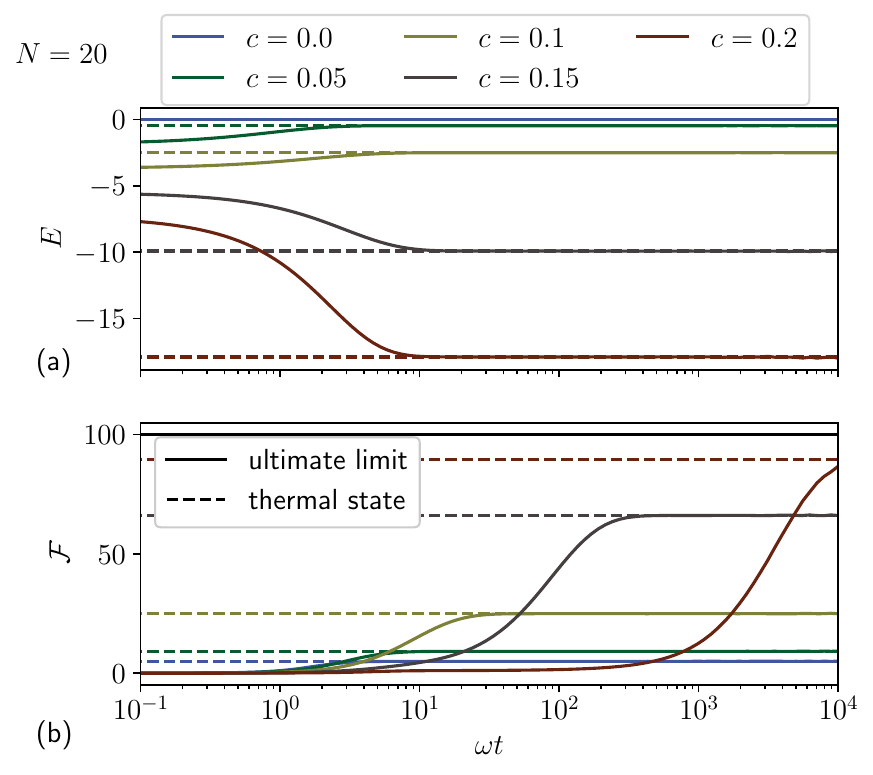}
	\caption{Energy \textbf{a)} and Fisher information \textbf{b)} as a function of time. Parameters are $N=20$, $\beta=1$, $\gamma=1$, $\theta=0$, initial state $p_n=1/(N+1)$.}
\label{fig:large_n}
\end{figure}

\newpage

In Fig.~\ref{fig:small_n} we plot the energy and the Fisher information, as a function of time. Every two panels correspond to $N=2,3,4$, respectively, and each curve to a different value of $c$ (see the legend). The colored dashed lines represent the thermal distribution, and the black line represents the ultimate limit to the Fisher information, which in this case is given by $\beta^2 \revadd{\addw^2} N^2/4$ (Result \ref{res:MaximalQFIGibbs}).

In Fig.~\ref{fig:large_n} we do the same plot, but we choose a large value of N, $N=20$, and, correspondingly we extend the time-simulation by a factor of 10 (we plot until $10^4$ and not $10^3$). Notice also that we choose smaller values of~$c$.

From Figs. \ref{fig:small_n} and \ref{fig:large_n}, we first observe that the QFI grows with $c$, reaching the value $\mathcal{F}\approx N^2/4$ in the limit $c\gg \theta$, which saturates bound~\eqref{eq:max_thermal_FI} (note that $\beta=1$ in  Fig.~\ref{fig:small_n}), as expected from Result~\ref{res:quantum-adv-twobody-gibbs}. Second, we observe a clear tradeoff between the enhancement in the QFI and the thermalization time, as the thermalization time of the QFI grows (exponentially) with $c$ and $N$. It is interesting to note that this slow thermalization does not affect the energy~$E$; this can be understood by noting that the QFI highly depends on small perturbations in the population of the thermal state $\rho_{\theta}$ induced by $\theta$. Such changes require crossing a free-energy barrier $\Delta F$ that grows linearly with $N$ and $c$ (the time for crossing the barrier via thermal fluctuations grows exponentially with $\Delta F$). Therefore, interaction-based enhancements in the QFI   come with the high price of long thermalization times. A similar behavior may arise in other models given the form of the optimal thermal state, $\rho_{\theta} \approx (\ket{0,...,0}\bra{0,...,0}+\ket{1,...,1}\bra{1,...,1})/2$, with the population concentrated in two locally stable states.  Indeed, in the case of thermometry, trade-offs between enhanced sensitivity and long thermalization times have been characterized~\cite{Anto-Sztrikacs2024}. 

\subsection{Global noise}
\label{subsec:global}

The previous case studies showed that interactions can enhance $\mathcal{F}$ in the presence of dephasing and thermalizing noise. Here we extend these considerations to another relevant case of (global) noise. 

We consider a global noise $S_x$ with a parameter strength $\gamma$, and $H_{\theta} = H_C + \theta \addw S_z$. That is, in Eq. \eqref{eq:limbladian_general}, $L_{\theta,i} = S_x$ and $\gamma_{i} = \gamma$. 
    
    We use $H_C = a S_x^2$ as the control Hamiltonian. This gives
    \begin{equation}\label{eq:global_h}
        H_\theta = \theta \addw S_z + a S_x^2\,.
    \end{equation}
    Then under the same conditions used to obtain Eq. \eqref{eq:fish2b} and measuring $S_x$ (as in Section \ref{sec:all_to_all_seciton}), we find that the CFI ($\mathcal{I}$) is 
    \begin{equation}\label{eq:cfi_noise}
        \mathcal{I}_{\rm control} =  4\revadd{\addw^2}\dfrac{(1-\mathrm{e}^{-t\addw {\gamma/2}})^2}{{\gamma^2}}\frac{N^{3/2}}{\sqrt{2\pi}} +\order{\frac{1}{a}}\,.
    \end{equation}
This can be contrasted to the QFI obtained in the absence of control:
 \begin{equation}\label{eq:qfi_no_control}
        \f_\text{no control} =  4\revadd{\addw^2}\dfrac{(1-\mathrm{e}^{-t \addw {\gamma/2}})^2}{{\gamma^2}}N\,.
    \end{equation}
Hence, we obtain an $N^{1/2}$ fold enhancement by appropriately tuning the interactions, which in this case does not come at the price of longer equilibration times.

\subsection{Local noise}

In the previous sections, we analyzed models of noise that involve global dissipators (in the thermalization case, even if the starting point is a local microscopic model with a thermal bath locally coupled to each qubit, the dissipators become global after performing the secular approximation). Many relevant noise models however involve local noise operators. To analyze this scenario, we consider the same model as in Section~\ref{subsec:global}, but replace the global noise operators $S_x$ with a collection of local noise operators $\sigma_x^{(i)}$ acting on every site $i$. The Hamiltonian is the same as Eq.~(\ref{eq:global_h}), and the initial state is
\begin{equation}
    \ket{\psi} = \ket{y}^{\otimes N}\,,
\end{equation}
with $\ket{y}$ the eigenstate of $\sigma_y\ket{y} = \ket{y}$. For details on the simulation, see Appendix~\ref{app:numerical_rotation}.

\begin{figure}[!tb]
	\centering
\includegraphics[width=0.99\columnwidth]{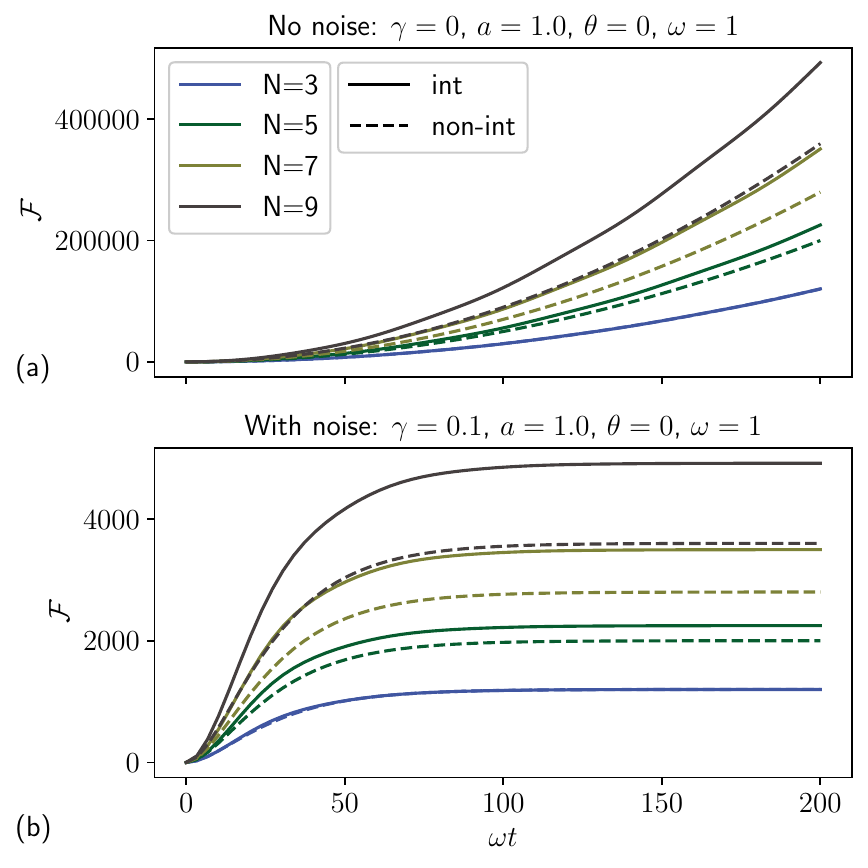}
	\caption{QFI, as a function of time, \textbf{a)} in the noiseless case  and \textbf{b)} in the presence of local noise. Each color corresponds to a different value of $N$ (see the legend). The dashed line corresponds to the noninteracting case and the solid line to the interacting case. Parameters are indicated in the panel title. }
\label{fig:local_noise_compare}
\end{figure}

In Fig.~\ref{fig:local_noise_compare} we plot the QFI, as a function of time, in the noiseless case (upper panel) and in the presence of local noise (lower panel). Each color corresponds to a different value of $N$, while the dashed line corresponds to the non-interacting case (i.e. $H_C = 0$), and the solid line to the interacting case.

As we expected, in the noiseless case (upper panel) the interacting case outperforms the noninteracting case, and the advantage visibly scales with $N$ (there is hardly any advantage for $N=3$, while it is quite pronounced for $N=9$). 

In the presence of noise (lower panel) the QFI does not increase indefinitely with time, but it saturates to a final value. Interestingly, such a final value is enhanced by the interaction, and the advantage increases also in this case with $N$.

\begin{figure}[!tb]
	\centering
\includegraphics[width=0.99\columnwidth]{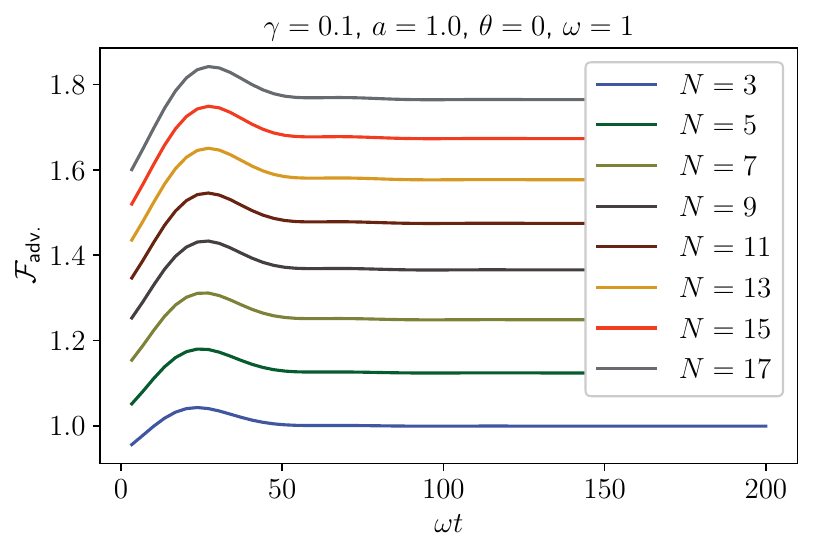}
	\caption{Ratio between the QFI in the interacting and non-interacting cases, as a function of $t$, in the presence of local noise. Parameters are indicated in the panel title. }
\label{fig:local_noise_int_advantage}
\end{figure}
In Fig.~\ref{fig:local_noise_int_advantage} we plot $\mathcal{F}_{\text{adv.}}$, defined as the ratio between the QFI in the interacting and noninteracting casea. As we can see, the advantage enabled by the interaction scales with $N$, and here we display values up to $N=17$. 

\begin{figure}[ht!]
	\centering
\includegraphics[width=0.99\columnwidth]{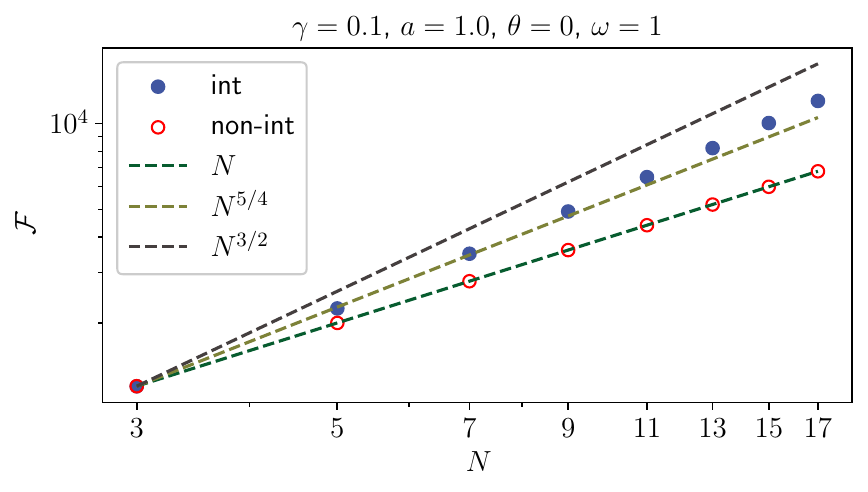}
	\caption{Log-log plot of the asymptotic (long-time) value of the QFI, in the presence of noise, as a function of $N$. The dashed lines \revadd{serve as} references to distinguish the $N$ scaling.}
\label{fig:local_noise_n_scaling}
\end{figure}
As clearly visible in Figs.~\ref{fig:local_noise_compare} and \ref{fig:local_noise_int_advantage}, the QFI saturates to a finite value for large times. In Fig.~\ref{fig:local_noise_n_scaling} we plot such an ``asymptotic'' value of the QFI as a function of $N$. \revadd{Guiding dashed lines differentiate the scalings of the plotted points, with $N$, $N^{5/4}$, and $N^{3/2}$ shown as visual references. As observed, the asymptotic QFI scales superlinearly with $N$.} As we can see, the asymptotic QFI scales superlinearly in $N$.

\section{Conclusion}
\label{sec:conclusions}

In this work, we considered the precision limits of many-body quantum probes. 
Given a  signal Hamiltonian that encodes an unknown parameter $\theta$ into the probe's state, 
see Eq.~\eqref{eq:paradigm}, we considered the maximization of the QFI over all 
control Hamiltonians.  
We considered two main scenarios: (i)~dynamical, i.e., unitary evolution driven by~$H_\theta$ given an initial eigenstate of $H_S$,  and (ii)~steady state, with a focus on the Gibbs~$\propto \exp(-\beta H_\theta)$ and diagonal~$\mathcal{D}_{H_\theta} (\rho)$ ensembles. For each case, we presented upper bounds on the QFI as well as corresponding interacting control Hamiltonians that can approach such bounds.  Our main results are summarized in Table~\ref{tab:1}. 

Having established general results for arbitrary encoding $H_S$, we focused on the estimation of the strength of a magnetic field via $N$-interacting spins with two-body interactions.  We showed that a carefully engineered central-spin model [Eq.~\eqref{eq: central spin}], as well as an spin-squeezing model [Eq.~\eqref{eq:all-to-allintro}], can saturate the dynamical upper bound $\f \sim t^2 \revadd{\addw^2} N^2$ when starting with an initial product state for the probe. Likewise, we also showed that the thermal bound $\f \sim \beta^2  \revadd{\addw^2} N^2$ can be saturated with the collective spin model as in Eq.~ \eqref{eq:all-to-allintro}, whereas it enables a QFI scaling as $\f \sim N^{3/2} \revadd{\addw^2}/E^2$ for the diagonal ensemble. Our results for the $N$-spin probe are summarized in Table~\ref{tab:2}. To establish  our result, in the Appendix~\ref{app:circleth} we derive upper bounds (Propositions~\ref{prop:gersh-cons} and \ref{prop:gersh-energy}) on the operator norm of the expressions $\sum_{j\neq k} f_{kj} \Pi_j\, H \,\Pi_k$, which can be of independent interest.

Finally, we explored connections between the two scenarios, studying the transient regime connecting the dynamical and steady-state scenarios for the $N$-spin probe. We observed trade-offs between an enhanced sensitivity of the steady state and the time required to reach it for both Gibbs and diagonal ensembles. We also considered other models of Markovian noise, namely local and global decoherence, showcasing advantages in the sensitivity arising due to the presence of interactions in the network. 

These results provide new insights into the potential and limits of many-body metrology~\cite{montenegro2024quantum}, and set the stage for the realization of highly sensitive interacting spin sensors.  For steady-state metrology, an important open question for the future is to generalize the QFI bounds to general non-equilibrium steady states of open quantum systems (see e.g.~\cite{FernandezLorenzo2017,fernandez2018heisenberg, Marzolino2017, Raghunandan2018, DiCandia2023, Ilias2024Criticality}) given arbitrary control on $H_C$. Characterizing optimal many-body Hamiltonian control for multiparameter parameter estimation~\cite{Hayes2018Making,Trnyi2024Activation,mihailescu2024uncertain,Mihailescu2024,Giovanni2022Multiparameter} is another exciting future challenge. 
Furthermore, understanding the time required to reach the steady state~\cite{Sahoo2024Localization, montenegro2023quantum} is crucial for characterizing how quickly dynamically-evolved states converge to the new QFI bounds and for comparing their finite-time performance with~\eqref{eq:boundCR}.
First steps in this direction were taken in Section~\ref{sec:transient}, but further research is still required. More generally, characterizing optimal many-body open  probes in the transient regime of open quantum dynamics remains an important open challenge. Other interesting future directions include exploring connections with other frameworks for many-body parameter estimation~\cite{Hayes2018Making, Trnyi2024Activation} as well as QFI optimization in many-body systems~\cite{Yang2022Variational, Yang2022Super,ban2022neural, Bai2023Floquet, Quentin2024}, and Hamiltonian learning~\cite{huang2023learning,dutkiewicz2023advantage}. 

\acknowledgments
R.P. acknowledges the support of the SNF Quantum Flagship Replacement Scheme (Grant No. 215933). M.P.-L. acknowledges support from the Grant RYC2022-036958-I funded by MICIU/AEI/10.13039/501100011033 and by ESF+. P.A.E. gratefully acknowledges funding from the Berlin Mathematics Center MATH+ (AA2-18). P.A. is supported by the QuantERA II programme that has received funding from the European Union’s Horizon 2020 research and innovation programme under Grant Agreement No. 101017733, and from the Austrian Science Fund (FWF), project ESP2889224. J.C. is supported by the Spanish MICINN AEI through Project No. PID2022-141283NB-I00, the MCIN with funding from the European Union NextGenerationEU (PRTR-C17.I1), and by Generalitat de Catalunya. J.C. also acknowledges support from an ICREA Academia award. P.S. acknowledges support from Swiss National Science Foundation (NCCR SwissMAP).

\bibliography{bib.bib}

\begin{thebibliography}{91}%
\makeatletter
\providecommand \@ifxundefined [1]{%
 \@ifx{#1\undefined}
}%
\providecommand \@ifnum [1]{%
 \ifnum #1\expandafter \@firstoftwo
 \else \expandafter \@secondoftwo
 \fi
}%
\providecommand \@ifx [1]{%
 \ifx #1\expandafter \@firstoftwo
 \else \expandafter \@secondoftwo
 \fi
}%
\providecommand \natexlab [1]{#1}%
\providecommand \enquote  [1]{``#1''}%
\providecommand \bibnamefont  [1]{#1}%
\providecommand \bibfnamefont [1]{#1}%
\providecommand \citenamefont [1]{#1}%
\providecommand \href@noop [0]{\@secondoftwo}%
\providecommand \href [0]{\begingroup \@sanitize@url \@href}%
\providecommand \@href[1]{\@@startlink{#1}\@@href}%
\providecommand \@@href[1]{\endgroup#1\@@endlink}%
\providecommand \@sanitize@url [0]{\catcode `\\12\catcode `\$12\catcode `\&12\catcode `\#12\catcode `\^12\catcode `\_12\catcode `\%12\relax}%
\providecommand \@@startlink[1]{}%
\providecommand \@@endlink[0]{}%
\providecommand \url  [0]{\begingroup\@sanitize@url \@url }%
\providecommand \@url [1]{\endgroup\@href {#1}{\urlprefix }}%
\providecommand \urlprefix  [0]{URL }%
\providecommand \Eprint [0]{\href }%
\providecommand \doibase [0]{https://doi.org/}%
\providecommand \selectlanguage [0]{\@gobble}%
\providecommand \bibinfo  [0]{\@secondoftwo}%
\providecommand \bibfield  [0]{\@secondoftwo}%
\providecommand \translation [1]{[#1]}%
\providecommand \BibitemOpen [0]{}%
\providecommand \bibitemStop [0]{}%
\providecommand \bibitemNoStop [0]{.\EOS\space}%
\providecommand \EOS [0]{\spacefactor3000\relax}%
\providecommand \BibitemShut  [1]{\csname bibitem#1\endcsname}%
\let\auto@bib@innerbib\@empty
%</preamble>
\bibitem [{\citenamefont {T{\'{o}}th}\ and\ \citenamefont {Apellaniz}(2014)}]{qmqi}%
  \BibitemOpen
  \bibfield  {author} {\bibinfo {author} {\bibfnamefont {G.}~\bibnamefont {T{\'{o}}th}}\ and\ \bibinfo {author} {\bibfnamefont {I.}~\bibnamefont {Apellaniz}},\ }\bibfield  {title} {\bibinfo {title} {Quantum metrology from a quantum information science perspective},\ }\href {https://doi.org/10.1088/1751-8113/47/42/424006} {\bibfield  {journal} {\bibinfo  {journal} {Journal of Physics A: Mathematical and Theoretical}\ }\textbf {\bibinfo {volume} {47}},\ \bibinfo {pages} {424006} (\bibinfo {year} {2014})}\BibitemShut {NoStop}%
\bibitem [{\citenamefont {Demkowicz-Dobrza{\'n}ski}\ \emph {et~al.}(2015)\citenamefont {Demkowicz-Dobrza{\'n}ski}, \citenamefont {Jarzyna},\ and\ \citenamefont {Ko{\l}ody{\'n}ski}}]{optical_interferometry_review}%
  \BibitemOpen
  \bibfield  {author} {\bibinfo {author} {\bibfnamefont {R.}~\bibnamefont {Demkowicz-Dobrza{\'n}ski}}, \bibinfo {author} {\bibfnamefont {M.}~\bibnamefont {Jarzyna}},\ and\ \bibinfo {author} {\bibfnamefont {J.}~\bibnamefont {Ko{\l}ody{\'n}ski}},\ }\bibfield  {title} {\bibinfo {title} {Quantum limits in optical interferometry},\ }\href {https://www.sciencedirect.com/science/article/pii/S0079663815000049} {\bibfield  {journal} {\bibinfo  {journal} {Progress in Optics}\ }\textbf {\bibinfo {volume} {60}},\ \bibinfo {pages} {345} (\bibinfo {year} {2015})}\BibitemShut {NoStop}%
\bibitem [{\citenamefont {Boixo}\ \emph {et~al.}(2007)\citenamefont {Boixo}, \citenamefont {Flammia}, \citenamefont {Caves},\ and\ \citenamefont {Geremia}}]{boixogeneralised}%
  \BibitemOpen
  \bibfield  {author} {\bibinfo {author} {\bibfnamefont {S.}~\bibnamefont {Boixo}}, \bibinfo {author} {\bibfnamefont {S.~T.}\ \bibnamefont {Flammia}}, \bibinfo {author} {\bibfnamefont {C.~M.}\ \bibnamefont {Caves}},\ and\ \bibinfo {author} {\bibfnamefont {J.}~\bibnamefont {Geremia}},\ }\bibfield  {title} {\bibinfo {title} {Generalized limits for single-parameter quantum estimation},\ }\href {https://doi.org/10.1103/PhysRevLett.98.090401} {\bibfield  {journal} {\bibinfo  {journal} {Physical Review Letters}\ }\textbf {\bibinfo {volume} {98}},\ \bibinfo {pages} {090401} (\bibinfo {year} {2007})}\BibitemShut {NoStop}%
\bibitem [{\citenamefont {Abiuso}\ \emph {et~al.}(2024{\natexlab{a}})\citenamefont {Abiuso}, \citenamefont {Sekatski}, \citenamefont {Calsamiglia},\ and\ \citenamefont {Perarnau-Llobet}}]{abiuso2024fundamental}%
  \BibitemOpen
  \bibfield  {author} {\bibinfo {author} {\bibfnamefont {P.}~\bibnamefont {Abiuso}}, \bibinfo {author} {\bibfnamefont {P.}~\bibnamefont {Sekatski}}, \bibinfo {author} {\bibfnamefont {J.}~\bibnamefont {Calsamiglia}},\ and\ \bibinfo {author} {\bibfnamefont {M.}~\bibnamefont {Perarnau-Llobet}},\ }\bibfield  {title} {\bibinfo {title} {Fundamental limits of metrology at thermal equilibrium},\ }\href {https://doi.org/10.48550/arXiv.2402.06582} {\bibfield  {journal} {\bibinfo  {journal} {arXiv preprint arXiv:2402.06582}\ } (\bibinfo {year} {2024}{\natexlab{a}})}\BibitemShut {NoStop}%
\bibitem [{\citenamefont {Montenegro}\ \emph {et~al.}(2024)\citenamefont {Montenegro}, \citenamefont {Mukhopadhyay}, \citenamefont {Yousefjani}, \citenamefont {Sarkar}, \citenamefont {Mishra}, \citenamefont {Paris},\ and\ \citenamefont {Bayat}}]{montenegro2024quantum}%
  \BibitemOpen
  \bibfield  {author} {\bibinfo {author} {\bibfnamefont {V.}~\bibnamefont {Montenegro}}, \bibinfo {author} {\bibfnamefont {C.}~\bibnamefont {Mukhopadhyay}}, \bibinfo {author} {\bibfnamefont {R.}~\bibnamefont {Yousefjani}}, \bibinfo {author} {\bibfnamefont {S.}~\bibnamefont {Sarkar}}, \bibinfo {author} {\bibfnamefont {U.}~\bibnamefont {Mishra}}, \bibinfo {author} {\bibfnamefont {M.~G.}\ \bibnamefont {Paris}},\ and\ \bibinfo {author} {\bibfnamefont {A.}~\bibnamefont {Bayat}},\ }\bibfield  {title} {\bibinfo {title} {Quantum metrology and sensing with many-body systems},\ }\href {https://arxiv.org/abs/2408.15323} {\bibfield  {journal} {\bibinfo  {journal} {arXiv preprint arXiv:2408.15323}\ } (\bibinfo {year} {2024})}\BibitemShut {NoStop}%
\bibitem [{\citenamefont {Macieszczak}\ \emph {et~al.}(2016)\citenamefont {Macieszczak}, \citenamefont {Gu{\c{t}}{\u{a}}}, \citenamefont {Lesanovsky},\ and\ \citenamefont {Garrahan}}]{macieszczak2016dynamical}%
  \BibitemOpen
  \bibfield  {author} {\bibinfo {author} {\bibfnamefont {K.}~\bibnamefont {Macieszczak}}, \bibinfo {author} {\bibfnamefont {M.}~\bibnamefont {Gu{\c{t}}{\u{a}}}}, \bibinfo {author} {\bibfnamefont {I.}~\bibnamefont {Lesanovsky}},\ and\ \bibinfo {author} {\bibfnamefont {J.~P.}\ \bibnamefont {Garrahan}},\ }\bibfield  {title} {\bibinfo {title} {Dynamical phase transitions as a resource for quantum enhanced metrology},\ }\href {http://dx.doi.org/10.1103/PhysRevA.93.022103} {\bibfield  {journal} {\bibinfo  {journal} {Physical Review A}\ }\textbf {\bibinfo {volume} {93}},\ \bibinfo {pages} {022103} (\bibinfo {year} {2016})}\BibitemShut {NoStop}%
\bibitem [{\citenamefont {Braun}\ \emph {et~al.}(2018)\citenamefont {Braun}, \citenamefont {Adesso}, \citenamefont {Benatti}, \citenamefont {Floreanini}, \citenamefont {Marzolino}, \citenamefont {Mitchell},\ and\ \citenamefont {Pirandola}}]{Braun2018}%
  \BibitemOpen
  \bibfield  {author} {\bibinfo {author} {\bibfnamefont {D.}~\bibnamefont {Braun}}, \bibinfo {author} {\bibfnamefont {G.}~\bibnamefont {Adesso}}, \bibinfo {author} {\bibfnamefont {F.}~\bibnamefont {Benatti}}, \bibinfo {author} {\bibfnamefont {R.}~\bibnamefont {Floreanini}}, \bibinfo {author} {\bibfnamefont {U.}~\bibnamefont {Marzolino}}, \bibinfo {author} {\bibfnamefont {M.~W.}\ \bibnamefont {Mitchell}},\ and\ \bibinfo {author} {\bibfnamefont {S.}~\bibnamefont {Pirandola}},\ }\bibfield  {title} {\bibinfo {title} {Quantum-enhanced measurements without entanglement},\ }\href {https://doi.org/10.1103/RevModPhys.90.035006} {\bibfield  {journal} {\bibinfo  {journal} {Rev. Mod. Phys.}\ }\textbf {\bibinfo {volume} {90}},\ \bibinfo {pages} {035006} (\bibinfo {year} {2018})}\BibitemShut {NoStop}%
\bibitem [{\citenamefont {Hotter}\ \emph {et~al.}(2024)\citenamefont {Hotter}, \citenamefont {Ritsch},\ and\ \citenamefont {Gietka}}]{Hotter2024Combining}%
  \BibitemOpen
  \bibfield  {author} {\bibinfo {author} {\bibfnamefont {C.}~\bibnamefont {Hotter}}, \bibinfo {author} {\bibfnamefont {H.}~\bibnamefont {Ritsch}},\ and\ \bibinfo {author} {\bibfnamefont {K.}~\bibnamefont {Gietka}},\ }\bibfield  {title} {\bibinfo {title} {Combining critical and quantum metrology},\ }\href {https://doi.org/10.1103/PhysRevLett.132.060801} {\bibfield  {journal} {\bibinfo  {journal} {Phys. Rev. Lett.}\ }\textbf {\bibinfo {volume} {132}},\ \bibinfo {pages} {060801} (\bibinfo {year} {2024})}\BibitemShut {NoStop}%
\bibitem [{\citenamefont {Baak}\ and\ \citenamefont {Fischer}(2024)}]{baak2024self}%
  \BibitemOpen
  \bibfield  {author} {\bibinfo {author} {\bibfnamefont {J.-G.}\ \bibnamefont {Baak}}\ and\ \bibinfo {author} {\bibfnamefont {U.~R.}\ \bibnamefont {Fischer}},\ }\bibfield  {title} {\bibinfo {title} {Self-consistent many-body metrology},\ }\href {https://journals.aps.org/prl/abstract/10.1103/PhysRevLett.132.240803} {\bibfield  {journal} {\bibinfo  {journal} {Physical Review Letters}\ }\textbf {\bibinfo {volume} {132}},\ \bibinfo {pages} {240803} (\bibinfo {year} {2024})}\BibitemShut {NoStop}%
\bibitem [{\citenamefont {Zhou}\ \emph {et~al.}(2020{\natexlab{a}})\citenamefont {Zhou}, \citenamefont {Choi}, \citenamefont {Choi}, \citenamefont {Landig}, \citenamefont {Douglas}, \citenamefont {Isoya}, \citenamefont {Jelezko}, \citenamefont {Onoda}, \citenamefont {Sumiya}, \citenamefont {Cappellaro} \emph {et~al.}}]{zhou2020quantum}%
  \BibitemOpen
  \bibfield  {author} {\bibinfo {author} {\bibfnamefont {H.}~\bibnamefont {Zhou}}, \bibinfo {author} {\bibfnamefont {J.}~\bibnamefont {Choi}}, \bibinfo {author} {\bibfnamefont {S.}~\bibnamefont {Choi}}, \bibinfo {author} {\bibfnamefont {R.}~\bibnamefont {Landig}}, \bibinfo {author} {\bibfnamefont {A.~M.}\ \bibnamefont {Douglas}}, \bibinfo {author} {\bibfnamefont {J.}~\bibnamefont {Isoya}}, \bibinfo {author} {\bibfnamefont {F.}~\bibnamefont {Jelezko}}, \bibinfo {author} {\bibfnamefont {S.}~\bibnamefont {Onoda}}, \bibinfo {author} {\bibfnamefont {H.}~\bibnamefont {Sumiya}}, \bibinfo {author} {\bibfnamefont {P.}~\bibnamefont {Cappellaro}}, \emph {et~al.},\ }\bibfield  {title} {\bibinfo {title} {Quantum metrology with strongly interacting spin systems},\ }\href {https://journals.aps.org/prx/abstract/10.1103/PhysRevX.10.031003} {\bibfield  {journal} {\bibinfo  {journal} {Physical review X}\ }\textbf {\bibinfo {volume} {10}},\ \bibinfo {pages} {031003} (\bibinfo {year} {2020}{\natexlab{a}})}\BibitemShut {NoStop}%
\bibitem [{\citenamefont {Tsang}(2013)}]{Tsang2013}%
  \BibitemOpen
  \bibfield  {author} {\bibinfo {author} {\bibfnamefont {M.}~\bibnamefont {Tsang}},\ }\bibfield  {title} {\bibinfo {title} {Quantum transition-edge detectors},\ }\href {https://doi.org/10.1103/PhysRevA.88.021801} {\bibfield  {journal} {\bibinfo  {journal} {Physical Review A}\ }\textbf {\bibinfo {volume} {88}},\ \bibinfo {pages} {021801} (\bibinfo {year} {2013})}\BibitemShut {NoStop}%
\bibitem [{\citenamefont {Chu}\ \emph {et~al.}(2021)\citenamefont {Chu}, \citenamefont {Zhang}, \citenamefont {Yu},\ and\ \citenamefont {Cai}}]{Chu2021Dynamic}%
  \BibitemOpen
  \bibfield  {author} {\bibinfo {author} {\bibfnamefont {Y.}~\bibnamefont {Chu}}, \bibinfo {author} {\bibfnamefont {S.}~\bibnamefont {Zhang}}, \bibinfo {author} {\bibfnamefont {B.}~\bibnamefont {Yu}},\ and\ \bibinfo {author} {\bibfnamefont {J.}~\bibnamefont {Cai}},\ }\bibfield  {title} {\bibinfo {title} {Dynamic framework for criticality-enhanced quantum sensing},\ }\href {https://doi.org/10.1103/PhysRevLett.126.010502} {\bibfield  {journal} {\bibinfo  {journal} {Physical Review Letters}\ }\textbf {\bibinfo {volume} {126}},\ \bibinfo {pages} {010502} (\bibinfo {year} {2021})}\BibitemShut {NoStop}%
\bibitem [{\citenamefont {Guan}\ and\ \citenamefont {Lewis-Swan}(2021)}]{Guan2021Identifying}%
  \BibitemOpen
  \bibfield  {author} {\bibinfo {author} {\bibfnamefont {Q.}~\bibnamefont {Guan}}\ and\ \bibinfo {author} {\bibfnamefont {R.~J.}\ \bibnamefont {Lewis-Swan}},\ }\bibfield  {title} {\bibinfo {title} {Identifying and harnessing dynamical phase transitions for quantum-enhanced sensing},\ }\href {https://doi.org/10.1103/PhysRevResearch.3.033199} {\bibfield  {journal} {\bibinfo  {journal} {Physical Review Research}\ }\textbf {\bibinfo {volume} {3}},\ \bibinfo {pages} {033199} (\bibinfo {year} {2021})}\BibitemShut {NoStop}%
\bibitem [{\citenamefont {Shi}\ \emph {et~al.}(2024)\citenamefont {Shi}, \citenamefont {Guan},\ and\ \citenamefont {Yang}}]{Shi2024Universal}%
  \BibitemOpen
  \bibfield  {author} {\bibinfo {author} {\bibfnamefont {H.-L.}\ \bibnamefont {Shi}}, \bibinfo {author} {\bibfnamefont {X.-W.}\ \bibnamefont {Guan}},\ and\ \bibinfo {author} {\bibfnamefont {J.}~\bibnamefont {Yang}},\ }\bibfield  {title} {\bibinfo {title} {Universal shot-noise limit for quantum metrology with local hamiltonians},\ }\href {http://dx.doi.org/10.1103/PhysRevLett.132.100803} {\bibfield  {journal} {\bibinfo  {journal} {Physical Review Letters}\ }\textbf {\bibinfo {volume} {132}},\ \bibinfo {pages} {100803} (\bibinfo {year} {2024})}\BibitemShut {NoStop}%
\bibitem [{\citenamefont {He}\ \emph {et~al.}(2023)\citenamefont {He}, \citenamefont {Yousefjani},\ and\ \citenamefont {Bayat}}]{He2023Stark}%
  \BibitemOpen
  \bibfield  {author} {\bibinfo {author} {\bibfnamefont {X.}~\bibnamefont {He}}, \bibinfo {author} {\bibfnamefont {R.}~\bibnamefont {Yousefjani}},\ and\ \bibinfo {author} {\bibfnamefont {A.}~\bibnamefont {Bayat}},\ }\bibfield  {title} {\bibinfo {title} {Stark localization as a resource for weak-field sensing with super-heisenberg precision},\ }\href {https://doi.org/10.1103/PhysRevLett.131.010801} {\bibfield  {journal} {\bibinfo  {journal} {Physical Review Letters}\ }\textbf {\bibinfo {volume} {131}},\ \bibinfo {pages} {010801} (\bibinfo {year} {2023})}\BibitemShut {NoStop}%
\bibitem [{\citenamefont {Manshouri}\ \emph {et~al.}(2024)\citenamefont {Manshouri}, \citenamefont {Zarei}, \citenamefont {Abdi}, \citenamefont {Bose},\ and\ \citenamefont {Bayat}}]{manshouri2024quantum}%
  \BibitemOpen
  \bibfield  {author} {\bibinfo {author} {\bibfnamefont {H.}~\bibnamefont {Manshouri}}, \bibinfo {author} {\bibfnamefont {M.}~\bibnamefont {Zarei}}, \bibinfo {author} {\bibfnamefont {M.}~\bibnamefont {Abdi}}, \bibinfo {author} {\bibfnamefont {S.}~\bibnamefont {Bose}},\ and\ \bibinfo {author} {\bibfnamefont {A.}~\bibnamefont {Bayat}},\ }\bibfield  {title} {\bibinfo {title} {Quantum enhanced sensitivity through many-body bloch oscillations},\ }\href {https://doi.org/10.48550/arXiv.2406.13921} {\bibfield  {journal} {\bibinfo  {journal} {arXiv preprint arXiv:2406.13921}\ } (\bibinfo {year} {2024})}\BibitemShut {NoStop}%
\bibitem [{\citenamefont {Sahoo}\ \emph {et~al.}(2024)\citenamefont {Sahoo}, \citenamefont {Mishra},\ and\ \citenamefont {Rakshit}}]{Sahoo2024Localization}%
  \BibitemOpen
  \bibfield  {author} {\bibinfo {author} {\bibfnamefont {A.}~\bibnamefont {Sahoo}}, \bibinfo {author} {\bibfnamefont {U.}~\bibnamefont {Mishra}},\ and\ \bibinfo {author} {\bibfnamefont {D.}~\bibnamefont {Rakshit}},\ }\bibfield  {title} {\bibinfo {title} {Localization-driven quantum sensing},\ }\href {https://doi.org/10.1103/PhysRevA.109.L030601} {\bibfield  {journal} {\bibinfo  {journal} {Physical Review A}\ }\textbf {\bibinfo {volume} {109}},\ \bibinfo {pages} {L030601} (\bibinfo {year} {2024})}\BibitemShut {NoStop}%
\bibitem [{\citenamefont {Garbe}\ \emph {et~al.}(2020)\citenamefont {Garbe}, \citenamefont {Bina}, \citenamefont {Keller}, \citenamefont {Paris},\ and\ \citenamefont {Felicetti}}]{Garbe2020Critical}%
  \BibitemOpen
  \bibfield  {author} {\bibinfo {author} {\bibfnamefont {L.}~\bibnamefont {Garbe}}, \bibinfo {author} {\bibfnamefont {M.}~\bibnamefont {Bina}}, \bibinfo {author} {\bibfnamefont {A.}~\bibnamefont {Keller}}, \bibinfo {author} {\bibfnamefont {M.~G.~A.}\ \bibnamefont {Paris}},\ and\ \bibinfo {author} {\bibfnamefont {S.}~\bibnamefont {Felicetti}},\ }\bibfield  {title} {\bibinfo {title} {Critical quantum metrology with a finite-component quantum phase transition},\ }\href {https://doi.org/10.1103/PhysRevLett.124.120504} {\bibfield  {journal} {\bibinfo  {journal} {Physical Review Letters}\ }\textbf {\bibinfo {volume} {124}},\ \bibinfo {pages} {120504} (\bibinfo {year} {2020})}\BibitemShut {NoStop}%
\bibitem [{\citenamefont {Garbe}\ \emph {et~al.}(2022)\citenamefont {Garbe}, \citenamefont {Abah}, \citenamefont {Felicetti},\ and\ \citenamefont {Puebla}}]{Garbe2022Critical}%
  \BibitemOpen
  \bibfield  {author} {\bibinfo {author} {\bibfnamefont {L.}~\bibnamefont {Garbe}}, \bibinfo {author} {\bibfnamefont {O.}~\bibnamefont {Abah}}, \bibinfo {author} {\bibfnamefont {S.}~\bibnamefont {Felicetti}},\ and\ \bibinfo {author} {\bibfnamefont {R.}~\bibnamefont {Puebla}},\ }\bibfield  {title} {\bibinfo {title} {Critical quantum metrology with fully-connected models: from heisenberg to kibble–zurek scaling},\ }\href {https://doi.org/10.1088/2058-9565/ac6ca5} {\bibfield  {journal} {\bibinfo  {journal} {Quantum Science and Technology}\ }\textbf {\bibinfo {volume} {7}},\ \bibinfo {pages} {035010} (\bibinfo {year} {2022})}\BibitemShut {NoStop}%
\bibitem [{\citenamefont {Ilias}\ \emph {et~al.}(2022)\citenamefont {Ilias}, \citenamefont {Yang}, \citenamefont {Huelga},\ and\ \citenamefont {Plenio}}]{Ilias2022Criticality}%
  \BibitemOpen
  \bibfield  {author} {\bibinfo {author} {\bibfnamefont {T.}~\bibnamefont {Ilias}}, \bibinfo {author} {\bibfnamefont {D.}~\bibnamefont {Yang}}, \bibinfo {author} {\bibfnamefont {S.~F.}\ \bibnamefont {Huelga}},\ and\ \bibinfo {author} {\bibfnamefont {M.~B.}\ \bibnamefont {Plenio}},\ }\bibfield  {title} {\bibinfo {title} {Criticality-enhanced quantum sensing via continuous measurement},\ }\href {https://doi.org/10.1103/PRXQuantum.3.010354} {\bibfield  {journal} {\bibinfo  {journal} {PRX Quantum}\ }\textbf {\bibinfo {volume} {3}},\ \bibinfo {pages} {010354} (\bibinfo {year} {2022})}\BibitemShut {NoStop}%
\bibitem [{\citenamefont {Gietka}\ \emph {et~al.}(2022)\citenamefont {Gietka}, \citenamefont {Ruks},\ and\ \citenamefont {Busch}}]{Gietka2022}%
  \BibitemOpen
  \bibfield  {author} {\bibinfo {author} {\bibfnamefont {K.}~\bibnamefont {Gietka}}, \bibinfo {author} {\bibfnamefont {L.}~\bibnamefont {Ruks}},\ and\ \bibinfo {author} {\bibfnamefont {T.}~\bibnamefont {Busch}},\ }\bibfield  {title} {\bibinfo {title} {Understanding and improving critical metrology. quenching superradiant light-matter systems beyond the critical point},\ }\href {https://doi.org/10.22331/q-2022-04-27-700} {\bibfield  {journal} {\bibinfo  {journal} {Quantum}\ }\textbf {\bibinfo {volume} {6}},\ \bibinfo {pages} {700} (\bibinfo {year} {2022})}\BibitemShut {NoStop}%
\bibitem [{\citenamefont {G{\'o}recki}\ \emph {et~al.}(2024)\citenamefont {G{\'o}recki}, \citenamefont {Albarelli}, \citenamefont {Felicetti}, \citenamefont {Di~Candia},\ and\ \citenamefont {Maccone}}]{gorecki2024interplay}%
  \BibitemOpen
  \bibfield  {author} {\bibinfo {author} {\bibfnamefont {W.}~\bibnamefont {G{\'o}recki}}, \bibinfo {author} {\bibfnamefont {F.}~\bibnamefont {Albarelli}}, \bibinfo {author} {\bibfnamefont {S.}~\bibnamefont {Felicetti}}, \bibinfo {author} {\bibfnamefont {R.}~\bibnamefont {Di~Candia}},\ and\ \bibinfo {author} {\bibfnamefont {L.}~\bibnamefont {Maccone}},\ }\bibfield  {title} {\bibinfo {title} {Interplay between time and energy in bosonic noisy quantum metrology},\ }\href {https://doi.org/10.48550/arXiv.2409.18791} {\bibfield  {journal} {\bibinfo  {journal} {arXiv preprint arXiv:2409.18791}\ } (\bibinfo {year} {2024})}\BibitemShut {NoStop}%
\bibitem [{\citenamefont {Alushi}\ \emph {et~al.}(2024)\citenamefont {Alushi}, \citenamefont {G\'orecki}, \citenamefont {Felicetti},\ and\ \citenamefont {Di~Candia}}]{Alushi2024Optimality}%
  \BibitemOpen
  \bibfield  {author} {\bibinfo {author} {\bibfnamefont {U.}~\bibnamefont {Alushi}}, \bibinfo {author} {\bibfnamefont {W.}~\bibnamefont {G\'orecki}}, \bibinfo {author} {\bibfnamefont {S.}~\bibnamefont {Felicetti}},\ and\ \bibinfo {author} {\bibfnamefont {R.}~\bibnamefont {Di~Candia}},\ }\bibfield  {title} {\bibinfo {title} {Optimality and noise resilience of critical quantum sensing},\ }\href {https://doi.org/10.1103/PhysRevLett.133.040801} {\bibfield  {journal} {\bibinfo  {journal} {Phys. Rev. Lett.}\ }\textbf {\bibinfo {volume} {133}},\ \bibinfo {pages} {040801} (\bibinfo {year} {2024})}\BibitemShut {NoStop}%
\bibitem [{\citenamefont {Chu}\ \emph {et~al.}(2023)\citenamefont {Chu}, \citenamefont {Li},\ and\ \citenamefont {Cai}}]{Chu2023Strong}%
  \BibitemOpen
  \bibfield  {author} {\bibinfo {author} {\bibfnamefont {Y.}~\bibnamefont {Chu}}, \bibinfo {author} {\bibfnamefont {X.}~\bibnamefont {Li}},\ and\ \bibinfo {author} {\bibfnamefont {J.}~\bibnamefont {Cai}},\ }\bibfield  {title} {\bibinfo {title} {Strong quantum metrological limit from many-body physics},\ }\href {http://dx.doi.org/10.1103/PhysRevLett.130.170801} {\bibfield  {journal} {\bibinfo  {journal} {Physical Review Letters}\ }\textbf {\bibinfo {volume} {130}},\ \bibinfo {pages} {170801} (\bibinfo {year} {2023})}\BibitemShut {NoStop}%
\bibitem [{\citenamefont {Zanardi}\ \emph {et~al.}(2008)\citenamefont {Zanardi}, \citenamefont {Paris},\ and\ \citenamefont {Venuti}}]{Zanardi2008}%
  \BibitemOpen
  \bibfield  {author} {\bibinfo {author} {\bibfnamefont {P.}~\bibnamefont {Zanardi}}, \bibinfo {author} {\bibfnamefont {M.~G.~A.}\ \bibnamefont {Paris}},\ and\ \bibinfo {author} {\bibfnamefont {L.~C.}\ \bibnamefont {Venuti}},\ }\bibfield  {title} {\bibinfo {title} {Quantum criticality as a resource for quantum estimation},\ }\href {https://doi.org/10.1103/physreva.78.042105} {\bibfield  {journal} {\bibinfo  {journal} {Physical Review A}\ }\textbf {\bibinfo {volume} {78}},\ \bibinfo {pages} {042105} (\bibinfo {year} {2008})}\BibitemShut {NoStop}%
\bibitem [{\citenamefont {Banchi}\ \emph {et~al.}(2014)\citenamefont {Banchi}, \citenamefont {Giorda},\ and\ \citenamefont {Zanardi}}]{Banchi2014}%
  \BibitemOpen
  \bibfield  {author} {\bibinfo {author} {\bibfnamefont {L.}~\bibnamefont {Banchi}}, \bibinfo {author} {\bibfnamefont {P.}~\bibnamefont {Giorda}},\ and\ \bibinfo {author} {\bibfnamefont {P.}~\bibnamefont {Zanardi}},\ }\bibfield  {title} {\bibinfo {title} {Quantum information-geometry of dissipative quantum phase transitions},\ }\href {https://doi.org/10.1103/PhysRevE.89.022102} {\bibfield  {journal} {\bibinfo  {journal} {Physical Review E}\ }\textbf {\bibinfo {volume} {89}},\ \bibinfo {pages} {022102} (\bibinfo {year} {2014})}\BibitemShut {NoStop}%
\bibitem [{\citenamefont {Rams}\ \emph {et~al.}(2018)\citenamefont {Rams}, \citenamefont {Sierant}, \citenamefont {Dutta}, \citenamefont {Horodecki},\ and\ \citenamefont {Zakrzewski}}]{Rams2018Limits}%
  \BibitemOpen
  \bibfield  {author} {\bibinfo {author} {\bibfnamefont {M.~M.}\ \bibnamefont {Rams}}, \bibinfo {author} {\bibfnamefont {P.}~\bibnamefont {Sierant}}, \bibinfo {author} {\bibfnamefont {O.}~\bibnamefont {Dutta}}, \bibinfo {author} {\bibfnamefont {P.}~\bibnamefont {Horodecki}},\ and\ \bibinfo {author} {\bibfnamefont {J.}~\bibnamefont {Zakrzewski}},\ }\bibfield  {title} {\bibinfo {title} {At the limits of criticality-based quantum metrology: Apparent super-heisenberg scaling revisited},\ }\href {https://doi.org/10.1103/PhysRevX.8.021022} {\bibfield  {journal} {\bibinfo  {journal} {Physical Review X}\ }\textbf {\bibinfo {volume} {8}},\ \bibinfo {pages} {021022} (\bibinfo {year} {2018})}\BibitemShut {NoStop}%
\bibitem [{\citenamefont {Gietka}\ \emph {et~al.}(2021)\citenamefont {Gietka}, \citenamefont {Metz}, \citenamefont {Keller},\ and\ \citenamefont {Li}}]{Gietka2021adiabaticcritical}%
  \BibitemOpen
  \bibfield  {author} {\bibinfo {author} {\bibfnamefont {K.}~\bibnamefont {Gietka}}, \bibinfo {author} {\bibfnamefont {F.}~\bibnamefont {Metz}}, \bibinfo {author} {\bibfnamefont {T.}~\bibnamefont {Keller}},\ and\ \bibinfo {author} {\bibfnamefont {J.}~\bibnamefont {Li}},\ }\bibfield  {title} {\bibinfo {title} {Adiabatic critical quantum metrology cannot reach the {H}eisenberg limit even when shortcuts to adiabaticity are applied},\ }\href {https://doi.org/10.22331/q-2021-07-01-489} {\bibfield  {journal} {\bibinfo  {journal} {{Quantum}}\ }\textbf {\bibinfo {volume} {5}},\ \bibinfo {pages} {489} (\bibinfo {year} {2021})}\BibitemShut {NoStop}%
\bibitem [{\citenamefont {Invernizzi}\ \emph {et~al.}(2008)\citenamefont {Invernizzi}, \citenamefont {Korbman}, \citenamefont {Campos~Venuti},\ and\ \citenamefont {Paris}}]{Invernizzi2008}%
  \BibitemOpen
  \bibfield  {author} {\bibinfo {author} {\bibfnamefont {C.}~\bibnamefont {Invernizzi}}, \bibinfo {author} {\bibfnamefont {M.}~\bibnamefont {Korbman}}, \bibinfo {author} {\bibfnamefont {L.}~\bibnamefont {Campos~Venuti}},\ and\ \bibinfo {author} {\bibfnamefont {M.~G.~A.}\ \bibnamefont {Paris}},\ }\bibfield  {title} {\bibinfo {title} {Optimal quantum estimation in spin systems at criticality},\ }\href {https://doi.org/10.1103/PhysRevA.78.042106} {\bibfield  {journal} {\bibinfo  {journal} {Physical Review A}\ }\textbf {\bibinfo {volume} {78}},\ \bibinfo {pages} {042106} (\bibinfo {year} {2008})}\BibitemShut {NoStop}%
\bibitem [{\citenamefont {Sarkar}\ \emph {et~al.}(2022)\citenamefont {Sarkar}, \citenamefont {Mukhopadhyay}, \citenamefont {Alase},\ and\ \citenamefont {Bayat}}]{Sarkar2022Free}%
  \BibitemOpen
  \bibfield  {author} {\bibinfo {author} {\bibfnamefont {S.}~\bibnamefont {Sarkar}}, \bibinfo {author} {\bibfnamefont {C.}~\bibnamefont {Mukhopadhyay}}, \bibinfo {author} {\bibfnamefont {A.}~\bibnamefont {Alase}},\ and\ \bibinfo {author} {\bibfnamefont {A.}~\bibnamefont {Bayat}},\ }\bibfield  {title} {\bibinfo {title} {Free-fermionic topological quantum sensors},\ }\href {https://doi.org/10.1103/PhysRevLett.129.090503} {\bibfield  {journal} {\bibinfo  {journal} {Physical Review Letters}\ }\textbf {\bibinfo {volume} {129}},\ \bibinfo {pages} {090503} (\bibinfo {year} {2022})}\BibitemShut {NoStop}%
\bibitem [{\citenamefont {Salvia}\ \emph {et~al.}(2023)\citenamefont {Salvia}, \citenamefont {Mehboudi},\ and\ \citenamefont {Perarnau-Llobet}}]{Salvia2023Critical}%
  \BibitemOpen
  \bibfield  {author} {\bibinfo {author} {\bibfnamefont {R.}~\bibnamefont {Salvia}}, \bibinfo {author} {\bibfnamefont {M.}~\bibnamefont {Mehboudi}},\ and\ \bibinfo {author} {\bibfnamefont {M.}~\bibnamefont {Perarnau-Llobet}},\ }\bibfield  {title} {\bibinfo {title} {Critical quantum metrology assisted by real-time feedback control},\ }\href {https://doi.org/10.1103/PhysRevLett.130.240803} {\bibfield  {journal} {\bibinfo  {journal} {Physical Review Letters}\ }\textbf {\bibinfo {volume} {130}},\ \bibinfo {pages} {240803} (\bibinfo {year} {2023})}\BibitemShut {NoStop}%
\bibitem [{\citenamefont {Mukhopadhyay}\ and\ \citenamefont {Bayat}(2024)}]{Mukhopadhyay2024Modular}%
  \BibitemOpen
  \bibfield  {author} {\bibinfo {author} {\bibfnamefont {C.}~\bibnamefont {Mukhopadhyay}}\ and\ \bibinfo {author} {\bibfnamefont {A.}~\bibnamefont {Bayat}},\ }\bibfield  {title} {\bibinfo {title} {Modular many-body quantum sensors},\ }\href {https://doi.org/10.1103/PhysRevLett.133.120601} {\bibfield  {journal} {\bibinfo  {journal} {Phys. Rev. Lett.}\ }\textbf {\bibinfo {volume} {133}},\ \bibinfo {pages} {120601} (\bibinfo {year} {2024})}\BibitemShut {NoStop}%
\bibitem [{\citenamefont {Zanardi}\ \emph {et~al.}(2007{\natexlab{a}})\citenamefont {Zanardi}, \citenamefont {Quan}, \citenamefont {Wang},\ and\ \citenamefont {Sun}}]{Zanardi2007Mixed}%
  \BibitemOpen
  \bibfield  {author} {\bibinfo {author} {\bibfnamefont {P.}~\bibnamefont {Zanardi}}, \bibinfo {author} {\bibfnamefont {H.~T.}\ \bibnamefont {Quan}}, \bibinfo {author} {\bibfnamefont {X.}~\bibnamefont {Wang}},\ and\ \bibinfo {author} {\bibfnamefont {C.~P.}\ \bibnamefont {Sun}},\ }\bibfield  {title} {\bibinfo {title} {Mixed-state fidelity and quantum criticality at finite temperature},\ }\href {https://doi.org/10.1103/PhysRevA.75.032109} {\bibfield  {journal} {\bibinfo  {journal} {Physical Review A}\ }\textbf {\bibinfo {volume} {75}},\ \bibinfo {pages} {032109} (\bibinfo {year} {2007}{\natexlab{a}})}\BibitemShut {NoStop}%
\bibitem [{\citenamefont {Gammelmark}\ and\ \citenamefont {Mølmer}(2011)}]{Gammelmark2011}%
  \BibitemOpen
  \bibfield  {author} {\bibinfo {author} {\bibfnamefont {S.}~\bibnamefont {Gammelmark}}\ and\ \bibinfo {author} {\bibfnamefont {K.}~\bibnamefont {Mølmer}},\ }\bibfield  {title} {\bibinfo {title} {Phase transitions and heisenberg limited metrology in an ising chain interacting with a single-mode cavity field},\ }\href {https://doi.org/10.1088/1367-2630/13/5/053035} {\bibfield  {journal} {\bibinfo  {journal} {New Journal of Physics}\ }\textbf {\bibinfo {volume} {13}},\ \bibinfo {pages} {053035} (\bibinfo {year} {2011})}\BibitemShut {NoStop}%
\bibitem [{\citenamefont {Mehboudi}\ \emph {et~al.}(2016)\citenamefont {Mehboudi}, \citenamefont {Correa},\ and\ \citenamefont {Sanpera}}]{Mehboudi2016}%
  \BibitemOpen
  \bibfield  {author} {\bibinfo {author} {\bibfnamefont {M.}~\bibnamefont {Mehboudi}}, \bibinfo {author} {\bibfnamefont {L.~A.}\ \bibnamefont {Correa}},\ and\ \bibinfo {author} {\bibfnamefont {A.}~\bibnamefont {Sanpera}},\ }\bibfield  {title} {\bibinfo {title} {Achieving sub-shot-noise sensing at finite temperatures},\ }\href {https://doi.org/10.1103/PhysRevA.94.042121} {\bibfield  {journal} {\bibinfo  {journal} {Physical Review A}\ }\textbf {\bibinfo {volume} {94}},\ \bibinfo {pages} {042121} (\bibinfo {year} {2016})}\BibitemShut {NoStop}%
\bibitem [{\citenamefont {Mehboudi}\ \emph {et~al.}(2019)\citenamefont {Mehboudi}, \citenamefont {Sanpera},\ and\ \citenamefont {Correa}}]{Mehboudi2019}%
  \BibitemOpen
  \bibfield  {author} {\bibinfo {author} {\bibfnamefont {M.}~\bibnamefont {Mehboudi}}, \bibinfo {author} {\bibfnamefont {A.}~\bibnamefont {Sanpera}},\ and\ \bibinfo {author} {\bibfnamefont {L.~A.}\ \bibnamefont {Correa}},\ }\bibfield  {title} {\bibinfo {title} {Thermometry in the quantum regime: recent theoretical progress},\ }\href {https://doi.org/10.1088/1751-8121/ab2828} {\bibfield  {journal} {\bibinfo  {journal} {Journal of Physics A: Mathematical and Theoretical}\ }\textbf {\bibinfo {volume} {52}},\ \bibinfo {pages} {303001} (\bibinfo {year} {2019})}\BibitemShut {NoStop}%
\bibitem [{\citenamefont {Abiuso}\ \emph {et~al.}(2024{\natexlab{b}})\citenamefont {Abiuso}, \citenamefont {Erdman}, \citenamefont {Ronen}, \citenamefont {Noé}, \citenamefont {Haack},\ and\ \citenamefont {Perarnau-Llobet}}]{Abiuso_2024Optimal}%
  \BibitemOpen
  \bibfield  {author} {\bibinfo {author} {\bibfnamefont {P.}~\bibnamefont {Abiuso}}, \bibinfo {author} {\bibfnamefont {P.~A.}\ \bibnamefont {Erdman}}, \bibinfo {author} {\bibfnamefont {M.}~\bibnamefont {Ronen}}, \bibinfo {author} {\bibfnamefont {F.}~\bibnamefont {Noé}}, \bibinfo {author} {\bibfnamefont {G.}~\bibnamefont {Haack}},\ and\ \bibinfo {author} {\bibfnamefont {M.}~\bibnamefont {Perarnau-Llobet}},\ }\bibfield  {title} {\bibinfo {title} {Optimal thermometers with spin networks},\ }\href {https://doi.org/10.1088/2058-9565/ad37d3} {\bibfield  {journal} {\bibinfo  {journal} {Quantum Science and Technology}\ }\textbf {\bibinfo {volume} {9}},\ \bibinfo {pages} {035008} (\bibinfo {year} {2024}{\natexlab{b}})}\BibitemShut {NoStop}%
\bibitem [{\citenamefont {Yu}\ \emph {et~al.}(2024)\citenamefont {Yu}, \citenamefont {Nguyen},\ and\ \citenamefont {Nimmrichter}}]{Yu2024Criticality}%
  \BibitemOpen
  \bibfield  {author} {\bibinfo {author} {\bibfnamefont {M.}~\bibnamefont {Yu}}, \bibinfo {author} {\bibfnamefont {H.~C.}\ \bibnamefont {Nguyen}},\ and\ \bibinfo {author} {\bibfnamefont {S.}~\bibnamefont {Nimmrichter}},\ }\bibfield  {title} {\bibinfo {title} {Criticality-enhanced precision in phase thermometry},\ }\href {https://doi.org/10.1103/PhysRevResearch.6.043094} {\bibfield  {journal} {\bibinfo  {journal} {Phys. Rev. Res.}\ }\textbf {\bibinfo {volume} {6}},\ \bibinfo {pages} {043094} (\bibinfo {year} {2024})}\BibitemShut {NoStop}%
\bibitem [{\citenamefont {Ostermann}\ and\ \citenamefont {Gietka}(2024)}]{Ostermann2024Temperature}%
  \BibitemOpen
  \bibfield  {author} {\bibinfo {author} {\bibfnamefont {L.}~\bibnamefont {Ostermann}}\ and\ \bibinfo {author} {\bibfnamefont {K.}~\bibnamefont {Gietka}},\ }\bibfield  {title} {\bibinfo {title} {Temperature-enhanced critical quantum metrology},\ }\href {https://doi.org/10.1103/PhysRevA.109.L050601} {\bibfield  {journal} {\bibinfo  {journal} {Phys. Rev. A}\ }\textbf {\bibinfo {volume} {109}},\ \bibinfo {pages} {L050601} (\bibinfo {year} {2024})}\BibitemShut {NoStop}%
\bibitem [{\citenamefont {Fern\'andez-Lorenzo}\ and\ \citenamefont {Porras}(2017)}]{FernandezLorenzo2017}%
  \BibitemOpen
  \bibfield  {author} {\bibinfo {author} {\bibfnamefont {S.}~\bibnamefont {Fern\'andez-Lorenzo}}\ and\ \bibinfo {author} {\bibfnamefont {D.}~\bibnamefont {Porras}},\ }\bibfield  {title} {\bibinfo {title} {Quantum sensing close to a dissipative phase transition: Symmetry breaking and criticality as metrological resources},\ }\href {https://doi.org/10.1103/PhysRevA.96.013817} {\bibfield  {journal} {\bibinfo  {journal} {Physical Review A}\ }\textbf {\bibinfo {volume} {96}},\ \bibinfo {pages} {013817} (\bibinfo {year} {2017})}\BibitemShut {NoStop}%
\bibitem [{\citenamefont {Fern\'andez-Lorenzo}\ \emph {et~al.}(2018)\citenamefont {Fern\'andez-Lorenzo}, \citenamefont {Dunningham},\ and\ \citenamefont {Porras}}]{fernandez2018heisenberg}%
  \BibitemOpen
  \bibfield  {author} {\bibinfo {author} {\bibfnamefont {S.}~\bibnamefont {Fern\'andez-Lorenzo}}, \bibinfo {author} {\bibfnamefont {J.~A.}\ \bibnamefont {Dunningham}},\ and\ \bibinfo {author} {\bibfnamefont {D.}~\bibnamefont {Porras}},\ }\bibfield  {title} {\bibinfo {title} {Heisenberg scaling with classical long-range correlations},\ }\href {https://doi.org/10.1103/PhysRevA.97.023843} {\bibfield  {journal} {\bibinfo  {journal} {Physical Review A}\ }\textbf {\bibinfo {volume} {97}},\ \bibinfo {pages} {023843} (\bibinfo {year} {2018})}\BibitemShut {NoStop}%
\bibitem [{\citenamefont {Marzolino}\ and\ \citenamefont {Prosen}(2017)}]{Marzolino2017}%
  \BibitemOpen
  \bibfield  {author} {\bibinfo {author} {\bibfnamefont {U.}~\bibnamefont {Marzolino}}\ and\ \bibinfo {author} {\bibfnamefont {T.}~\bibnamefont {Prosen}},\ }\bibfield  {title} {\bibinfo {title} {Fisher information approach to nonequilibrium phase transitions in a quantum xxz spin chain with boundary noise},\ }\href {https://doi.org/10.1103/PhysRevB.96.104402} {\bibfield  {journal} {\bibinfo  {journal} {Physical Review B}\ }\textbf {\bibinfo {volume} {96}},\ \bibinfo {pages} {104402} (\bibinfo {year} {2017})}\BibitemShut {NoStop}%
\bibitem [{\citenamefont {Raghunandan}\ \emph {et~al.}(2018)\citenamefont {Raghunandan}, \citenamefont {Wrachtrup},\ and\ \citenamefont {Weimer}}]{Raghunandan2018}%
  \BibitemOpen
  \bibfield  {author} {\bibinfo {author} {\bibfnamefont {M.}~\bibnamefont {Raghunandan}}, \bibinfo {author} {\bibfnamefont {J.}~\bibnamefont {Wrachtrup}},\ and\ \bibinfo {author} {\bibfnamefont {H.}~\bibnamefont {Weimer}},\ }\bibfield  {title} {\bibinfo {title} {High-density quantum sensing with dissipative first order transitions},\ }\href {https://doi.org/10.1103/PhysRevLett.120.150501} {\bibfield  {journal} {\bibinfo  {journal} {Physical Review Letters}\ }\textbf {\bibinfo {volume} {120}},\ \bibinfo {pages} {150501} (\bibinfo {year} {2018})}\BibitemShut {NoStop}%
\bibitem [{\citenamefont {Di~Candia}\ \emph {et~al.}(2023)\citenamefont {Di~Candia}, \citenamefont {Minganti}, \citenamefont {Petrovnin}, \citenamefont {Paraoanu},\ and\ \citenamefont {Felicetti}}]{DiCandia2023}%
  \BibitemOpen
  \bibfield  {author} {\bibinfo {author} {\bibfnamefont {R.}~\bibnamefont {Di~Candia}}, \bibinfo {author} {\bibfnamefont {F.}~\bibnamefont {Minganti}}, \bibinfo {author} {\bibfnamefont {K.}~\bibnamefont {Petrovnin}}, \bibinfo {author} {\bibfnamefont {G.}~\bibnamefont {Paraoanu}},\ and\ \bibinfo {author} {\bibfnamefont {S.}~\bibnamefont {Felicetti}},\ }\bibfield  {title} {\bibinfo {title} {Critical parametric quantum sensing},\ }\href {http://dx.doi.org/10.1038/s41534-023-00690-z} {\bibfield  {journal} {\bibinfo  {journal} {npj Quantum Information}\ }\textbf {\bibinfo {volume} {9}},\ \bibinfo {pages} {23} (\bibinfo {year} {2023})}\BibitemShut {NoStop}%
\bibitem [{\citenamefont {Ilias}\ \emph {et~al.}(2024)\citenamefont {Ilias}, \citenamefont {Yang}, \citenamefont {Huelga},\ and\ \citenamefont {Plenio}}]{Ilias2024Criticality}%
  \BibitemOpen
  \bibfield  {author} {\bibinfo {author} {\bibfnamefont {T.}~\bibnamefont {Ilias}}, \bibinfo {author} {\bibfnamefont {D.}~\bibnamefont {Yang}}, \bibinfo {author} {\bibfnamefont {S.~F.}\ \bibnamefont {Huelga}},\ and\ \bibinfo {author} {\bibfnamefont {M.~B.}\ \bibnamefont {Plenio}},\ }\bibfield  {title} {\bibinfo {title} {Criticality-enhanced electric field gradient sensor with single trapped ions},\ }\href {http://dx.doi.org/10.1038/s41534-024-00833-w} {\bibfield  {journal} {\bibinfo  {journal} {npj Quantum Information}\ }\textbf {\bibinfo {volume} {10}},\ \bibinfo {pages} {36} (\bibinfo {year} {2024})}\BibitemShut {NoStop}%
\bibitem [{\citenamefont {Paris}(2008)}]{paris2008quantum}%
  \BibitemOpen
  \bibfield  {author} {\bibinfo {author} {\bibfnamefont {M.~G.}\ \bibnamefont {Paris}},\ }\bibfield  {title} {\bibinfo {title} {Quantum estimation for quantum technology},\ }\href {https://arxiv.org/abs/0804.2981} {\bibfield  {journal} {\bibinfo  {journal} {arXiv preprint arXiv:0804.2981}\ } (\bibinfo {year} {2008})}\BibitemShut {NoStop}%
\bibitem [{\citenamefont {Rigol}\ \emph {et~al.}(2008)\citenamefont {Rigol}, \citenamefont {Dunjko},\ and\ \citenamefont {Olshanii}}]{Rigol2008}%
  \BibitemOpen
  \bibfield  {author} {\bibinfo {author} {\bibfnamefont {M.}~\bibnamefont {Rigol}}, \bibinfo {author} {\bibfnamefont {V.}~\bibnamefont {Dunjko}},\ and\ \bibinfo {author} {\bibfnamefont {M.}~\bibnamefont {Olshanii}},\ }\bibfield  {title} {\bibinfo {title} {Thermalization and its mechanism for generic isolated quantum systems},\ }\href {https://doi.org/10.1038/nature06838} {\bibfield  {journal} {\bibinfo  {journal} {Nature}\ }\textbf {\bibinfo {volume} {452}},\ \bibinfo {pages} {854–858} (\bibinfo {year} {2008})}\BibitemShut {NoStop}%
\bibitem [{\citenamefont {Kollar}\ and\ \citenamefont {Eckstein}(2008)}]{Kollar2008}%
  \BibitemOpen
  \bibfield  {author} {\bibinfo {author} {\bibfnamefont {M.}~\bibnamefont {Kollar}}\ and\ \bibinfo {author} {\bibfnamefont {M.}~\bibnamefont {Eckstein}},\ }\bibfield  {title} {\bibinfo {title} {Relaxation of a one-dimensional mott insulator after an interaction quench},\ }\href {https://doi.org/10.1103/PhysRevA.78.013626} {\bibfield  {journal} {\bibinfo  {journal} {Physical Review A}\ }\textbf {\bibinfo {volume} {78}},\ \bibinfo {pages} {013626} (\bibinfo {year} {2008})}\BibitemShut {NoStop}%
\bibitem [{\citenamefont {Cassidy}\ \emph {et~al.}(2011)\citenamefont {Cassidy}, \citenamefont {Clark},\ and\ \citenamefont {Rigol}}]{Cassidy2011Generalized}%
  \BibitemOpen
  \bibfield  {author} {\bibinfo {author} {\bibfnamefont {A.~C.}\ \bibnamefont {Cassidy}}, \bibinfo {author} {\bibfnamefont {C.~W.}\ \bibnamefont {Clark}},\ and\ \bibinfo {author} {\bibfnamefont {M.}~\bibnamefont {Rigol}},\ }\bibfield  {title} {\bibinfo {title} {Generalized thermalization in an integrable lattice system},\ }\href {https://doi.org/10.1103/PhysRevLett.106.140405} {\bibfield  {journal} {\bibinfo  {journal} {Physical Review Letters}\ }\textbf {\bibinfo {volume} {106}},\ \bibinfo {pages} {140405} (\bibinfo {year} {2011})}\BibitemShut {NoStop}%
\bibitem [{\citenamefont {Eisert}\ \emph {et~al.}(2015)\citenamefont {Eisert}, \citenamefont {Friesdorf},\ and\ \citenamefont {Gogolin}}]{Eisert2015}%
  \BibitemOpen
  \bibfield  {author} {\bibinfo {author} {\bibfnamefont {J.}~\bibnamefont {Eisert}}, \bibinfo {author} {\bibfnamefont {M.}~\bibnamefont {Friesdorf}},\ and\ \bibinfo {author} {\bibfnamefont {C.}~\bibnamefont {Gogolin}},\ }\bibfield  {title} {\bibinfo {title} {Quantum many-body systems out of equilibrium},\ }\href {https://doi.org/10.1038/nphys3215} {\bibfield  {journal} {\bibinfo  {journal} {Nature Physics}\ }\textbf {\bibinfo {volume} {11}},\ \bibinfo {pages} {124–130} (\bibinfo {year} {2015})}\BibitemShut {NoStop}%
\bibitem [{\citenamefont {Baamara}\ \emph {et~al.}(2021)\citenamefont {Baamara}, \citenamefont {Sinatra},\ and\ \citenamefont {Gessner}}]{baamara2021squeezingnonlinearspinobservables}%
  \BibitemOpen
  \bibfield  {author} {\bibinfo {author} {\bibfnamefont {Y.}~\bibnamefont {Baamara}}, \bibinfo {author} {\bibfnamefont {A.}~\bibnamefont {Sinatra}},\ and\ \bibinfo {author} {\bibfnamefont {M.}~\bibnamefont {Gessner}},\ }\bibfield  {title} {\bibinfo {title} {Squeezing of nonlinear spin observables by one axis twisting in the presence of decoherence: An analytical study},\ }\href {https://arxiv.org/abs/2112.01786} {\bibfield  {journal} {\bibinfo  {journal} {arXiv preprint arXiv:2112.01786}\ } (\bibinfo {year} {2021})}\BibitemShut {NoStop}%
\bibitem [{\citenamefont {Chalopin}\ \emph {et~al.}(2018)\citenamefont {Chalopin}, \citenamefont {Bouazza}, \citenamefont {Evrard}, \citenamefont {Makhalov}, \citenamefont {Dreon}, \citenamefont {Dalibard}, \citenamefont {Sidorenkov},\ and\ \citenamefont {Nascimbene}}]{Chalopin_2018}%
  \BibitemOpen
  \bibfield  {author} {\bibinfo {author} {\bibfnamefont {T.}~\bibnamefont {Chalopin}}, \bibinfo {author} {\bibfnamefont {C.}~\bibnamefont {Bouazza}}, \bibinfo {author} {\bibfnamefont {A.}~\bibnamefont {Evrard}}, \bibinfo {author} {\bibfnamefont {V.}~\bibnamefont {Makhalov}}, \bibinfo {author} {\bibfnamefont {D.}~\bibnamefont {Dreon}}, \bibinfo {author} {\bibfnamefont {J.}~\bibnamefont {Dalibard}}, \bibinfo {author} {\bibfnamefont {L.~A.}\ \bibnamefont {Sidorenkov}},\ and\ \bibinfo {author} {\bibfnamefont {S.}~\bibnamefont {Nascimbene}},\ }\bibfield  {title} {\bibinfo {title} {Quantum-enhanced sensing using non-classical spin states of a highly magnetic atom},\ }\href {http://dx.doi.org/10.1038/s41467-018-07433-1} {\bibfield  {journal} {\bibinfo  {journal} {Nature communications}\ }\textbf {\bibinfo {volume} {9}},\ \bibinfo {pages} {4955} (\bibinfo {year} {2018})}\BibitemShut {NoStop}%
\bibitem [{\citenamefont {Gross}\ \emph {et~al.}(2010)\citenamefont {Gross}, \citenamefont {Zibold}, \citenamefont {Nicklas}, \citenamefont {Estève},\ and\ \citenamefont {Oberthaler}}]{Gross_2010}%
  \BibitemOpen
  \bibfield  {author} {\bibinfo {author} {\bibfnamefont {C.}~\bibnamefont {Gross}}, \bibinfo {author} {\bibfnamefont {T.}~\bibnamefont {Zibold}}, \bibinfo {author} {\bibfnamefont {E.}~\bibnamefont {Nicklas}}, \bibinfo {author} {\bibfnamefont {J.}~\bibnamefont {Estève}},\ and\ \bibinfo {author} {\bibfnamefont {M.~K.}\ \bibnamefont {Oberthaler}},\ }\bibfield  {title} {\bibinfo {title} {Nonlinear atom interferometer surpasses classical precision limit},\ }\href {https://doi.org/10.1038/nature08919} {\bibfield  {journal} {\bibinfo  {journal} {Nature}\ }\textbf {\bibinfo {volume} {464}},\ \bibinfo {pages} {1165–1169} (\bibinfo {year} {2010})}\BibitemShut {NoStop}%
\bibitem [{\citenamefont {Riedel}\ \emph {et~al.}(2010)\citenamefont {Riedel}, \citenamefont {Böhi}, \citenamefont {Li}, \citenamefont {Hänsch}, \citenamefont {Sinatra},\ and\ \citenamefont {Treutlein}}]{Riedel_2010}%
  \BibitemOpen
  \bibfield  {author} {\bibinfo {author} {\bibfnamefont {M.~F.}\ \bibnamefont {Riedel}}, \bibinfo {author} {\bibfnamefont {P.}~\bibnamefont {Böhi}}, \bibinfo {author} {\bibfnamefont {Y.}~\bibnamefont {Li}}, \bibinfo {author} {\bibfnamefont {T.~W.}\ \bibnamefont {Hänsch}}, \bibinfo {author} {\bibfnamefont {A.}~\bibnamefont {Sinatra}},\ and\ \bibinfo {author} {\bibfnamefont {P.}~\bibnamefont {Treutlein}},\ }\bibfield  {title} {\bibinfo {title} {Atom-chip-based generation of entanglement for quantum metrology},\ }\href {https://doi.org/10.1038/nature08988} {\bibfield  {journal} {\bibinfo  {journal} {Nature}\ }\textbf {\bibinfo {volume} {464}},\ \bibinfo {pages} {1170–1173} (\bibinfo {year} {2010})}\BibitemShut {NoStop}%
\bibitem [{\citenamefont {Braunstein}\ and\ \citenamefont {Caves}(1994)}]{first_quantum_fisher}%
  \BibitemOpen
  \bibfield  {author} {\bibinfo {author} {\bibfnamefont {S.~L.}\ \bibnamefont {Braunstein}}\ and\ \bibinfo {author} {\bibfnamefont {C.~M.}\ \bibnamefont {Caves}},\ }\bibfield  {title} {\bibinfo {title} {Statistical distance and the geometry of quantum states},\ }\href {https://doi.org/10.1103/PhysRevLett.72.3439} {\bibfield  {journal} {\bibinfo  {journal} {Physical Review Letters}\ }\textbf {\bibinfo {volume} {72}},\ \bibinfo {pages} {3439} (\bibinfo {year} {1994})}\BibitemShut {NoStop}%
\bibitem [{\citenamefont {Barnett}(1966)}]{assymptotic}%
  \BibitemOpen
  \bibfield  {author} {\bibinfo {author} {\bibfnamefont {V.~D.}\ \bibnamefont {Barnett}},\ }\bibfield  {title} {\bibinfo {title} {Evaluation of the maximum-likelihood estimator where the likelihood equation has multiple roots},\ }\href {http://www.jstor.org/stable/2334061} {\bibfield  {journal} {\bibinfo  {journal} {Biometrika}\ }\textbf {\bibinfo {volume} {53}},\ \bibinfo {pages} {151} (\bibinfo {year} {1966})}\BibitemShut {NoStop}%
\bibitem [{\citenamefont {Li}\ \emph {et~al.}(2018)\citenamefont {Li}, \citenamefont {Pezz{\`e}}, \citenamefont {Gessner}, \citenamefont {Ren}, \citenamefont {Li},\ and\ \citenamefont {Smerzi}}]{li2018frequentist}%
  \BibitemOpen
  \bibfield  {author} {\bibinfo {author} {\bibfnamefont {Y.}~\bibnamefont {Li}}, \bibinfo {author} {\bibfnamefont {L.}~\bibnamefont {Pezz{\`e}}}, \bibinfo {author} {\bibfnamefont {M.}~\bibnamefont {Gessner}}, \bibinfo {author} {\bibfnamefont {Z.}~\bibnamefont {Ren}}, \bibinfo {author} {\bibfnamefont {W.}~\bibnamefont {Li}},\ and\ \bibinfo {author} {\bibfnamefont {A.}~\bibnamefont {Smerzi}},\ }\bibfield  {title} {\bibinfo {title} {Frequentist and bayesian quantum phase estimation},\ }\href {https://doi.org/10.3390/e20090628} {\bibfield  {journal} {\bibinfo  {journal} {Entropy}\ }\textbf {\bibinfo {volume} {20}},\ \bibinfo {pages} {628} (\bibinfo {year} {2018})}\BibitemShut {NoStop}%
\bibitem [{\citenamefont {Pang}\ and\ \citenamefont {Brun}(2014)}]{Pang_2014}%
  \BibitemOpen
  \bibfield  {author} {\bibinfo {author} {\bibfnamefont {S.}~\bibnamefont {Pang}}\ and\ \bibinfo {author} {\bibfnamefont {T.~A.}\ \bibnamefont {Brun}},\ }\bibfield  {title} {\bibinfo {title} {Quantum metrology for a general hamiltonian parameter},\ }\href {http://dx.doi.org/10.1103/PhysRevA.90.022117} {\bibfield  {journal} {\bibinfo  {journal} {Physical Review A}\ }\textbf {\bibinfo {volume} {90}},\ \bibinfo {pages} {022117} (\bibinfo {year} {2014})}\BibitemShut {NoStop}%
\bibitem [{\citenamefont {Horn}\ and\ \citenamefont {Johnson}(2012)}]{circle}%
  \BibitemOpen
  \bibfield  {author} {\bibinfo {author} {\bibfnamefont {R.~A.}\ \bibnamefont {Horn}}\ and\ \bibinfo {author} {\bibfnamefont {C.~R.}\ \bibnamefont {Johnson}},\ }\href@noop {} {\emph {\bibinfo {title} {Matrix Analysis}}},\ \bibinfo {edition} {2nd}\ ed.\ (\bibinfo  {publisher} {Cambridge University Press},\ \bibinfo {year} {2012})\BibitemShut {NoStop}%
\bibitem [{\citenamefont {Giovannetti}\ \emph {et~al.}(2006)\citenamefont {Giovannetti}, \citenamefont {Lloyd},\ and\ \citenamefont {Maccone}}]{original_heisenberg}%
  \BibitemOpen
  \bibfield  {author} {\bibinfo {author} {\bibfnamefont {V.}~\bibnamefont {Giovannetti}}, \bibinfo {author} {\bibfnamefont {S.}~\bibnamefont {Lloyd}},\ and\ \bibinfo {author} {\bibfnamefont {L.}~\bibnamefont {Maccone}},\ }\bibfield  {title} {\bibinfo {title} {Quantum metrology},\ }\href {https://doi.org/10.1103%2Fphysrevlett.96.010401} {\bibfield  {journal} {\bibinfo  {journal} {Physical review letters}\ }\textbf {\bibinfo {volume} {96}},\ \bibinfo {pages} {010401} (\bibinfo {year} {2006})}\BibitemShut {NoStop}%
\bibitem [{\citenamefont {Hayes}\ \emph {et~al.}(2018)\citenamefont {Hayes}, \citenamefont {Dooley}, \citenamefont {Munro}, \citenamefont {Nemoto},\ and\ \citenamefont {Dunningham}}]{Hayes2018Making}%
  \BibitemOpen
  \bibfield  {author} {\bibinfo {author} {\bibfnamefont {A.~J.}\ \bibnamefont {Hayes}}, \bibinfo {author} {\bibfnamefont {S.}~\bibnamefont {Dooley}}, \bibinfo {author} {\bibfnamefont {W.~J.}\ \bibnamefont {Munro}}, \bibinfo {author} {\bibfnamefont {K.}~\bibnamefont {Nemoto}},\ and\ \bibinfo {author} {\bibfnamefont {J.}~\bibnamefont {Dunningham}},\ }\bibfield  {title} {\bibinfo {title} {Making the most of time in quantum metrology: concurrent state preparation and sensing},\ }\href {https://doi.org/10.1088/2058-9565/aac30b} {\bibfield  {journal} {\bibinfo  {journal} {Quantum Science and Technology}\ }\textbf {\bibinfo {volume} {3}},\ \bibinfo {pages} {035007} (\bibinfo {year} {2018})}\BibitemShut {NoStop}%
\bibitem [{\citenamefont {Zhou}\ \emph {et~al.}(2020{\natexlab{b}})\citenamefont {Zhou}, \citenamefont {Zou},\ and\ \citenamefont {Jiang}}]{Zhou_2020}%
  \BibitemOpen
  \bibfield  {author} {\bibinfo {author} {\bibfnamefont {S.}~\bibnamefont {Zhou}}, \bibinfo {author} {\bibfnamefont {C.-L.}\ \bibnamefont {Zou}},\ and\ \bibinfo {author} {\bibfnamefont {L.}~\bibnamefont {Jiang}},\ }\bibfield  {title} {\bibinfo {title} {Saturating the quantum cramér–rao bound using locc},\ }\href {https://doi.org/10.1088/2058-9565/ab71f8} {\bibfield  {journal} {\bibinfo  {journal} {Quantum Science and Technology}\ }\textbf {\bibinfo {volume} {5}},\ \bibinfo {pages} {025005} (\bibinfo {year} {2020}{\natexlab{b}})}\BibitemShut {NoStop}%
\bibitem [{\citenamefont {{Gaudin, M.}}(1976)}]{gaudin1976diagonalisation}%
  \BibitemOpen
  \bibfield  {author} {\bibinfo {author} {\bibnamefont {{Gaudin, M.}}},\ }\bibfield  {title} {\bibinfo {title} {Diagonalisation d'une classe d'hamiltoniens de spin},\ }\href {https://doi.org/10.1051/jphys:0197600370100108700} {\bibfield  {journal} {\bibinfo  {journal} {J. Phys. France}\ }\textbf {\bibinfo {volume} {37}},\ \bibinfo {pages} {1087} (\bibinfo {year} {1976})}\BibitemShut {NoStop}%
\bibitem [{\citenamefont {Breuer}\ \emph {et~al.}(2004)\citenamefont {Breuer}, \citenamefont {Burgarth},\ and\ \citenamefont {Petruccione}}]{breuer2004non-markovian}%
  \BibitemOpen
  \bibfield  {author} {\bibinfo {author} {\bibfnamefont {H.-P.}\ \bibnamefont {Breuer}}, \bibinfo {author} {\bibfnamefont {D.}~\bibnamefont {Burgarth}},\ and\ \bibinfo {author} {\bibfnamefont {F.}~\bibnamefont {Petruccione}},\ }\bibfield  {title} {\bibinfo {title} {Non-markovian dynamics in a spin star system: Exact solution and approximation techniques},\ }\href {https://doi.org/10.1103/PhysRevB.70.045323} {\bibfield  {journal} {\bibinfo  {journal} {Physical Review B}\ }\textbf {\bibinfo {volume} {70}},\ \bibinfo {pages} {045323} (\bibinfo {year} {2004})}\BibitemShut {NoStop}%
\bibitem [{\citenamefont {Hutton}\ and\ \citenamefont {Bose}(2004)}]{hutton2004mediated}%
  \BibitemOpen
  \bibfield  {author} {\bibinfo {author} {\bibfnamefont {A.}~\bibnamefont {Hutton}}\ and\ \bibinfo {author} {\bibfnamefont {S.}~\bibnamefont {Bose}},\ }\bibfield  {title} {\bibinfo {title} {Mediated entanglement and correlations in a star network of interacting spins},\ }\href {https://doi.org/10.1103/PhysRevA.69.042312} {\bibfield  {journal} {\bibinfo  {journal} {Physical Review A}\ }\textbf {\bibinfo {volume} {69}},\ \bibinfo {pages} {042312} (\bibinfo {year} {2004})}\BibitemShut {NoStop}%
\bibitem [{\citenamefont {Bortz}\ and\ \citenamefont {Stolze}(2007)}]{bortz2007exact}%
  \BibitemOpen
  \bibfield  {author} {\bibinfo {author} {\bibfnamefont {M.}~\bibnamefont {Bortz}}\ and\ \bibinfo {author} {\bibfnamefont {J.}~\bibnamefont {Stolze}},\ }\bibfield  {title} {\bibinfo {title} {Exact dynamics in the inhomogeneous central-spin model},\ }\href {https://doi.org/10.1103/PhysRevB.76.014304} {\bibfield  {journal} {\bibinfo  {journal} {Physical Review B}\ }\textbf {\bibinfo {volume} {76}},\ \bibinfo {pages} {014304} (\bibinfo {year} {2007})}\BibitemShut {NoStop}%
\bibitem [{\citenamefont {Schliemann}\ \emph {et~al.}(2003)\citenamefont {Schliemann}, \citenamefont {Khaetskii},\ and\ \citenamefont {Loss}}]{schliemann2003electron}%
  \BibitemOpen
  \bibfield  {author} {\bibinfo {author} {\bibfnamefont {J.}~\bibnamefont {Schliemann}}, \bibinfo {author} {\bibfnamefont {A.}~\bibnamefont {Khaetskii}},\ and\ \bibinfo {author} {\bibfnamefont {D.}~\bibnamefont {Loss}},\ }\bibfield  {title} {\bibinfo {title} {Electron spin dynamics in quantum dots and related nanostructures due to hyperfine interaction with nuclei},\ }\href {https://doi.org/10.1088/0953-8984/15/50/R01} {\bibfield  {journal} {\bibinfo  {journal} {Journal of Physics: Condensed Matter}\ }\textbf {\bibinfo {volume} {15}},\ \bibinfo {pages} {R1809} (\bibinfo {year} {2003})}\BibitemShut {NoStop}%
\bibitem [{\citenamefont {Dutt}\ \emph {et~al.}(2007)\citenamefont {Dutt}, \citenamefont {Childress}, \citenamefont {Jiang}, \citenamefont {Togan}, \citenamefont {Maze}, \citenamefont {Jelezko}, \citenamefont {Zibrov}, \citenamefont {Hemmer},\ and\ \citenamefont {Lukin}}]{gurudev2007quantum}%
  \BibitemOpen
  \bibfield  {author} {\bibinfo {author} {\bibfnamefont {M.~G.}\ \bibnamefont {Dutt}}, \bibinfo {author} {\bibfnamefont {L.}~\bibnamefont {Childress}}, \bibinfo {author} {\bibfnamefont {L.}~\bibnamefont {Jiang}}, \bibinfo {author} {\bibfnamefont {E.}~\bibnamefont {Togan}}, \bibinfo {author} {\bibfnamefont {J.}~\bibnamefont {Maze}}, \bibinfo {author} {\bibfnamefont {F.}~\bibnamefont {Jelezko}}, \bibinfo {author} {\bibfnamefont {A.}~\bibnamefont {Zibrov}}, \bibinfo {author} {\bibfnamefont {P.}~\bibnamefont {Hemmer}},\ and\ \bibinfo {author} {\bibfnamefont {M.}~\bibnamefont {Lukin}},\ }\bibfield  {title} {\bibinfo {title} {Quantum register based on individual electronic and nuclear spin qubits in diamond},\ }\href {https://www.science.org/doi/abs/10.1126/science.1139831} {\bibfield  {journal} {\bibinfo  {journal} {Science}\ }\textbf {\bibinfo {volume} {316}},\ \bibinfo {pages} {1312} (\bibinfo {year} {2007})}\BibitemShut {NoStop}%
\bibitem [{\citenamefont {Arenz}\ \emph {et~al.}(2014)\citenamefont {Arenz}, \citenamefont {Gualdi},\ and\ \citenamefont {Burgarth}}]{arenz2014control}%
  \BibitemOpen
  \bibfield  {author} {\bibinfo {author} {\bibfnamefont {C.}~\bibnamefont {Arenz}}, \bibinfo {author} {\bibfnamefont {G.}~\bibnamefont {Gualdi}},\ and\ \bibinfo {author} {\bibfnamefont {D.}~\bibnamefont {Burgarth}},\ }\bibfield  {title} {\bibinfo {title} {Control of open quantum systems: case study of the central spin model},\ }\href {https://doi.org/10.1088/1367-2630/16/6/065023} {\bibfield  {journal} {\bibinfo  {journal} {New Journal of Physics}\ }\textbf {\bibinfo {volume} {16}},\ \bibinfo {pages} {065023} (\bibinfo {year} {2014})}\BibitemShut {NoStop}%
\bibitem [{\citenamefont {Denning}\ \emph {et~al.}(2019)\citenamefont {Denning}, \citenamefont {Gangloff}, \citenamefont {Atat\"ure}, \citenamefont {M\o{}rk},\ and\ \citenamefont {Le~Gall}}]{denning2019collective}%
  \BibitemOpen
  \bibfield  {author} {\bibinfo {author} {\bibfnamefont {E.~V.}\ \bibnamefont {Denning}}, \bibinfo {author} {\bibfnamefont {D.~A.}\ \bibnamefont {Gangloff}}, \bibinfo {author} {\bibfnamefont {M.}~\bibnamefont {Atat\"ure}}, \bibinfo {author} {\bibfnamefont {J.}~\bibnamefont {M\o{}rk}},\ and\ \bibinfo {author} {\bibfnamefont {C.}~\bibnamefont {Le~Gall}},\ }\bibfield  {title} {\bibinfo {title} {Collective quantum memory activated by a driven central spin},\ }\href {https://doi.org/10.1103/PhysRevLett.123.140502} {\bibfield  {journal} {\bibinfo  {journal} {Physical Review Letters}\ }\textbf {\bibinfo {volume} {123}},\ \bibinfo {pages} {140502} (\bibinfo {year} {2019})}\BibitemShut {NoStop}%
\bibitem [{\citenamefont {Liu}\ \emph {et~al.}(2021)\citenamefont {Liu}, \citenamefont {Shi}, \citenamefont {Shi}, \citenamefont {Wang},\ and\ \citenamefont {Yang}}]{liu2021entanglement}%
  \BibitemOpen
  \bibfield  {author} {\bibinfo {author} {\bibfnamefont {J.-X.}\ \bibnamefont {Liu}}, \bibinfo {author} {\bibfnamefont {H.-L.}\ \bibnamefont {Shi}}, \bibinfo {author} {\bibfnamefont {Y.-H.}\ \bibnamefont {Shi}}, \bibinfo {author} {\bibfnamefont {X.-H.}\ \bibnamefont {Wang}},\ and\ \bibinfo {author} {\bibfnamefont {W.-L.}\ \bibnamefont {Yang}},\ }\bibfield  {title} {\bibinfo {title} {Entanglement and work extraction in the central-spin quantum battery},\ }\href {https://doi.org/10.1103/PhysRevB.104.245418} {\bibfield  {journal} {\bibinfo  {journal} {Physical Review B}\ }\textbf {\bibinfo {volume} {104}},\ \bibinfo {pages} {245418} (\bibinfo {year} {2021})}\BibitemShut {NoStop}%
\bibitem [{\citenamefont {Rolandi}\ \emph {et~al.}(2023)\citenamefont {Rolandi}, \citenamefont {Abiuso},\ and\ \citenamefont {Perarnau-Llobet}}]{rolandi2023collective}%
  \BibitemOpen
  \bibfield  {author} {\bibinfo {author} {\bibfnamefont {A.}~\bibnamefont {Rolandi}}, \bibinfo {author} {\bibfnamefont {P.}~\bibnamefont {Abiuso}},\ and\ \bibinfo {author} {\bibfnamefont {M.}~\bibnamefont {Perarnau-Llobet}},\ }\bibfield  {title} {\bibinfo {title} {Collective advantages in finite-time thermodynamics},\ }\href {https://doi.org/10.1103/PhysRevLett.131.210401} {\bibfield  {journal} {\bibinfo  {journal} {Physical Review Letters}\ }\textbf {\bibinfo {volume} {131}},\ \bibinfo {pages} {210401} (\bibinfo {year} {2023})}\BibitemShut {NoStop}%
\bibitem [{\citenamefont {Zanardi}\ \emph {et~al.}(2007{\natexlab{b}})\citenamefont {Zanardi}, \citenamefont {Campos~Venuti},\ and\ \citenamefont {Giorda}}]{zanardi2007bures}%
  \BibitemOpen
  \bibfield  {author} {\bibinfo {author} {\bibfnamefont {P.}~\bibnamefont {Zanardi}}, \bibinfo {author} {\bibfnamefont {L.}~\bibnamefont {Campos~Venuti}},\ and\ \bibinfo {author} {\bibfnamefont {P.}~\bibnamefont {Giorda}},\ }\bibfield  {title} {\bibinfo {title} {Bures metric over thermal state manifolds and quantum criticality},\ }\href {https://doi.org/10.1103/PhysRevA.76.062318} {\bibfield  {journal} {\bibinfo  {journal} {Physical Review A}\ }\textbf {\bibinfo {volume} {76}},\ \bibinfo {pages} {062318} (\bibinfo {year} {2007}{\natexlab{b}})}\BibitemShut {NoStop}%
\bibitem [{\citenamefont {Scandi}\ \emph {et~al.}(2025)\citenamefont {Scandi}, \citenamefont {Abiuso}, \citenamefont {Surace},\ and\ \citenamefont {De~Santis}}]{scandi2023quantum}%
  \BibitemOpen
  \bibfield  {author} {\bibinfo {author} {\bibfnamefont {M.}~\bibnamefont {Scandi}}, \bibinfo {author} {\bibfnamefont {P.}~\bibnamefont {Abiuso}}, \bibinfo {author} {\bibfnamefont {J.}~\bibnamefont {Surace}},\ and\ \bibinfo {author} {\bibfnamefont {D.}~\bibnamefont {De~Santis}},\ }\bibfield  {title} {\bibinfo {title} {Quantum fisher information and its dynamical nature},\ }\href {https://doi.org/10.1088/1361-6633/ade453} {\bibfield  {journal} {\bibinfo  {journal} {Reports on Progress in Physics}\ } (\bibinfo {year} {2025})}\BibitemShut {NoStop}%
\bibitem [{\citenamefont {Anto-Sztrikacs}\ \emph {et~al.}(2024)\citenamefont {Anto-Sztrikacs}, \citenamefont {Miller}, \citenamefont {Nazir},\ and\ \citenamefont {Segal}}]{Anto-Sztrikacs2024}%
  \BibitemOpen
  \bibfield  {author} {\bibinfo {author} {\bibfnamefont {N.}~\bibnamefont {Anto-Sztrikacs}}, \bibinfo {author} {\bibfnamefont {H.~J.~D.}\ \bibnamefont {Miller}}, \bibinfo {author} {\bibfnamefont {A.}~\bibnamefont {Nazir}},\ and\ \bibinfo {author} {\bibfnamefont {D.}~\bibnamefont {Segal}},\ }\bibfield  {title} {\bibinfo {title} {Bypassing thermalization timescales in temperature estimation using prethermal probes},\ }\href {https://doi.org/10.1103/PhysRevA.109.L060201} {\bibfield  {journal} {\bibinfo  {journal} {Physical Review A}\ }\textbf {\bibinfo {volume} {109}},\ \bibinfo {pages} {L060201} (\bibinfo {year} {2024})}\BibitemShut {NoStop}%
\bibitem [{\citenamefont {Tr{\'e}nyi}\ \emph {et~al.}(2024)\citenamefont {Tr{\'e}nyi}, \citenamefont {Luk{\'a}cs}, \citenamefont {Horodecki}, \citenamefont {Horodecki}, \citenamefont {V{\'e}rtesi},\ and\ \citenamefont {T{\'o}th}}]{Trnyi2024Activation}%
  \BibitemOpen
  \bibfield  {author} {\bibinfo {author} {\bibfnamefont {R.}~\bibnamefont {Tr{\'e}nyi}}, \bibinfo {author} {\bibfnamefont {{\'A}.}~\bibnamefont {Luk{\'a}cs}}, \bibinfo {author} {\bibfnamefont {P.}~\bibnamefont {Horodecki}}, \bibinfo {author} {\bibfnamefont {R.}~\bibnamefont {Horodecki}}, \bibinfo {author} {\bibfnamefont {T.}~\bibnamefont {V{\'e}rtesi}},\ and\ \bibinfo {author} {\bibfnamefont {G.}~\bibnamefont {T{\'o}th}},\ }\bibfield  {title} {\bibinfo {title} {Activation of metrologically useful genuine multipartite entanglement},\ }\href {https://iopscience.iop.org/article/10.1088/1367-2630/ad1e93} {\bibfield  {journal} {\bibinfo  {journal} {New Journal of Physics}\ }\textbf {\bibinfo {volume} {26}},\ \bibinfo {pages} {023034} (\bibinfo {year} {2024})}\BibitemShut {NoStop}%
\bibitem [{\citenamefont {Mihailescu}\ \emph {et~al.}(2024{\natexlab{a}})\citenamefont {Mihailescu}, \citenamefont {Campbell},\ and\ \citenamefont {Gietka}}]{mihailescu2024uncertain}%
  \BibitemOpen
  \bibfield  {author} {\bibinfo {author} {\bibfnamefont {G.}~\bibnamefont {Mihailescu}}, \bibinfo {author} {\bibfnamefont {S.}~\bibnamefont {Campbell}},\ and\ \bibinfo {author} {\bibfnamefont {K.}~\bibnamefont {Gietka}},\ }\bibfield  {title} {\bibinfo {title} {Uncertain quantum critical metrology: From single to multi parameter sensing},\ }\href {https://doi.org/10.48550/arXiv.2407.19917} {\bibfield  {journal} {\bibinfo  {journal} {arXiv preprint arXiv:2407.19917}\ } (\bibinfo {year} {2024}{\natexlab{a}})}\BibitemShut {NoStop}%
\bibitem [{\citenamefont {Mihailescu}\ \emph {et~al.}(2024{\natexlab{b}})\citenamefont {Mihailescu}, \citenamefont {Bayat}, \citenamefont {Campbell},\ and\ \citenamefont {Mitchell}}]{Mihailescu2024}%
  \BibitemOpen
  \bibfield  {author} {\bibinfo {author} {\bibfnamefont {G.}~\bibnamefont {Mihailescu}}, \bibinfo {author} {\bibfnamefont {A.}~\bibnamefont {Bayat}}, \bibinfo {author} {\bibfnamefont {S.}~\bibnamefont {Campbell}},\ and\ \bibinfo {author} {\bibfnamefont {A.~K.}\ \bibnamefont {Mitchell}},\ }\bibfield  {title} {\bibinfo {title} {Multiparameter critical quantum metrology with impurity probes},\ }\href {https://doi.org/10.1088/2058-9565/ad438d} {\bibfield  {journal} {\bibinfo  {journal} {Quantum Science and Technology}\ }\textbf {\bibinfo {volume} {9}},\ \bibinfo {pages} {035033} (\bibinfo {year} {2024}{\natexlab{b}})}\BibitemShut {NoStop}%
\bibitem [{\citenamefont {Fresco}\ \emph {et~al.}(2022)\citenamefont {Fresco}, \citenamefont {Spagnolo}, \citenamefont {Valenti},\ and\ \citenamefont {Carollo}}]{Giovanni2022Multiparameter}%
  \BibitemOpen
  \bibfield  {author} {\bibinfo {author} {\bibfnamefont {G.~D.}\ \bibnamefont {Fresco}}, \bibinfo {author} {\bibfnamefont {B.}~\bibnamefont {Spagnolo}}, \bibinfo {author} {\bibfnamefont {D.}~\bibnamefont {Valenti}},\ and\ \bibinfo {author} {\bibfnamefont {A.}~\bibnamefont {Carollo}},\ }\bibfield  {title} {\bibinfo {title} {{Multiparameter quantum critical metrology}},\ }\href {https://doi.org/10.21468/SciPostPhys.13.4.077} {\bibfield  {journal} {\bibinfo  {journal} {SciPost Phys.}\ }\textbf {\bibinfo {volume} {13}},\ \bibinfo {pages} {077} (\bibinfo {year} {2022})}\BibitemShut {NoStop}%
\bibitem [{\citenamefont {Montenegro}\ \emph {et~al.}(2023)\citenamefont {Montenegro}, \citenamefont {Genoni}, \citenamefont {Bayat},\ and\ \citenamefont {Paris}}]{montenegro2023quantum}%
  \BibitemOpen
  \bibfield  {author} {\bibinfo {author} {\bibfnamefont {V.}~\bibnamefont {Montenegro}}, \bibinfo {author} {\bibfnamefont {M.~G.}\ \bibnamefont {Genoni}}, \bibinfo {author} {\bibfnamefont {A.}~\bibnamefont {Bayat}},\ and\ \bibinfo {author} {\bibfnamefont {M.~G.}\ \bibnamefont {Paris}},\ }\bibfield  {title} {\bibinfo {title} {Quantum metrology with boundary time crystals},\ }\href {https://doi.org/10.1038/s42005-023-01423-6} {\bibfield  {journal} {\bibinfo  {journal} {Communications Physics}\ }\textbf {\bibinfo {volume} {6}},\ \bibinfo {pages} {304} (\bibinfo {year} {2023})}\BibitemShut {NoStop}%
\bibitem [{\citenamefont {Yang}\ \emph {et~al.}(2022{\natexlab{a}})\citenamefont {Yang}, \citenamefont {Pang}, \citenamefont {Chen}, \citenamefont {Jordan},\ and\ \citenamefont {del Campo}}]{Yang2022Variational}%
  \BibitemOpen
  \bibfield  {author} {\bibinfo {author} {\bibfnamefont {J.}~\bibnamefont {Yang}}, \bibinfo {author} {\bibfnamefont {S.}~\bibnamefont {Pang}}, \bibinfo {author} {\bibfnamefont {Z.}~\bibnamefont {Chen}}, \bibinfo {author} {\bibfnamefont {A.~N.}\ \bibnamefont {Jordan}},\ and\ \bibinfo {author} {\bibfnamefont {A.}~\bibnamefont {del Campo}},\ }\bibfield  {title} {\bibinfo {title} {Variational principle for optimal quantum controls in quantum metrology},\ }\href {https://doi.org/10.1103/PhysRevLett.128.160505} {\bibfield  {journal} {\bibinfo  {journal} {Physical Review Letters}\ }\textbf {\bibinfo {volume} {128}},\ \bibinfo {pages} {160505} (\bibinfo {year} {2022}{\natexlab{a}})}\BibitemShut {NoStop}%
\bibitem [{\citenamefont {Yang}\ \emph {et~al.}(2022{\natexlab{b}})\citenamefont {Yang}, \citenamefont {Pang}, \citenamefont {del Campo},\ and\ \citenamefont {Jordan}}]{Yang2022Super}%
  \BibitemOpen
  \bibfield  {author} {\bibinfo {author} {\bibfnamefont {J.}~\bibnamefont {Yang}}, \bibinfo {author} {\bibfnamefont {S.}~\bibnamefont {Pang}}, \bibinfo {author} {\bibfnamefont {A.}~\bibnamefont {del Campo}},\ and\ \bibinfo {author} {\bibfnamefont {A.~N.}\ \bibnamefont {Jordan}},\ }\bibfield  {title} {\bibinfo {title} {Super-heisenberg scaling in hamiltonian parameter estimation in the long-range kitaev chain},\ }\href {https://doi.org/10.1103/PhysRevResearch.4.013133} {\bibfield  {journal} {\bibinfo  {journal} {Physical Review Research}\ }\textbf {\bibinfo {volume} {4}},\ \bibinfo {pages} {013133} (\bibinfo {year} {2022}{\natexlab{b}})}\BibitemShut {NoStop}%
\bibitem [{\citenamefont {Ban}\ \emph {et~al.}(2022)\citenamefont {Ban}, \citenamefont {Casanova},\ and\ \citenamefont {Puebla}}]{ban2022neural}%
  \BibitemOpen
  \bibfield  {author} {\bibinfo {author} {\bibfnamefont {Y.}~\bibnamefont {Ban}}, \bibinfo {author} {\bibfnamefont {J.}~\bibnamefont {Casanova}},\ and\ \bibinfo {author} {\bibfnamefont {R.}~\bibnamefont {Puebla}},\ }\bibfield  {title} {\bibinfo {title} {Neural networks for bayesian quantum many-body magnetometry},\ }\href {https://doi.org/10.48550/arXiv.2212.12058} {\bibfield  {journal} {\bibinfo  {journal} {arXiv preprint arXiv:2212.12058}\ } (\bibinfo {year} {2022})}\BibitemShut {NoStop}%
\bibitem [{\citenamefont {Bai}\ and\ \citenamefont {An}(2023)}]{Bai2023Floquet}%
  \BibitemOpen
  \bibfield  {author} {\bibinfo {author} {\bibfnamefont {S.-Y.}\ \bibnamefont {Bai}}\ and\ \bibinfo {author} {\bibfnamefont {J.-H.}\ \bibnamefont {An}},\ }\bibfield  {title} {\bibinfo {title} {Floquet engineering to overcome no-go theorem of noisy quantum metrology},\ }\href {https://doi.org/10.1103/PhysRevLett.131.050801} {\bibfield  {journal} {\bibinfo  {journal} {Physical Review Letters}\ }\textbf {\bibinfo {volume} {131}},\ \bibinfo {pages} {050801} (\bibinfo {year} {2023})}\BibitemShut {NoStop}%
\bibitem [{\citenamefont {Ansel}\ \emph {et~al.}(2024)\citenamefont {Ansel}, \citenamefont {Dionis},\ and\ \citenamefont {Sugny}}]{Quentin2024}%
  \BibitemOpen
  \bibfield  {author} {\bibinfo {author} {\bibfnamefont {Q.}~\bibnamefont {Ansel}}, \bibinfo {author} {\bibfnamefont {E.}~\bibnamefont {Dionis}},\ and\ \bibinfo {author} {\bibfnamefont {D.}~\bibnamefont {Sugny}},\ }\bibfield  {title} {\bibinfo {title} {Optimal control strategies for parameter estimation of quantum systems},\ }\href {https://doi.org/10.21468/SciPostPhys.16.1.013} {\bibfield  {journal} {\bibinfo  {journal} {SciPost Phys.}\ }\textbf {\bibinfo {volume} {16}},\ \bibinfo {pages} {013} (\bibinfo {year} {2024})}\BibitemShut {NoStop}%
\bibitem [{\citenamefont {Huang}\ \emph {et~al.}(2023)\citenamefont {Huang}, \citenamefont {Tong}, \citenamefont {Fang},\ and\ \citenamefont {Su}}]{huang2023learning}%
  \BibitemOpen
  \bibfield  {author} {\bibinfo {author} {\bibfnamefont {H.-Y.}\ \bibnamefont {Huang}}, \bibinfo {author} {\bibfnamefont {Y.}~\bibnamefont {Tong}}, \bibinfo {author} {\bibfnamefont {D.}~\bibnamefont {Fang}},\ and\ \bibinfo {author} {\bibfnamefont {Y.}~\bibnamefont {Su}},\ }\bibfield  {title} {\bibinfo {title} {Learning many-body hamiltonians with heisenberg-limited scaling},\ }\href {https://doi.org/10.1103/PhysRevLett.130.200403} {\bibfield  {journal} {\bibinfo  {journal} {Physical Review Letters}\ }\textbf {\bibinfo {volume} {130}},\ \bibinfo {pages} {200403} (\bibinfo {year} {2023})}\BibitemShut {NoStop}%
\bibitem [{\citenamefont {Dutkiewicz}\ \emph {et~al.}(2024)\citenamefont {Dutkiewicz}, \citenamefont {O'Brien},\ and\ \citenamefont {Schuster}}]{dutkiewicz2023advantage}%
  \BibitemOpen
  \bibfield  {author} {\bibinfo {author} {\bibfnamefont {A.}~\bibnamefont {Dutkiewicz}}, \bibinfo {author} {\bibfnamefont {T.~E.}\ \bibnamefont {O'Brien}},\ and\ \bibinfo {author} {\bibfnamefont {T.}~\bibnamefont {Schuster}},\ }\bibfield  {title} {\bibinfo {title} {The advantage of quantum control in many-body hamiltonian learning},\ }\href {https://doi.org/10.22331/q-2024-11-26-1537} {\bibfield  {journal} {\bibinfo  {journal} {Quantum}\ }\textbf {\bibinfo {volume} {8}},\ \bibinfo {pages} {1537} (\bibinfo {year} {2024})}\BibitemShut {NoStop}%
\bibitem [{\citenamefont {Wilcox}(1967)}]{exponential_differentiaion}%
  \BibitemOpen
  \bibfield  {author} {\bibinfo {author} {\bibfnamefont {R.~M.}\ \bibnamefont {Wilcox}},\ }\bibfield  {title} {\bibinfo {title} {{Exponential Operators and Parameter Differentiation in Quantum Physics}},\ }\href {https://doi.org/10.1063/1.1705306} {\bibfield  {journal} {\bibinfo  {journal} {J. Math. Phys.}\ }\textbf {\bibinfo {volume} {8}},\ \bibinfo {pages} {962} (\bibinfo {year} {1967})}\BibitemShut {NoStop}%
\bibitem [{\citenamefont {Binney}\ and\ \citenamefont {Skinner}(2013)}]{binney2013physics}%
  \BibitemOpen
  \bibfield  {author} {\bibinfo {author} {\bibfnamefont {J.}~\bibnamefont {Binney}}\ and\ \bibinfo {author} {\bibfnamefont {D.}~\bibnamefont {Skinner}},\ }\href@noop {} {\emph {\bibinfo {title} {The Physics of Quantum Mechanics}}}\ (\bibinfo  {publisher} {OUP Oxford},\ \bibinfo {year} {2013})\BibitemShut {NoStop}%
\bibitem [{\citenamefont {Ma}\ \emph {et~al.}(2011)\citenamefont {Ma}, \citenamefont {Wang}, \citenamefont {Sun},\ and\ \citenamefont {Nori}}]{MA201189}%
  \BibitemOpen
  \bibfield  {author} {\bibinfo {author} {\bibfnamefont {J.}~\bibnamefont {Ma}}, \bibinfo {author} {\bibfnamefont {X.}~\bibnamefont {Wang}}, \bibinfo {author} {\bibfnamefont {C.}~\bibnamefont {Sun}},\ and\ \bibinfo {author} {\bibfnamefont {F.}~\bibnamefont {Nori}},\ }\bibfield  {title} {\bibinfo {title} {Quantum spin squeezing},\ }\href {https://doi.org/https://doi.org/10.1016/j.physrep.2011.08.003} {\bibfield  {journal} {\bibinfo  {journal} {Physics Reports}\ }\textbf {\bibinfo {volume} {509}},\ \bibinfo {pages} {89} (\bibinfo {year} {2011})}\BibitemShut {NoStop}%
\bibitem [{\citenamefont {Mele}(2024)}]{Mele_2024}%
  \BibitemOpen
  \bibfield  {author} {\bibinfo {author} {\bibfnamefont {A.~A.}\ \bibnamefont {Mele}},\ }\bibfield  {title} {\bibinfo {title} {Introduction to haar measure tools in quantum information: A beginner's tutorial},\ }\href {https://doi.org/10.22331/q-2024-05-08-1340} {\bibfield  {journal} {\bibinfo  {journal} {Quantum}\ }\textbf {\bibinfo {volume} {8}},\ \bibinfo {pages} {1340} (\bibinfo {year} {2024})}\BibitemShut {NoStop}%
\end{thebibliography}%

\onecolumngrid

\newpage

\appendix

\section{Preliminaries}
In this section, we provide a concise overview of key analytical tools and concepts that will be referenced throughout the subsequent sections.

\subsection{Derivative of the exponential operator}\label{app:derivative_exp}

Here we briefly summarize how to take the derivative of the exponential of a Hamiltonian that depends on a parameter $x$. We refer the reader to Ref.~\cite{exponential_differentiaion} for further details. 

Assume we have a Hamiltonian $H$ that depends on a parameter ${x}$, i.e. $H = H({x})$. Then we can compute the derivative of the exponential of this operator with respect to $x$ as follows
\begin{equation}\label{eq:derivative_matrixexp}
    \frac{d}{d x}e^{-\beta H({x})} = -\beta \int_{0}^1  e^{- s \beta H({x})} \frac{d}{d x_i} H({x}) e^{- (1-s) \beta  H({x})} ds\,,
\end{equation}
for some $\beta\in	\mathbb{C}$. 

\subsection{Time-independent perturbation theory}\label{app:perturbationtheory}
Here we explain the basic concepts necessary to understand how perturbation theory is used in this paper to derive results. For a more in-depth analysis check Ref.~\cite{binney2013physics}.

We start by tackling the non-degenerate case. Let us consider a physical system in the form of 
\begin{equation}
    H = H_0 + x V\,,
\end{equation}
where $x$ is a small parameter. Then $V$ can be treated as a perturbation of the system. In this limit, we can approximate the eigenvectors and eigenvalues of $H$ with the ones of $H_0$ and some corrections. Indeed, we can expand the eigenvalues/eigenvectors of $H$  (denoted as $\overline{E_i}, \ \ket{\overline{E_i}}$) as a power series of $x$ and the eigenvalues/eigenvectors of $H_0$ (denoted as $E_i$, $\ket{E_i}$). In the limit of $x\to 0$, we can approximate this power series at just the first order. Under this approximation, the eigenvectors of $H$ are

\begin{equation}\label{eq:perturbation_nodeg}
    \ket{\overline{E_i}} = \ket{E_i} + x\sum_{j\neq i}\frac{\bra{E_i}V\ket{E_j}}{E_i-E_j}\ket{E_j} + \order{x^2}\,,
\end{equation}
and similarly, the energies are
\begin{equation}
    \overline{E_i} = E_i + x\expval{V}{E_n} + \order{x^2}\,.
\end{equation}

When $H_0$ is degenerate, the previous sum in Eq. \eqref{eq:perturbation_nodeg} can diverge. Indeed, it will only converge if the term $\bra{E_i}V\ket{E_j}$ is zero when $E_i-E_j = 0$. This means that the perturbative expansion has a preferred basis. And this is the one that fulfills
\begin{equation}
    \bra{E_{i,1}}V\ket{E_{i,2}} = 0\,,
\end{equation}
where we denote $\ket{E_{i,j}}$ as the $i$-th excited state and the index $j$ accounts for the different states with the same energy. Therefore, the preferred basis is the one in which the unperturbed states are orthogonal under the action of $V$, or in other words, the one which ensures that Eq.~\eqref{eq:perturbation_nodeg} does not diverge.

\subsection{Tensor product of N spin-$\frac{1}{2}$ systems as a linear combination of Dicke states}\label{app:block_dycke}

We first define a Dicke state. The Dicke states are eigenvectors of $S_z$ and $S^2$. Recall that
\begin{align}
    S_z =\frac{1}{2}\sum_i \sigma_z^{(i)}\,,
\end{align}
where $\sigma_z^{(i)}$ represents the Pauli matrix z acting in the $i$-th particle. Furthermore, $S^2$ is the total spin of the state. These states are usually written as $\ket{S, m_z}$ where $S$ represents the total spin (proportional to the eigenvalues of $S^2$) and $m_z\in[-S, S]$ are the eigenvalues of $S_z$, the $z$ component of this spin. 

With this, we are ready to define spin-coherent states. 
Take one vector in the Bloch sphere to be 
$\ket{\vartheta,\phi}=\cos\tfrac{\theta}{2} \ket{0}+e^{-i\phi}\sin\tfrac{\theta}{2} \ket{1}$ where $\vartheta,\phi$ are polar and azimuthal angles  respectively.
Then if we take the tensor product of $N$ of this state we can rewrite the state as follows~\cite{MA201189}:
\begin{equation}\label{eq:cssz}
   \ket{\vartheta,\phi}^{\otimes N}=\frac{1}{(1+|\eta|^2)^S}\sum_{m_z=-S}^S \binom{2S}{S+m_z}^{1/2} \eta^{S+m_z} \ket{S,m_z} \, .
\end{equation}
where $\eta = -\tan\left(\frac{\vartheta}{2}\right)e^{-\ii \phi}$.
Furthermore, without loss of generality, we can take $\vartheta, \phi$ with respect to the $x$-axis of the Block sphere. By doing so, we can move from expressing the state in the basis of $S_z$, $\{\ket{S,m_z}\}_{m_z = -S}^{S}$ to the basis of $S_x$, $\{\ket{S,m_x}\}_{m_x = -S}^S$. Notice that here we write $m_x$ to emphasize that we are on the basis of $S_x$. Then we can rewrite Eq. \ref{eq:cssz} in this basis by
\begin{equation}\label{eq:cssx}
   \ket{\vartheta_x,\phi_x}^{\otimes N}=\frac{1}{(1+|\eta|^2)^{S}}\sum_{m_x=-{S}}^{S} \binom{2{S}}{{S}+m_x}^{1/2} \eta^{{S}+m_x} \ket{{S},m_x} \,,
\end{equation}
where we also emphasize  that the reference for $\vartheta_x,\phi_x$ is the $x-$axis in the Block sphere\footnote{  This means that $\ket{\vartheta_x = 0,\phi = 0}$ is not the usual eigenstate $\ket{0}$ of $\sigma_z$, i.e. $\sigma_z\ket{0} = \ket{0}$, but would be the eigenstate of $\sigma_x$. Indeed, $\sigma_x\ket{\vartheta_x = 0,\phi = 0} = \ket{\vartheta_x = 0,\phi = 0}$ }. 

\subsection{Vectorization formalism}\label{app:vec_formalism}

Here we will cover the basics of the vectorization formalism that are useful to understanding this manuscript. For a more in-depth review please refer to Ref.~\cite{Mele_2024}. Vectorization is a linear \revadd{map} defined as follows
\revadd{\begin{align}
       \begin{matrix}
            \mathrm{vec}:& L\left(\mathbb{C}^d\right)&\to& \mathbb{C}^d \otimes \mathbb{C}^d  \\
            \quad\\
        &A = \sum_{i,j}^d A_{i,j}\ketbra{i}{j}&\mapsto& \vecc{A} = \sum_{i,j} A_{i,j}\ket{i}\ket{j^*}\,,
       \end{matrix}
\end{align}}
where $\ket{j^*}$ denotes the conjugate of $\ket{j}$ \revadd{(in the computational basis)}.  

A basic property of the vectorization formalism that is very useful is the following. Given $A,B,C$ three matrices that can be multiplied as follows $ABC$, then
\begin{equation}
\vecc{ABC}  = A\otimes C^T\vecc{B}\,,
\end{equation}
where we emphasize that $C^T$ is the transpose of $C$ \revadd{(in the computational basis)}. Furthermore, we can see that 
\begin{align}
    \langle\!\langle A|B\rangle\!\rangle = \langle\!\langle \1|A^\dagger B \otimes \1 |\1\rangle\!\rangle = \sum_{i,j}\bra{i}\bra{{i}^*}A^\dagger B \otimes \1\ket{j}\ket{{j}^*}=\Tr[A^\dagger B]\, .
\end{align}

Finally, we \revadd{discuss} the form of a super-operator in the vectorized form. We denote the vectorized form of a super-operator $\Phi(\cdot) = \sum A_i (\cdot) B_j^\dagger$ as
\begin{align}\label{eq:vectorised_op}
    \hat{\Phi}:={\rm Vec}\left[\Phi(\cdot)\right] = A_i\otimes {B}_j^* \, ,
\end{align}
where $B^*$ is the conjugate of $B$. \revadd{With this one easily verifies that 
\begin{equation}
    \vecc{\Phi(C)} = \hat{\Phi} \vecc{C}.
\end{equation}
} 

\subsection{Gershgorin circle theorem and its implication on operator eigenvalues}\label{app:circleth}
We present a simplified version of Gershgorin circle theorem~\cite{circle} that we use in order to upper bound the eigenvalues of an hermitian matrix.

\begin{proposition}\label{prop:upper-eigen}
 \revadd{Consider a $n\times n$ Hermitian matrix $M$ with zeroes on the diagonal $M_{ii}=0$. If the sum of elements in any row (or column) is upper bounded by some value $E$, that is $\sum_{j} |M_{ij}| \leq E$ for any $i \in \{1,..., n\}$, then the eigenvalues of the matrix satisfy
 }
 \begin{equation}
     -E \leq {\rm Eig}(M)\leq E.
     \end{equation}
\end{proposition}
This theorem gives a simple way to upper-bound the eigenvalues of any Hermitian matrix. \revadd{Moreover it turns out very useful to bound its eigenvalues after the action of specific "off-diagonal" superoperators, as we now show.}

\revadd{
\begin{proposition}\label{prop:gersh-cons} 
For a set or $n$ orthogonal projectors $\{\Pi_j\}$, and an Hermitian  operator $H$, let  
\begin{equation}
O = \sum_{j,k|j\neq k }^n f_{k,j} \, \Pi_k H \Pi_j,
\end{equation}
for some complex coefficients $f_{k,j}$ such that $O$ is Hermitian.
Its operator norm is bounded by
\begin{equation}
    \| O \|_\infty \leq \|H\|_\infty \, \max_k \sqrt{\sum_{j\neq k}^n |f_{k,j}|^2 }
\end{equation}
\end{proposition}
\begin{proof}
    Since $O$ is Hermitian we have $\| O \|_\infty = \max_{\ket{\Psi}} \bra{\Psi} O \ket{\Psi}$. For any state $\ket{\Psi}$ we have
    \begin{equation}
        \bra{\Psi} O \ket{\Psi} = \sum_{j,k|j\neq k }^n \sqrt{p_k p_j} f_{k,j} \bra{\varphi_k} H \ket{\varphi_j} = \sum_{j,k=1}^n \sqrt{p_k p_j} \, M_{kj}
    \end{equation}
    where we defined $\sqrt{p_j} \ket{\varphi_j} := \Pi_j \ket{\Psi}$ and the Hermitian $n\times n$ matrix 
    \begin{equation}
        M_{kj}= \begin{cases}
        0 & k=j \\
        f_{k,j} \bra{\varphi_k} H \ket{\varphi_j} & k\neq j
        \end{cases}.
    \end{equation}
From the last equality we see that 
\begin{equation}
    \bra{\Psi} O \ket{\Psi} \leq \lambda_{\max}^{(M)} \leq \max_k \sum_j |M_{kj}|
\end{equation}
where we applied the Gershgorin circle theorem (Proposition~\ref{prop:upper-eigen}). Next, for a given $k$ rewrite the last term as
\begin{align}
    &\sum_j |M_{kj}| = \sum_{j\neq k}|\bra{\varphi_k} H f_{k,j}\ket{\varphi_j} |  = |\bra{\varphi_k}H \ket{\xi_k}| \\
    &\text{where} \quad \ket{\xi_k} = \sum_{j\neq k} e^{-\ii\,  \text{arg}\bra{\varphi_k}H f_{k,j} \ket{\varphi_j}} f_{k,j} \ket{\varphi_j}
\end{align}
where the phase is defined such that $\bra{\varphi_k} H f_{k,j}\ket{\varphi_j} =|\bra{\varphi_k} H f_{k,j}\ket{\varphi_j} |$. Note that the ket $\ket{\zeta_k}$ is not normalized in contrast to $\ket{\varphi_k}$, nevertheless we have
\begin{equation}
     |\bra{\varphi_k}H \ket{\xi_k}| \leq \| H \|_\infty \, \| \ket{\xi_k}\|_2
\end{equation}
with 
\begin{align}
     \| \ket{\xi_k}\|_2^2 = \braket{\xi_k} = \sum_{j\neq k} |f_{k,j}|^2 \braket{\varphi_j} = \sum_{j\neq k} |f_{k,j}|^2.
\end{align}
Hence combining all the above inequalities we find $\|O\|_\infty \leq \max_k \sum_j |M_{kj}| \leq \|H\|_\infty \, \max_k \sqrt{\sum_{j\neq k}^n |f_{k,j}|^2 }$ proving the proposition.
\end{proof}

Now we apply Proposition~\ref{prop:gersh-cons} to a particular case which is directly useful for the cases studied in this paper, where the coefficients $f_{k,j}$ are inversely proportional to the energy gap between the energy subspaces associated to the projectors $\Pi_k$ and $\Pi_j$.

\begin{proposition}
    \label{prop:gersh-energy} 
Let $\{\Pi_j\}$ with $j=1,\dots,n$ be a set of orthogonal projectors ordered in the associated "energies" $E_j$, i.e. $E_j< E_{j+1}$, and with a finite energy gap $E_{j+1}-E_j \geq \Delta$. For some Hermitian  operator $H$ let  
\begin{equation}
O = \sum_{j,k|j\neq k }^n f_{k,j}\, \Pi_k H \Pi_j,
\end{equation}
for some complex coefficients $f_{k,j}$ such that $O$ is Hermitian, and $|f_{k,j}|\leq \frac{\mu }{|E_k-E_j|}$ for $k\neq j$ and some positive real number $\mu$.
The operator norm of $O$ is bounded by
\begin{align}
    &\| O \|_\infty \leq \|H\|_\infty \, \frac{\mu}{\Delta} \sqrt{{\rm s}_n} \leq   \|H\|_\infty \, \frac{\mu}{\Delta} \frac{\pi}{\sqrt{3}}
\\
&\text{where}\qquad
    {\rm s}_n := \begin{cases} 2 h_{\frac{n-1}{2}}^{(2)}& n \text{ odd}\\
h_{\frac{n}{2}}^{(2)}+h_{\frac{n}{2}-1}^{(2)}& n \text{ even}
\end{cases}
\end{align}
and $h_m^{(2)}=\sum_{k=1}^m \frac{1}{k^2}$ is the harmonic number of order two.
\end{proposition}
\begin{proof}
By Proposition~\ref{prop:gersh-cons} we know that 
\begin{equation}
    \| O \|_\infty \leq \|H\|_\infty \, \max_k \sqrt{\sum_{j\neq k}^n |f_{k,j}|^2 },
\end{equation}
our goal here is to upper-bound the last term. By assumption we have that for all $k$
\begin{align}
   \sum_{j\neq k}^n |f_{k,j}|^2  \leq \sum_{j\neq k}^n \frac{\mu^2}{|E_k-E_j|^2}  
\end{align}
Since all energies are gaped we furthermore have $|E_k-E_j|\geq \Delta |k-j|$ implying
\begin{align}
   \sum_{j\neq k}^n |f_{k,j}|^2  \leq \frac{\mu^2}{\Delta^2} \sum_{j\neq k}^n \frac{1}{|k-j|^2}.
\end{align}
To prove the corollary it remains to upper-bound the harmonic sum in the last expression. By direct inspection one sees that it is maximized by taking $k$ in the middle of the spectrum such that the gaps to all other energy levels are as small as possible. Hence, for the the harmonic number of order two $h_m^{(2)} := \sum_{k=1}^m \frac{1}{k^2}$ we obtain the following tight bound
\begin{equation}
  \sum_{j\neq k}^n \frac{1}{|k-j|^2}  \leq \begin{cases} 2 h_{\frac{n-1}{2}}^{(2)}& n \text{ odd}\\
h_{\frac{n}{2}}^{(2)} + h_{\frac{n}{2}-1}^{(2)}& n \text{ even}
\end{cases}.
\end{equation}
Finally, note that the harmonic number is increasing in $m$ and satisfy $\lim_{m\to \infty} h^{(2)}_m=\frac{\pi^2}{6}$, giving the $n$-independent upper bound 
\begin{equation}
  \sum_{j\neq k}^n \frac{1}{|k-j|^2}  \leq \frac{\pi^2}{3},
  \end{equation}
and concluding the proof.
\end{proof}
}

\subsection{Simulations of local noise.}\label{app:numerical_rotation}
The simulations of Figs.~\ref{fig:local_noise_compare},~\ref{fig:local_noise_int_advantage},~\ref{fig:local_noise_n_scaling} have been done with QuTip and the ``Permutational Invariant Quantum Solver (PIQS)''. For numerical reasons we do this analysis in a rotated (but equivalent) frame: it is more convenient to have a dissipator that depends on $\sigma_z^{(i)}$ rather than $\sigma_x^{(i)}$. We therefore consider a unitary transformation such that
\begin{align}
    \sigma_z^{(i)} &\to \sigma_x^{(i)},\\
    \sigma_x^{(i)} &\to \sigma_z^{(i)},\\
    \sigma_y^{(i)} &\to -\sigma_y^{(i)}.
\end{align}
With this transformation, the Hamiltonian $H_{\theta}^{\rm original} = \theta S_z + a S_x^2$ becomes
\begin{equation}
    H_\theta=\theta S_x + a S_z^2,
\end{equation}
and the Lindblad equation takes the desired form
\begin{equation}
    \dot{\rho}_{\theta}(t)= -\ii [H_\theta,{\rho}_{\theta}(t)] + \frac{\gamma}{4}\sum_i \left(\sigma_z^{(i)} {\rho}_{\theta}(t) \sigma_z^{(i)} - {\rho}_{\theta}(t)\right) \ .
\end{equation}
In this ``rotated reference frame'', we choose as initial state
\begin{equation}
    \ket{\psi} = \ket{-y}^{\otimes N},
\end{equation}
where $\ket{-y}$ is an eigenstate of $\sigma_y\ket{-y} =-\ket{-y}$.

\section{Dynamical Quantum Fisher Information}
\label{app: proof 1}
We devote this section to proving Result \ref{res:pinched}. 
\setcounter{result}{0}
\begin{result}[Dynamical quantum Fisher information]\label{res:pinched_app}
   Let $H_\theta$ be the Hamiltonian of the form of Eq.~\eqref{eq:paradigm}, $H_\theta =  \theta H_S + H_C$ where $\theta$ is the unknown parameter and $H_S$ is the signal and $H_C$ is a controlled term. Let the final state after the evolution be $\ket{\psi_\theta(t)} = e^{-\ii t H_\theta}\ket{\psi}$, for some initial pure state $\ket{\psi}$. The QFI of this state is given by the variance
\begin{equation}
    \f = 4 t^2 \var( H_P )_{\ket{\psi}} + \order{t\revadd{ \|H_S\|^2/\Delta_g}} \ 
\label{eq:Fisher_Pinched_app}
\end{equation}
of the pinched signal Hamiltonian
\begin{equation}\label{eq:refpinched_app}
H_P = \mathcal{D}_{H_\theta}(H_S)= \sum_k \Pi_k H_S \Pi_{\revadd{k}}\, .
\end{equation}
Here, $\mathcal{D}_{H_\theta}$ is the self-adjoint pinching (dephasing) map in the eigenbasis of $H_\theta$ introduced in Eq. \eqref{eq:pinching_map}, and $\Pi_k$ are the projectors on the eigenspaces of $H_\theta = \sum_{k} E_k \Pi_k$ associated to different eigenvalues. 

\revadd{In Eq.~\eqref{eq:Fisher_Pinched_app}, we define 
$\Delta_g = \min_{i,j}|E_i-E_j|$ 
as the minimum non-zero energy difference between energy levels of $H_\theta$. 
} 
\end{result}

\begin{proof}
    We start this proof by recalling the definition of the QFI ($\f$). For some pure state evolving under some Hamiltonian $H_\theta = H_C + \theta H_S$ that depends on a parameter $\theta$, i.e. $e^{-\ii t H_\theta}\ket{\psi}$ we write the QFI ($\f$) as in Eq.~\eqref{eq:fishd}
    \begin{equation}
        \f = 4 \left(\braket{\Dot{\psi}_\theta(t)} - \left| \braket{\Dot{\psi}_\theta(t)}{{\psi}_\theta(t)} \right|^2\right)\,.
        \label{eq:fishd_app}
    \end{equation}
    where we are using the following notation:
    \begin{align}
    \ket{\psi_\theta(t)} =&  e^{-\ii t H_\theta}\ket{\psi}\,,\\
        \ket{\dot\psi_\theta(t)} =&  \frac{d}{d\theta}e^{-\ii t H_\theta}\ket{\psi}\,.
    \end{align}

    This derivative can be computed using  Appendix \ref{app:derivative_exp}. That is 
    \begin{equation}\label{eq:derivative_htheta}
        \ket{\dot\psi_\theta(t)} = -\ii  t\int_{0}^1 e^{-\ii tsH_\theta}H_S e^{-\ii t(1-s)H_\theta} ds \ket{\psi} =  -\ii  t\int_{0}^1 e^{-\ii tsH_\theta}H_S e^{\ii   tsH_\theta} ds \ket{\psi_\theta(t)}\, .
    \end{equation}

    We can rewrite this equation is terms of the generator or effective Hamiltonian:$H_{\rm eff}$:
    \begin{equation}
        H_{\rm eff} = \int_{0}^1 e^{-\ii tsH_\theta}H_S e^{-\ii t(1-s)H_\theta} ds\,,
    \end{equation}
as 
 \begin{equation}
\ket{\dot\psi_\theta(t)} = -\ii t H_{\rm eff} \ket{\psi_\theta(t)}\, .
\end{equation}
    With this, the expression of the QFI in Eq. \eqref{eq:fishd_app} becomes
     \begin{equation}\label{eq:fisher_as_varianceHeff}
        \f = 4 t^2\expval{H_{\rm eff}^2}{\psi_{\theta}(t)} - 4t^2\left(\expval{H_{\rm eff}}{\psi_\theta(t)}\right)^2= 4t^2 \var (H_{\rm eff})_{\ket{\psi_\theta(t)}}\, .
    \end{equation}

    We can now work with this equation to simplify it. We start by tackling the $E_{\rm eff}$. We write $H_\theta = \sum_k^N E_k \Pi_k$, where $E_k$ are the energies of $H_\theta$ and $\Pi_k$ the projectors into the eigen-basis of $H_\theta$. \revadd{Furthermore, we define $\widehat{H}_\theta = \sum_k^M E_k \Pi_k$ as the part of the Hamiltonian that interacts non-trivially with $\ket{\psi}$, i.e. $\Pi_i\ket{\psi}\neq 0$. Note that $M\leq N$, and in general. Furthermore, to ease the notation, and because the full $H_\theta$ is never relevant, we will use $H_\theta, \, \widehat{H}_\theta$ interchangeably.} With this, we can simplify the effective Hamiltonian as follows:
    
    \begin{align}
        H_{\rm eff} =&  \int_{0}^1 e^{-\ii tsH_\theta}H_S e^{\ii   tsH_\theta} ds 
        =  \sum_{k,j}^{M} \int_0^1 e^{-\ii ts E_k}\Pi_k H_S \Pi_j e^{\ii   ts E_j}ds \label{eq:exp-to-eigen}\\
        =& \sum_{k,j}^{M} \Pi_k H_S \Pi_j \int_0^1 e^{-\ii ts (E_k-E_j)}ds \nonumber\\
        =& \sum_k^{M} \Pi_k H_S \Pi_k  +\ii \sum_{j,k| j\neq k}^{M}\frac{1-e^{\ii    t(E_j - E_k)}}{t(E_j - E_k)}\Pi_k H_S \Pi_j \label{eq:exp-to-eigen-group}\\
        =& H_P +\ii \sum_{j,k| j\neq k}^{M}\frac{1-e^{\ii    t(E_j - E_k)}}{t(E_j - E_k)}\Pi_k H_S \Pi_j\,, \label{eq:exp-to-eigen-pinched}
    \end{align}
    where in Eq. \eqref{eq:exp-to-eigen} we rewrote $H_\theta$ as a sum of its energies and projectors, i.e. $e^{-\ii tH_\theta} = e^{-\ii t\sum_k E_k \Pi_k} = \sum _k e^{-\ii tE_k}\Pi_k$. In Eq. \eqref{eq:exp-to-eigen-group} we separated the previous sum in the terms where $k = j$ and $k\neq j$ and computed the integral. Note that when $k = j$ the terms in the complex exponential cancel. In Eq. \eqref{eq:exp-to-eigen-pinched} we define the pinched Hamiltonian $H_P = \mathcal{D}_{H_\theta}(H_S)= \sum_k \Pi_k H_S \Pi_k$, \revadd{note that this implies $\|H_P\|_\infty \leq \|H_S\|_\infty $ since $\mathcal{D}_{H_\theta}$ is a self-adjoint CPTP map.}

\revadd{
 Next we focus on bounding the operator norm of the second term in Eq~\eqref{eq:exp-to-eigen-pinched}, i.e. 
    \begin{align}\label{eq:original_problem}
       O = \sum_{j,k| j\neq k}^{M} \ii\frac{1-e^{\ii t(E_j - E_k)}}{t(E_j - E_k)}\Pi_k H_S \Pi_j = \sum_{j,k| j\neq k}^{M} f_{k,j}\,  \Pi_k H_S \Pi_j .
    \end{align}
To this end we note that 
\begin{equation}
    |f_{k,j}| = \frac{2 |\sin(\frac{t(E_k-E_j)}{2})|}{t |E_k-E_j|} \leq \frac{2}{t} \frac{1}{|E_k-E_j|}.
\end{equation}
Hence, by virtue of Proposition~\ref{prop:gersh-energy} we obtain 
\begin{align}\label{eq:original_problem_2}
    \left \| O \right\|_\infty  = \left\| \sum_{j,k| j\neq k}^{M} \ii\frac{1-e^{\ii t(E_j - E_k)}}{t(E_j - E_k)}\Pi_k H_S \Pi_j\right\|_\infty \leq \frac{2\pi }{ \sqrt 3}\frac{\|H_S\|_\infty}{t\Delta_g}  = \zeta \|H_S\|_\infty
    \end{align}
    where $\Delta_g  \leq |E_k-E_j|$ is minimal energy gap, and $ \zeta =\frac{2\pi }{ \sqrt 3\, t\Delta_g}$ is introduced to shorten the following expressions. 

    With this bound we obtain for $H_{\rm eff}= H_P + O$ and any state $\ket{\psi}$ we get 
\begin{align}
        \expval{H_{\rm eff}^2}{{\psi}} &= \expval{H_{P}^2}{{\psi}} +\expval{H_{P}O +OH_P + O^2}{{\psi}}\\
         \expval{H_{\rm eff}}{{\psi}}^2 &= \expval{H_{P}}{{\psi}}^2 +2 \expval{H_{P}}{{\psi}} \expval{O}{{\psi}} + \expval{O}{{\psi}}^2.
\end{align}
Implying 
\begin{align}
        |\expval{H_{\rm eff}^2}{{\psi}} - \expval{H_{P}^2}{{\psi}}| &\leq | \expval{H_{P}O +OH_P + O^2}{{\psi}}| \\
        &\leq \|O\|_\infty \, (2\|H_S\|_\infty + \|O\|_\infty ) 
        \\
        &\leq \|H_S\|^2_\infty \zeta(2 +\zeta)\\
         |\expval{H_{\rm eff}}{{\psi}}^2 - \expval{H_{P}}{{\psi}}^2| & \leq \|H_S\|^2_\infty \zeta(2 +\zeta),
\end{align}
and
\begin{align}
    |{\rm Var}(H_{\rm eff})_{\ket{\psi}} - {\rm Var}(H_{P})_{\ket{\psi}}|\leq 2 \|H_S\|^2_\infty \zeta(2 +\zeta) = 4 \,\|H_S\|^2_\infty \left(\zeta +\frac{\zeta^2}{2}\right).
\end{align}
Finally note that without loss of generality we can consider that $H_S$ to have a symmetric spectrum $\lambda_{\rm min}=\lambda_{\rm max}$ since the unitary encoding is invariant under the addition/subtraction of a term proportional to identity to $H_\theta$. Hence we find $\|H_S\|_{\infty} = \frac{1}{2} \|H_S\|$ and 
\begin{align}
    |\f  - 4 t^2 {\rm Var}(H_{P})_{\ket{\psi}}|\leq  4 t^2 \,\|H_S\|^2 \left(\zeta +\frac{\zeta^2}{2}\right) = \|H_S\|^2 4 t  \left( \frac{2\pi}{ \sqrt 3\, \Delta_g} +\frac{2\pi^2}{ 3 t\, \Delta_g^2} \right).
\end{align}
In particular, this implies the desired relation 
\begin{equation}
  \f =  4 t^2 \, {\rm Var}(H_{P})_{\ket{\psi}} + \mathcal{O}\left(t \frac{ \|H_S\|^2}{\Delta_g}\right),
\end{equation}
in the limit of large $t$.
}
\end{proof}

\section{Attaining Result \ref{res:pinchedlocal} with local measurements}\label{app:proof_measurement_general}
We devote this section to prove Result \ref{res:pinchedlocal} can be attained via local measurements if the signal is local $H_S = S_z = \frac{1}{2}\sum_{i=1}^N\sigma_z^{(i)}$. We start by recalling result 2.
\begin{result}\label{res:pinchedlocal_app}
    Let $H_\theta$ be the Hamiltonian of the form $H_\theta =\theta H_S + H_C $ where $\theta$ is the unknown parameter, $H_S$ is the signal and $H_C$ a control Hamiltonian on which we have full control. Let the sensor $\ket{\psi}$ be the ground state of the signal $H_S$, $\ket{\psi} = \ket{\Phi_\downarrow}$. Let it interact coherently with $H_\theta$ for a time $t$. The final state after the evolution is $\ket{\Phi_{\downarrow,\theta}(t)} = e^{-\ii t H_\theta} \ket{\Phi_\downarrow}$. Then for a good choice of $H_C$ the following QFI is attainable
    \begin{equation}
    \f = t^2\frac{\|H_S\|^2}{4} + \revadd{\order{\frac{t\|H_S\|^2}{\Delta_g}}}\,,
    \label{eq:finch_res2_app}
\end{equation}
where $\|H_S\| = \lambda_{\rm max}^{S} - \lambda_{\rm min}^{S}$ with $\lambda_{\rm max}^{S}$ ($\lambda_{\rm min}^{S}$) the maximal (minimal) eigenvalue of $H_S$.
\end{result}

In Section~\ref{sec:spin_network}, we claim that under the assumption that the signal $H_S$, for $H_S=\sum_i \sigma_{z}^{(i)}$ is a local Hamiltonian, then we can readily saturate the QFI via local measurements. That is we can reach the optimal mean squared error precision limit given by the Cram\' er-Rao bound with the QFI of  Eq.~\eqref{eq:optfisher}. 

\begin{proof} 

We start by  defining the (local) Pauli matrices $\sigma_{\vec{n}}=n_x \sigma_x+n_y\sigma_y+n_z\sigma_z$, with $|\vec{n}|=1$, written in the basis $\{\ket{\lambda_{\max}},\ket{\lambda_{\min}}\}$ of the local Hamiltonian. Similarly, we define the (collective) Pauli matrices with a hat, $\widehat{\sigma}_{\vec{n}}$, when they are defined in the basis of collective spin states $\{\ket{\Phi_\uparrow}=\ket{\lambda_{\max}}^{\otimes{N}},
\ket{\Phi_\downarrow}=\ket{\lambda_{\min}}^{\otimes N}\}$. Letting the initial state $\ket{\psi} = \ket{\Phi_\downarrow}$ evolve for a time $t$ and measuring the observable
\begin{equation}
\widehat\sigma_{\vec{m}(t)}={U}_t\widehat\sigma_y{U}_t^\dagger
\;\text{ with }\; {U}_t=e^{-\ii t H_\theta}
\label{eq:meassig}
\end{equation}
renders a precision given by by the QFI in Eq.~\eqref{eq:optfisher}. Indeed, by the simple error propagation formula~\cite{qmqi} we get
\begin{equation}
(\Delta\hat\theta)^2=\frac{1}{n}\frac{\var(\widehat\sigma_{\vec{m}(t)})_{\Phi_\uparrow(t)}}{(\partial_{\theta}\langle \sigma_{\vec{m}(t)}\rangle_{\Phi_\uparrow(t)}) ^{2}}=\frac{4}{n N^2 t^2 \Delta_-^2}\,.
\end{equation}

This is easily seen by computing the two terms that appear in the error propagation formula. First we focus on the variance of $\widehat\sigma_{\bm{m}}$ with respect to the state $\Phi_{\uparrow}(t)$. The expected value of the Pauli is
\begin{align}
\langle \widehat\sigma_{\vec{m}(t)}\rangle_{\Phi_\uparrow(t)}&=
\Tr(\widehat\sigma_{\vec{m}(t)}U_t \Phi_{\uparrow} U_t^\dagger)
=\Tr(\widehat\sigma_{y}\Phi_{\uparrow})=0\,.
\end{align}

And thus the variance is 
\begin{align}
    \var({\widehat\sigma_{\vec{m}(t)}})_{\Phi_\uparrow(t)} &=\langle \widehat\sigma_{\vec{m}(t)}^2\rangle_{\Phi_\uparrow(t)} -\langle \widehat\sigma_{\vec{m}(t)}\rangle_{\Phi_\uparrow(t)}^2=1\,,\\
\end{align}
where we use that $\widehat\sigma_{\bm{m}(t)}^2 = \1$ in the subspace of $\{\Phi_\uparrow(t), \Phi_\downarrow(t)\}$.

Next, we focus on the derivative of the expected value with respect to the unknown parameter. This can be computed as follows
\begin{align}
\frac{d\langle \sigma_{\vec{m}(t)}\rangle_{\Phi_\uparrow(t)}}{d\theta}&
\Tr\left[\widehat\sigma_{\vec{m}(t)}\partial_{\theta}\left(e^{-\ii t H_\theta} \Phi_{\uparrow} e^{\ii t H_\theta}\right)\right]=\nonumber\\
=& \ii t \Tr(\widehat\sigma_{\vec{m}(t)}[H_P,e^{-\ii t H_\theta} \Phi_{\uparrow}e^{\ii t H_\theta}]) + \revadd{\order{\|H_S\|^2/\Delta_g}}
\label{eq:effham2}\\
=& \ii t \Tr(\widehat\sigma_{y}[h_P,\Phi_{\uparrow}])+ \revadd{\order{\|H_S\|^2/\Delta_g}}\label{eq:comm}\\
=& \ii t \frac{N\Delta_-}{8} \Tr(\widehat\sigma_{y}[\widehat\sigma_x,\widehat\sigma_z])+ \revadd{\order{\|H_S\|^2/\Delta_g}}
\label{eq:comm2}\\
=&  t \frac{N\Delta_-}{2}+\revadd{\order{\|H_S\|^2/\Delta_g}}\,,
\end{align}
where in Eq. \eqref{eq:effham2}
we have introduced effective pinched Hamiltonian by means of Eq. \eqref{eq:A_as_sum}. Furthermore, in \eqref{eq:comm} we have used that  
 $[e^{-\ii t H_\theta},H_P]=0$, as they diagonalise in the same basis. Finally, in Eq. \eqref{eq:comm2}
we have used Eq. \eqref{eq:hpPauli}
and written the initial state as $\Phi_\uparrow=\frac{1}{2}(\1+\widehat{\sigma}_z)$.

Next we note that the action of the same local Pauli on all spins can be mapped to the action of a collective Pauli operator on the subspace 
$\{\ket{\Phi_\uparrow},\ket{\Phi_\downarrow}\}$:
\begin{align}
\widehat\sigma_{x}=
\widehat{P}\sigma_{x}^{\otimes N}\widehat{P} \;\mbox{ for } \;  N\in \mathbb{N}\,,\\
\widehat\sigma_{y}=
\widehat{P}\sigma_{y}^{\otimes N}\widehat{P} \;\mbox{ for } \; N=4k+1  \mbox{ and } k\in \mathbb{N}\,,\\
\widehat\sigma_{z}=
\widehat{P}\sigma_{z}^{\otimes N}\widehat{P} \;\mbox{ for } \; N=2k+1  \mbox{ and } k\in \mathbb{N}\,,
\end{align}
where $\widehat{P}=\ketbra{\Phi_\uparrow}{\Phi_\uparrow}+
\ketbra{\Phi_\downarrow}{\Phi_\downarrow}$.

This means that, for $N=4k+1$, at time $t$ we can measure the observable $\hat{O}_t=O_t^{\otimes N}$ where $O_t=U_t \sigma_y U_t^\dagger$ with $U_t=\exp{-\ii t \sigma_{\scriptscriptstyle\nearrow}}$ to efectively measure the collective observable $\widehat\sigma_{\vec{m}(t)}$ in
\eqref{eq:meassig} and thereby  saturate the QFI. Note that $\hat{O}_t$ is the product of identical Pauli operators, and hence we have shown that the optimal precision bound can be attained by performing a fixed Stern-Gerlach type measurement on every spin. 

For arbitrary $N$ we can also attain the QFI by strategy that local strategy that requires an adaptive measurement in the last step. To show this note that at any  time $t$ the state of the system can be written
as $\ket{\Psi}=\alpha \ket{\Phi_\uparrow}+
\beta \ket{\Phi_\downarrow}$ of $N$ spins, which after performing a sequence of $\sigma_x$
 on the first $(N-1)$ spins, the last spin will be found in the conditional state
$\ket{\tilde\Psi}=\alpha \ket{\uparrow}+(-1)^{p}\beta \ket{\downarrow}$, where ${p}$ is the number of $-1$ obtained in the sequence. It is now clear that all the  information encoded in the collective state $\ket{\Psi}$ can be accessed by a local measurement in the last qubit. In particular, for $p$ even one can directly measure the obervable $O_t$ on the last qubit. For $p$ odd this measurement should be preceeded by a $sigma_z$ unitary evolution in order to correct the negative sign.
 
\end{proof}

\section{Quantum advantage with two body interactions}

We devote this section to prove the remaining claims in Section~\ref{sec:all_to_all_seciton}. In Appendix~\ref{app:oneaxistwisting} we prove that the QFI scales as Eq.~\eqref{eq:fish2b} when the following conditions are met: the control term is $H_C = a S_x^2$ (where $a$ is much larger than the unknown signal $\theta$), the initial state is a product of the eigenstates of the Pauli matrix $\sigma_y$ and we have an odd number of particles $N$. In Appendix~\ref{app:measurement} we show that the QFI can be attained by measuring $S_x$.

\subsection{Proof of Eq.~\eqref{eq:fish2b}}\label{app:oneaxistwisting}
We devote this section to prove that having a one-axis twisting Hamiltonian ($H_C = S_x^2$) as a control term is enough to have a quantum advantage when the signal is a magnetic field in the $z$ direction ($H_S = S_z$). Namely, assuming the following system:
\begin{equation}
    H_\theta = \theta S_z + a S_x^2\,,
\end{equation}
where $a$ is just a term that dictates the strength of our interaction. With this system, we can achieve a QFI in the long-time regime of
\begin{equation}
        \f = t^2 \frac{N^{3/2}}{\sqrt{2\pi}}+\order{ t\revadd{\addw^2 N^2/a} }\,,
        \label{eq:fish2b_app}
\end{equation}
\revadd{assuming that we can} choose an odd number of particles and assume that the initial state of such particles is that all of them are pointing in the $+y$ direction in the Bloch sphere. That is, assume that the particles are in an initial state $\ket{\psi} = \ket{+y}^{\otimes N}$. \revadd{Finally, we set $\theta = 0$.}

\begin{proof} We start the proof by recalling and introducing some notation. As in the main text, we define $\ket{\psi_\theta(t)}$ as
\begin{equation}
    \ket{\psi_\theta(t)} = e^{-\ii t H_\theta} \ket{\psi}\,.
\end{equation}

As in Appendix~\ref{app:block_dycke} we define the eigenbasis $\{\ket{S, m_x}\}_{m_x=-S/2}^{S/2}$ fulfilling  $S^2\ket{S, m_x}=S(S+1)\ket{S, m_x}$
and
\begin{equation}\label{eq:basis_sx_app}
    S_x\ket{S, m_x} = m_x \ket{S, m_x}\,.
\end{equation}

In the regime of $\theta/a \approx 0$, we can use perturbation theory introduced in Appendix~\ref{app:perturbationtheory} in order to approximate the pinched Hamiltonian. More succinctly, we will treat $S_x^2$ as the non-perturbed Hamiltonian, and compute the eigenstates under the perturbation $\tfrac{\theta}{a} S_z$. With this at hand we will be able to compute the corresponding projectors $\Pi_k$ and the pinched Hamiltonian $H_P=\sum_k \Pi_k S_z \Pi_k$.

For $N$ odd all eigenvalues of unperturbed Hamiltonian $S_x^2$ 
are double degenerate $\ket{S,\pm m_x}$.
At first order perturbation theory 
the degeneracy is only broken for $|m|=\tfrac{1}{2}$. For all the other 
levels $|m|>\tfrac{1}{2}$ we can write the perturbed eigenstates as
\begin{align}\label{eq:use_ladder_ops}
    \ket{\overline{S, m_x}} =& \ket{S,m_x} + \frac{\theta}{a} \sum_{l_x\neq m_x} \frac{\bra{S,l_x}S_z\ket{S,m_x}}{m_x^2-l_x^2}\ket{S,l_x} +\order{\frac{\theta^2}{a^2}}\\
    =&\ket{S,m_x} + \frac{\theta}{2a} \left( \frac{\sqrt{(S+1)S - m_x(m_x-1)}}{2m_x-1}\ket{S,m_x-1} - \frac{\sqrt{(S+1)S - m_x(m_x+1)}}{2m_x+1}\ket{S,m_x+1} \right)\,+\order{\frac{\theta^2}{a^2}}\nonumber,
\end{align}
where have used that $S_z = ({S_+^{(x)} + S_-^{(x)}})/{2}$, with $S_+^{(x)},S_-^{(x)}$ are the ladder operators for the eigenstates of $S_x$, with $S_{\pm}^{(x)}\ket{S, m_x} = \sqrt{(S+1)S - m_x(m_x\pm1)}\ket{S,m_x\pm 1}$. Note that this also implies that for $|m_x|>1/2$, $\bra{S,\mp m_x}S_z\ket{S,\pm m_x}=0$ which was a prerequisite to apply perturbation theory in the degenerate case.
At first order in $\theta/a \approx 0$ these doublets $|m|>1/2$ remain degenerate.
The corresponding term of the pinched Hamiltonian $\Pi_{|m_x|} S_z \Pi_{|m_x|}$ is at most of order $\theta/a$ ($\forall m_x\neq \pm1/2$). Indeed,
\begin{equation}
    \left(\ketbra{\overline{S, m_x}}+\ketbra{\overline{S, -m_x}}\right) S_z\left(\ketbra{\overline{S, m_x}}+\ketbra{\overline{S, -m_x}}\right) = \order{\frac{\theta}{a}}\,.
\end{equation}
Note that at higher orders of perturbation theory, the degeneracy of these subspaces might break, rendering one-dimensional pinching projectors. 
However, the ``good'' perturbation basis  will always be of the form $\ket{S,|m_x|} = k_1 \ket{S,+m_x}+k_2 \ket{S,-m_x}$. Thus it is trivial to see that the $0$-th order of the pinched Hamiltonian will not change, as 
\begin{equation}
    \expval{S_z}{\overline{S,m_x}} = \order{\frac{\theta}{a}}\, \, \forall\,\, |m_x|>\frac{1}{2}
\end{equation}

We can now pay attention to the degenerate subspace with $|m| = \frac{1}{2}$. In this case the perturbation couples the two degenerate levels, $\bra{S,\mp m_x}S_z\ket{S,\pm m_x}\neq 0$, and one needs to find the basis that diagonalizes the perturbation in this subspace (see Appendix \ref{app:perturbationtheory}):
\begin{equation}
    \ket{S,\pm_x} = \frac{1}{\sqrt{2}}\left(\ket{S,m_x = \frac{1}{2}}\pm\ket{S,m_x = -\frac{1}{2}}\right)\;.
\end{equation}
Indeed, the states that fulfill $  \bra{S,\pm_x}S_z\ket{S,\pm_x} = 1, \, \bra{S,\pm_x}S_z\ket{S,\mp_x} = 0 $ . 

Now we can compute the terms of the pinched Hamiltonian corresponding to this subspace. This is
\begin{equation}\label{eq:use_sz_ladder}
    \overline{\Pi}_{\pm_x}S_z\overline{\Pi}_{\pm_x} = \ketbra{\overline{S,\pm_x}}S_z\ketbra{\overline{S,\pm_x}} = \ketbra{S,\pm_x} S_z \ketbra{S,\pm_x} + \order{\frac{\theta}{a}}\, ,
\end{equation}
where we have again used that $S_z = (S_+^{(x)}+S_-^{(x)})/2$.

Next, we can write the pinched Hamiltonian. Here we see the $0$-th order contributions of the $H_P$. That is
\begin{align}\label{eq:hp_oneaxis}
    H_P =& \ketbra{S,+_x} S_z \ketbra{S,+_x}+ \ketbra{S,-_x} S_z \ketbra{S,-_x} + \order{\frac{\theta}{a}}\\
    =& \frac{1}{2}\sqrt{(S + 1)S-\frac{1}{4}}(\ketbra{S,+_x} - \ketbra{S,-_x}) + \order{\frac{\theta}{a}}\,.
\end{align}

By choosing our initial state $\ket{\psi}$ as a tensor product of vectors in the Bloch sphere pointing in the same direction we can finalize the computation of the variance of $H_P$. We start by expressing this state as a linear combination of Dicke states as in Appendix~\ref{app:block_dycke}
\begin{equation}
    \ket{\psi} = \ket{\vartheta_x, \phi_x}^{\otimes N} = \frac{1}{(1+|\eta|^2)^{S}}\sum_{m_x=-S}^{S} \binom{2{S}}{{S}+m_x}^{1/2} \eta^{S+m_x} \ket{S,m_x}\,,
\end{equation}
where $\eta = -\tan\left(\frac{\vartheta_x}{2}\right)e^{-\ii \phi_x}$. 

Using Result \ref{res:pinched}, we can write the QFI for long times as the variance of such pinched Hamiltonian:
\begin{equation}\label{eq:pinched_here}
    \f = 4 t^2 \var(H_P)_{\ket{\psi}} +\order{t}\,.
\end{equation}

We separate this computation in two. First we will compute $\expval{H_P^2}{\psi}$ and later $\expval{H_P}{\psi}$. Just by direct computation, we find that
\begin{align}\label{eq:var_sx_1}
    \expval{H_P^2}{\psi} =& \frac{1}{4}\left( S(S+1)-\frac{1}{4} \right)\bra{\psi}\left( \ketbra{S,+_x} + \ketbra{S,-_x} \right)\ket{\psi}\\
    =&\frac{1}{4}\left( k(k+2)-\frac{1}{2} \right)\frac{|\eta|^{2k}}{(1+|\eta|^{2})^{2k}}\binom{2k+1}{k}\,,\label{eq:first_use_of_k}
\end{align}
where we have imposed that $S$ is odd by the following change of variable $S = \frac{2k+1}{2}$. Similarly, we can compute the other term of the variance
\begin{align}\label{eq:var_sx_2}
    \expval{H_P}{\psi} = -\sqrt{k(k+2)-\frac{1}{2}}\frac{|\eta|^{2k}}{(1+|\eta|^2)^{2k+1}}\tan\left(\frac{\revadd{\vartheta_x}}{2}\right)\cos (\phi_x)\,.
\end{align}

Then we can substitute Eqs.~(\ref{eq:var_sx_1},\ref{eq:var_sx_2}) into Eq.~\eqref{eq:pinched_here} \revadd{for $\theta = 0$}. Furthermore, we can let $\phi_x =\pm \frac{\pi}{2}$, to find that 
\begin{equation}
    \f = 4t^2 \frac{1}{4}\left( k(k+2)-\frac{1}{2} \right)\frac{|\eta|^{2k}}{(1+|\eta|^{2})^{2k}}\binom{2k+1}{k} + \revadd{\order{t\|H_S\|^2/\Delta_g}}\,.
\end{equation}
where we can see that $\Delta_g = \min_{j,k|j\neq k} |E_j-E_k| = \order{a}$. Indeed, we have that the minimum difference in energies of $J_x^2$ for different eigenvalues, is the difference between two subsequent energy levels, i.e. $9/4 - 1/4$. 

We see that the previous equation is maximized when $\vartheta_x = \pi$. If we substitute it, we obtain
\begin{equation}\label{eq:QFI_twisting}
    \f = t^2 \left( k(k+2)-\frac{1}{2} \right)\frac{1}{4^{k}}\binom{2k+1}{k}  + \order{ t\revadd{\addw ^2 N^2/a}} \approx t^2 \frac{N^{3/2}}{\sqrt{2\pi}} \,,
\end{equation}
\revadd{where we used that $\Delta_g = \order{a}$}.

Finally, we can check that $\phi_x = -\frac{\pi}{2}, \, \vartheta_x = \pi$ corresponds to the state $\ket{+y}$. We can write this state as follows
\begin{align}
    \ket{\vartheta_x = \pi, \phi_x = -\frac{\pi}{4}} =& \cos\frac{\pi}{4} \ket{+}+\sin\frac{\pi}{2}e^{-\ii \pi/2} \ket{-}\\
      =&\frac{e^{-\ii \pi/4}}{\sqrt{2}}\left( \ket{0}+ \ii\ket{1}\right) = \ket{+y}\,.
\end{align}

\end{proof}

\subsection{Comment on the measurement for the one-axis twisting Hamiltonian}\label{app:measurement}
Here we prove that under the QFI of Appendix~\ref{app:oneaxistwisting} can be attained by measuring $S_x$, the $x$ component of the spin.

\begin{proof} We start the proof by recalling some notation. We consider the setting $H_\theta = \theta S_z + a S_x^2$. As in the main text, we define $\ket{\psi_\theta(t)}$ as
\begin{equation}
    \ket{\psi_\theta(t)} = e^{-\ii t H_\theta} \ket{\psi}\,.
\end{equation}

Moreover, we define the eigenstates of $S_x$ in the symmetric space. This are $\{\ket{S, m_x}\}_{m_x=-S}^{S}$. These are defined as in the previous proof. They fulfill that
\begin{equation}\label{eq:basis_sx_app_measuremet}
    S_x\ket{S, m_x} = m_x \ket{S, m_x}\,.
\end{equation}

The measurement we perform is $M = S_x$, i.e. we measure the $x$ component of the spin. We also define the Classical Fisher Information (CFI):
\begin{equation}
        \mathcal{I} = \sum_x\frac{\Tr[\Pi_x\partial_\theta \rho_\theta]^2}{\Tr[\Pi_x\rho_\theta]} \,,
        \label{eq:classical_fisher_app}\,
\end{equation}
where $\Pi_x$ are the projectors of the measurement $M$, i.e. the possible outcomes of the spin in the $x$ component, $S_x$. 

First of all, we can compute the derivative of the evolved state with respect to the unknown parameter $\theta$ 
\begin{equation}
    \partial_\theta \ketbra{\psi_\theta(t)} = \ii t (\ketbra{\psi_\theta(t)}  H_P- H_P \ketbra{\psi_\theta(t)}) + \order{\revadd{\|H_S\|^2/a}}\,,
\end{equation}
\revadd{. We note that the term $\order{{\|H_S\|^2/a}}$ is independent of time}, and that thus in the long time regime will not be relevant. Furthermore, we recall that $H_P$ is the same pinched Hamiltonian as the one in Eq. \eqref{eq:hp_oneaxis}:
\begin{equation}
     H_P = \frac{1}{2}\sqrt{(S + 1)S-\frac{1}{4}}(\ketbra{S,+_x} - \ketbra{S,-_x}) + \order{\frac{\theta}{a}}\, .
\end{equation}

We now have all the elements to compute the CFI. When $\theta/a$ is around 0, the pinched Hamiltonian ($H_P$) only has non-zero eigenvalues in the sub-spaces $\ket{S,m_x = \pm1/2}$. In this regime, the computation of the CFI is quite simple. Indeed, we can readily apply Eq. \eqref{eq:classical_fisher_app} to obtain that \revadd{for $\theta = 0$}
\begin{align}\label{eq:classical_fisher_twisting}
    \mathcal{I}  = &\frac{t^2\Tr\bigg[\ii \left[\ketbra{S,m_x = +1/2},H_p\right]\ketbra{\psi_\theta}\bigg]^2}{ |\langle \psi_\theta\ket{S,m_x = + 1/2}|^2}\nonumber\\
    &+\frac{t^2\Tr\bigg[\ii \left[\ketbra{S,m_x = -1/2},H_p\right]\ketbra{\psi_\theta}\bigg]^2}{ |\langle \psi_\theta\ket{S,m_x = - 1/2}|^2}  +\order{t\revadd{\|H_S\|^2/a}}\,,
\end{align}
where we stress that the only non-vanishing terms corresponding to the outcome of the measurement being of $m_x= \pm 1/2$. Next is just a matter of computing the terms obtained. 

By recalling that the initial state is $\ket{+y}^{\otimes N}$, a tensor product of a Bloch-vectors pointing in the $y$ direction, we can just compute the two necessary terms as follows. Firstly, we compute the infinity norm of $H_P$, to simplify the notation moving forward. We find $\|H_P\|_{\infty} = \frac{1}{2}\sqrt{(S + 1)S-\frac{1}{4}}$. With this, we can proceed to do the following tedious calculations. Firstly we compute the terms in the numerator of Eq. \eqref{eq:classical_fisher_twisting}
\begin{equation}\label{eq:trace_upFI}
    \Tr[\ii \left[\ketbra{S,m_x = \pm 1/2},H_p\right]\ketbra{\psi_\theta}] =\|H_P\|_{\infty} \frac{1}{2^{2s}}\binom{2k+1}{k}2\cos\left( t\frac{\theta}{a}\|H_P\|_{\infty} \right)\,,
\end{equation}
where we have imposed that $S$ is an odd number via the change of variable $S = (2k+1)/2$ as used previously in Eq. \eqref{eq:first_use_of_k}. Secondly, we focus on the denominator of the same equation
\begin{equation}\label{eq:trace_downFI}
   |\langle \psi_\theta\ket{S,m_x = \pm 1/2}|^2 = \frac{2}{2^{2s}}\binom{2k+1}{k}\left[1\pm \sin\left( t\frac{\theta}{a}\|H_P\|_{\infty} \right)\right]\,.
\end{equation}

With these two calculations, we can compute Eq. \eqref{eq:classical_fisher_twisting}, to find that \revadd{for $\theta = 0$} the CFI under this measurement is 
\begin{align}
    \mathcal{I} =& \|H_P\|_{\infty} \frac{2}{2^{2s}}\binom{2k+1}{k}\cos^2\left( t\frac{\theta}{a}\|H_P\|_{\infty} \right) \left(\frac{1}{1-\sin\left( t\frac{\theta}{a}\|H_P\|_{\infty} \right)} + \frac{1}{1+ \sin\left( t\frac{\theta}{a}\|H_P\|_{\infty} \right)}\right)+\order{t\revadd{\|H_S\|^2/a}}\\
    =&t^2 \left( k(k+2)-\frac{1}{2} \right)\frac{1}{4^{k}}\binom{2k+1}{k} + \order{t\revadd{\|H_S\|^2/a}} \approx t^2 \frac{N^{3/2}}{\sqrt{2\pi}} 
\end{align}
and thus we see that this CFI is equal to the QFI in Eq \eqref{eq:QFI_twisting}, i.e. $\f = \mathcal{I}$.

\end{proof}

\section{Dephasing metrology with Hamiltonian control}\label{app:dephasing}

This section is devoted to proving Results~\ref{res: dephasing upperbound} and \ref{res: dephasing attainable} of the main text.

\subsection{Proof of the upper bound (Result~\ref{res: dephasing upperbound})}\label{subsection:proofupperbound}

\setcounter{result}{4}
\begin{result}\label{res:upperbound_app}
For any input state $\initial$ let  $\rho_\theta=\mathcal{D}_{H_\theta}(\initial)$ be the state resulting from the application of the dephasing Eq.~\eqref{eq: rho dephasing} with a nondegenerate Hamiltonian $H_\theta=H_C +\theta H_S$ with all energies gapped by at least $E$. The QFI of the state $\rho_\theta$ is upper-bounded by
\begin{align}
     \f \leq \frac{(3+\pi^2)}{3} \frac{\|H_S\|^2}{E^2}\;,
     \label{eq:dephasing_bound_app}
\end{align}
where $\|H_S\| = \lambda_{\rm max}^{S} - \lambda_{\rm min}^{S}$, as in Section~\ref{sec:dynamical_metro}.
\end{result}

\begin{proof}
We start this proof by emphasizing that the dephasing map $\mathcal{D}_{H_{\theta}}$ is invariant under any uniform shift Hamiltonian $H_\theta+f(\theta)\mathbf{1}$ because the eigenbasis does not change. Therefore, we can choose to have $H_S = H_S' + c\1$ such that we have the smallest $\|H_S\|_{\infty}$, i.e. $\min_{c} \|H_S' + c\1\|_\infty = \frac{1}{2}|\lambda_{\min}^S-\lambda_{\min}^S|$ (obtained choosing $c=\frac{1}{2}(\lambda_{\rm min}-\lambda_{\rm max})$). \revadd{This gives $\|H_S\|_{\infty}=\frac{1}{2} \|H_S\|$}. Hence, it suffices to prove that 
\begin{equation}\label{eqapp:deph_bound}
    \f\leq 4\frac{(3 +\pi^2)}{3}\frac{\|H_S\|^2_{\infty}}{E^2} \, ,
\end{equation}
to prove Result~\ref{res:upperbound_app}.

We now apply the Result~\ref{res: dephasing general} to the Hermitian operator $S$ in the specific form of Eq.~\eqref{eq: S of Hs}. We consider the two contributions to the QFI in Eq.~\eqref{eq: UB1 dephasing} separately.

Let us start with $\f_{\rm ext}$, which is the QFI of the state
\begin{equation}
\rho_\theta = \mathcal{D}_{H_\theta}(\psi) = \sum_k p_k \ketbra{\varphi_k},
\end{equation}
with $p_k = \bra{\varphi_k} \psi \ket{\varphi_k}$, under the unitary rotation generated by $S$. By convexity of the Fisher information, it is upper bounded $\f\leq \sum_k p_k \f_k$ by the average QFI of the trajectories where the same rotation is applied on the states $\ket{\varphi_k}$. For a fixed $k$ this is given by
\begin{equation}
\f_k = 4 \|\ket{\dot \varphi_k} \|_2^2=4{\rm Var}_{\ket{\varphi_k}}(S)= 4\|S \ket{\varphi_k}\|_2^2\, ,
\end{equation}
where we used the fact that $\bra{\varphi_k}S \ket{\varphi_k}=0$ to get the last equality. Furtehrmore, we used the definition of $S$ given in Eq.~\eqref{eq: S of Hs}. Now, from this expression we get
\begin{align}\nonumber
\| \ket{\varphi_k}\|_2^2 &= \sum_{j\neq k} \frac{|\bra{\varphi_j} H_S \ket{\varphi_k}|^2}{(E_k-E_j)^2} \\ &\leq \frac{\sum_{j\neq k} |\bra{\varphi_j} H_S \ket{\varphi_k}|^2} {\min_{j\neq k}(E_k-E_j)^2}
\leq \frac{\|H_S\|_{\infty}^2}{E^2}\, , 
\end{align}
where we used that the sum $ \sum_{j\neq k} |\bra{\varphi_j} H_S \ket{\varphi_k}|^2 = \expval{H_S \ketbra{\phi_k} H_S}{\phi_k}$,
and therefore $\f_{\rm ext}\leq 4 \frac{\|H_S\|_{\infty}^2}{E^2}$.

Next, consider $\f_{\rm int}$, which is the Fisher information of the distribution $p_k= \bra{\varphi_k} \psi \ket{\varphi_k}$ with $\dot p_k =\bra{\varphi_k} \ii [S,\psi] \ket{\varphi_k}$. It is thus bounded by the QFI of $\psi$ rotated with the generator $-S$, and we have
\begin{equation}
\f_{\rm int} \leq 4\|S\|_{\infty}^2 \qquad \revadd{\text{where} \qquad S:= \ii \sum_{j\neq k} \frac{\ketbra{\varphi_j} H_S \ketbra{\varphi_k}}{E_k-E_j}}
\end{equation}
\revadd{was derived in the main text. The operator can be written as 
\begin{equation}
    S=  \sum_{j\neq k} f_{k,j}\ketbra{\varphi_j} H_S \ketbra{\varphi_k} \qquad \text{with} \qquad |f_{k,j}| = \frac{1}{|E_k-E_j|}.
\end{equation}
Hence by virtue of the Proposition~\ref{prop:gersh-energy} we obtain 
\begin{equation}
  \|S\|_{\infty}^2 \leq \frac{s_d}{E^2} \|H_S\|_\infty^2 \leq \frac{\pi^2}{3 E^2} \|H_S\|_\infty^2 \quad \implies \quad \f_{\rm int} \leq \frac{4 \pi^2}{3 E^2} \|H_S\|_\infty^2
\end{equation}
where $d$ is the dimension of the Hilbert space.}

Bringing the two contributions together,  the total QFI satisfies
\begin{equation}
\f \leq \frac{4(3+\pi^2)}{3} \frac{\|H_S\|_{\infty}^2}{E^2}
\label{eq:decoh_bound}
\end{equation}
and complete the proof.
\end{proof}

\subsection{Proof of the attainable value (Result~\ref{res: dephasing attainable})}
\label{app: dephasing example}

\setcounter{result}{5}
\begin{result}
For any signal Hamiltonian $H_S$ with extremal eigenvalues $\lambda_{\min}^S$ and $\lambda_{\max}^S$ and the initial state $\psi$ satisfying $H_S \ket{\psi} =\frac{\lambda_\text{min}^S+\lambda_\text{max}^S}{2}\ket{\psi}$, there exists a control Hamiltonian $H_C$ such that the QFI of the dephased state $\rho_\theta=\mathcal{D}_{H_\theta}(\psi)$ in Eq.~\eqref{eq: rho dephasing} equals 
\begin{equation}
\cF = \frac{3}{2} \frac{\|H_S \|^2}{E^2} 
\end{equation}
where $E$ is the minimal energy gap of the (nondegenerate) Hamiltonian $H_\theta= H_C+ \theta H_S$, and $\|H_S\| = \lambda_{\rm max}^{S} - \lambda_{\rm min}^{S}$.
\end{result}

\begin{proof}
Let us denote by $\ket{\Phi_\downarrow}$ and $\ket{\Phi_\uparrow}$ the extremal eigenstates of  $H_S$ corresponding to the eigenvalue $\lambda_\text{min}^S$ and  $\lambda_\text{max}^S$ respectively. As discussed at the beginning of Section~\ref{subsection:proofupperbound}, without loss of generality we can consider $H_S$ with symmetric extremal eigenvalues given by $\lambda_\text{min}^S=-\|H_S\|_{\infty}$ and $\lambda_\text{max}^S=\|H_S\|_{\infty}$ (because shifting the Hamiltonian by a constant does not change the dephasing map). Then, by assumptions of the result, the initial state $\psi$ is in the kernel $H_S\ket{\psi}= 0$  of $H_S$.

It is also sufficient to consider the estimation around $\theta=0$ since any offset by a known value can be absorbed in the control Hamiltonian that we are free to choose. With this convention, at the parameter value $\theta=0$ the Hamiltonians $H_{\theta}=H_C$ coincide. Thus, bounding the energy gap of $H_\theta$ is equivalent to bounding the energy gap of $H_C$.

Since we are free to choose the control Hamiltonian $H_C$, we will choose it to be block-diagonal with respect to the qutrit subspace spanned by the basis $\{\ket{\Phi_\uparrow}, \ket{\psi}, \ket{\Phi_\downarrow}\}$. With this choice the dephased state $\mathcal{D}_{H_\theta}(\psi)$ remains in the subspace for all values of $\theta$. We can now restrict our analysis to the subspace, where the signal Hamiltonian reads  
\begin{equation}
H_S = \left(\begin{array}{ccc}
\|H_S\|_{\infty} &&\\
&0&\\
&& -\|H_S\|_{\infty}
\end{array}\right).
\end{equation}
For simplicity we restrict the control Hamiltonian $H_C=\sum_{i=1}^3 E_i \ketbra{\bm v_i}$ to be real in our basis and represent its eigenstates $\ket{\bm v_i}= x_i \ket{\Phi_\uparrow}+y_i\ket{\psi}+z_i\ket{\Phi_\downarrow}$ by orthonormal real vectors $\bm v_i = (x_i,y_i,z_i)$. By virtue of the Result~\ref{res: dephasing general} the QFI of the final state is the sum of two contributions
\begin{equation}
\cF = \cF_\text{int}+\cF_\text{ext},
\end{equation}
where $\cF_\text{int}$ is the Fisher information of the distribution $p_k = |\braket{\psi}{\bm v_k}|^2$ (recall that $\theta=0$, hence $\ket{\varphi_k}=\ket{\bm v_k}$), and $\cF_\text{ext}$ is the quantum Fisher information of the state $\psi$ rotated by the generator $S$ in Eq.~\eqref{eq: S of Hs}.

We start by analyzing the contribution $\cF_\text{int}$. With our notation, the probability distribution reads 
\begin{equation}
p_k = |\braket{\psi}{\bm v_k}|^2=y_k^2.
\end{equation}
Accordingly to first-order eigenvalue perturbation, for infinitesimal $\theta$ the eigenstates of $H_\theta$ (and of the dephasing map) change accordingly to 
\begin{align}
    \ket{\dot{\bm v}_k} &=  \sum_{j\neq k}\ket{\bm v_j} \frac{\bra{\bm v_j} H_S \ket{\bm v_k}}{E_k-E_j} = \|H_S\|_{\infty} \sum_{j\neq k}\ket{\bm v_j} \frac{x_j x_k - z_j z_k}{E_k-E_j}.
\end{align}
This allows us to compute the derivatives of output probabilities
\begin{align}
\dot p_k &= 2\, \text{Re}(\braket{\dot{\bm v}_k}{\psi}\braket{\psi}{\bm v_k}) = 2 \|H_S\|_{\infty} \,y_k \sum_{j\neq k} y_j \frac{x_j x_k - z_j z_k}{E_k-E_j},
\end{align}
and its Fisher information
\begin{align}
\cF_\text{int}&= \sum_{k=1}^3 \frac{(\dot p_k)^2}{p_k} 
= 4 \|H_S\|_{\infty}^2 \sum_{k=1}^3\left( \sum_{j\neq k} y_j \frac{x_j x_k - z_j z_k}{E_k-E_j}\right)^2.
\end{align}
We now choose the energies of $H_C$ to be arranged as $(E_1, E_2, E_3)=(E,0,-E)$ (satisfying the gap assumption) and parameterize the orthonormal vectors with the Euler angles $\phi',\theta',\psi'$ (as lines of a general rotation matrix) 
\begin{align}
\bm v_1 &=\Big(\cos (\psi' ) \cos (\phi' )-\cos (\theta' ) \sin (\psi' ) \sin (\phi' ),-\cos (\theta' ) \cos (\psi' ) \sin (\phi' )-\sin (\psi' ) \cos (\phi' ),\sin (\theta') \sin (\phi' )\Big) \\
\bm v_2 &=\Big(\cos (\theta' ) \sin (\psi' ) \cos (\phi' )+\cos (\psi' ) \sin (\phi' ),\cos (\theta' ) \cos (\psi' ) \cos (\phi' )-\sin (\psi' ) \sin (\phi' ), -\sin (\theta ')-\cos (\phi' ) \Big)\\
\bm v_3 & =\Big(\sin (\theta' ) \sin (\psi' ),\sin (\theta' ) \cos (\psi' ),\cos (\theta' )\Big).
\end{align}
This allows us to express $\cF_\text{int}(\phi',\theta',\psi')$ in terms of the three Euler angles. The expression is cumbersome, so we do not report it here.
Nevertheless, we numerically maximize it with respect to the free parameters $\phi',\theta',\psi'$. We find that the value  
\begin{equation}
\cF_\text{int} =\max_{\phi',\theta',\psi'} \cF_\text{int}(\phi',\theta',\psi') = 4 \frac{\|H_S\|_{\infty}^2}{E^2}
\end{equation}
is attained for
\begin{align}
\bm v_1 =\left(\frac{1}{2},\frac{1}{\sqrt{2}},\frac{1}{2}\right),\quad \bm v_2= \left(-\frac{1}{\sqrt{2}},0,\frac{1}{\sqrt{2}}\right), \quad \bm v_3= \left(\frac{1}{2},-\frac{1}{\sqrt{2}},\frac{1}{2}\right)
\end{align}
or for the control Hamiltonian
\begin{equation}
H_C = \frac{E}{\sqrt 2} (\ketbra{\bm v_1} - \ketbra{\bm v_3})= \frac{E}{\sqrt{2}}\left(
\begin{array}{ccc}
 0 & 1 & 0 \\
 1 & 0 & 1 \\
 0 & 1 & 0 \\
\end{array}
\right).
\end{equation}

For the choice of $H_C$ let us now also compute the expression of $\cF_\text{ext}$. For the dephased state we find
\begin{equation}\rho = \sum_{i} \ketbra{\bm v_i} \psi \ketbra{\bm v_i} =\frac{1}{2}(\ketbra{\bm v_1}+\ketbra{\bm v_3})=\frac{1}{2} \left(
\begin{array}{ccc}
 \frac{1}{2} & 0 & \frac{1}{2} \\
 0 & 1 & 0 \\
 \frac{1}{2} & 0 & \frac{1}{2} \\
\end{array}
\right).
\end{equation}
While the operator $S$ in Eq.~\eqref{eq: S of Hs} we find
\begin{equation}
S =  \ii \sum_{j\neq k} \frac{\ketbra{\bm v_j} H_S \ketbra{\bm v_k}}{E_k-E_j} = \frac{\|H_S\|_{\infty}}{\sqrt 2 E} \left(
\begin{array}{ccc}
 0 & -\ii & 0 \\
 \ii & 0 & - \ii \\
 0 & \ii & 0 \\
\end{array}
\right).
\end{equation}
$\cF_\text{ext}$ is the QFI for the rotation $\dot \rho = -\ii [S,\rho]$ which we compute using the spectral decomposition of $\rho = \sum_{k} q_k \ketbra{\rho_k}$ and the standard expression
\begin{equation}
\cF_\text{ext} = 2 \sum_{\substack{j,k\\q_j+q_k>0}} \frac{(q_k-q_j)^2}{q_k+q_j}|\bra{\rho_k}S\ket{\rho_j}|^2 = 2\frac{\|H_S\|_{\infty}^2}{E^2}
\end{equation}
Summing up the two contributions we obtain the final expression of the total QFI
\begin{equation}
\cF=\cF_\text{int}+\cF_\text{ext}= 6 \frac{\|H_S\|_{\infty}^2}{E^2}=\frac{3}{2} \frac{|\lambda^S_{\rm max}-\lambda^S_{\rm min}|^2}{E^2} .
\end{equation}
where we have used that $\|H_S\|_{\infty} = \frac{1}{2}(\lambda^{S}_{\max}-\lambda^{S}_{\min})$.

\end{proof}

\section{Transient regime and global noise}

\subsection{Proof of Eqs.~\eqref{eq:cfi_noise} and~\eqref{eq:qfi_no_control}}
We devote this section to proving the claims in Section~\ref{subsec:global}. We consider a global noise $S_x$ with a parameter strength $\gamma$. That is, in Eq. \eqref{eq:limbladian_general}, $L_{\theta, i} = S_x$ and $\gamma_{i} = \gamma$  such that the differential equation that governs the dynamics of our system is 
\begin{equation}
        \dot{\psi}_{\theta}(t) = -\ii [H_\theta,{\psi}_{\theta}(t)] + \gamma\Big(S_x {\psi}_{\theta}(t) S_x -\frac{1}{2}\{S_x^2, {\psi}_{\theta}(t) 
        \}\Big) \ ,
    \end{equation}
    where $H_{\theta} = H_C + \theta S_z$. 
    
    We use $H_C = a S_x^2$ as the control Hamiltonian. \revadd{For large $a$, and $\theta=0$}, then under the same conditions used to obtain Eq. \eqref{eq:fish2b} and measuring $S_x$, we find that the CFI 
    \begin{equation}
        \mathcal{I}_{\rm control} =  4\dfrac{(1-\mathrm{e}^{-t{\gamma/2}})^2}{{\gamma^2}}\frac{N^{3/2}}{\sqrt{2\pi}} +\order{\frac{1}{a}}\,.
    \end{equation}
This can be contrasted to the QFI obtained in the absence of control ($H_C = 0$):
 \begin{equation}
        \f_{\rm{no \,\, control}} =  4\dfrac{(1-\mathrm{e}^{-t{\gamma/2}})^2}{{\gamma^2}}N.
    \end{equation}
Hence, we obtain a $N^{1/2}$ fold enhancement by appropriately tuning the interactions, which in this case does not come at the price of higher equilibration times.

\begin{proof} 
We divide this into two steps. First, we compute the CFI with control, i.e. $\mathcal{I}_{\rm cotrol}$. 

\vspace{0.5mm}
\noindent\textbf{\underline{Fisher with control, i.e. $\mathcal{I}_{\rm cotrol}$.} } We will prove this by computing the CFI for this setting. That is we assume a Hamiltonian for the coherent dynamics of the form 
\begin{equation}
    H_\theta = a S_x^2 + \theta S_z \,.
\end{equation}

We assume that we can prepare a product state of spins in the $y$ direction in the Bloch sphere, i.e. $\ket{\psi} = \ket{y}^{\otimes N}$. We have shown that measuring $S_x$ is the best measurement when using the one-axis twisting control term without noise. Thus, we use it to compute the CFI in the noisy case.  Then we assume we have a collective noise that is $S_x$. We model this with a Limbladian equation. That is 
\begin{equation}
        \dot{\psi}_{\theta}(t) = -\ii [H_\theta,{\psi}_{\theta}(t)] + \gamma\Big(S_x {\psi}_{\theta}(t) S_x -\frac{1}{2}\{S_x^2, {\psi}_{\theta}(t) 
        \}\Big) \, ,
    \end{equation}
where $\gamma$ is the strength of the noise. We emphasise that the initial state $\ket{\psi}= \ket{y}^{\otimes N}$. 

By solving this differential equation one obtains 
\begin{equation}
     \psi_\theta(t) = e^{-t\mathcal{L}_\theta }\left( \ketbra{y}^{\otimes N} \right)\,.
\end{equation}

To compute the Fisher Information we need to compute the derivative $\partial_\theta \psi_\theta(t)$. To do this we use the vectorization formalism (for more information look at Appendix \ref{app:vec_formalism}). The vectorised form of $\psi_\theta(t)$ is denoted as $\vecc{\psi_\theta(t)}$, with this
\begin{equation}
    \vecc{\psi_\theta(t)} = e^{-\ii t H_\theta \otimes \1 + \ii t \1\otimes H_\theta ^T + \frac{t\gamma}{2}\left( 2 S_x\otimes S_x^T - S_x^2\otimes\1 -\1\otimes \left(S_x^2\right)^T \right)}\left(\ket{y}\ket{y^*}\right)^{\otimes N}\,,
\end{equation}
where $\ket{y^*}$ indicates the conjugate version of $\ket{y}$ that appears when vectorising $\ketbra{y}$. From now on, and to ease the notation, we will denote $\vecc{y} = \left(\ket{y}\ket{y^*}\right)^{\otimes N}$. We also used that the vectorised form of the limbladian $\mathcal{L}_\theta$ is $\hat{\mathcal{L}}_\theta = -\ii t H_\theta \otimes \1 + \ii t \1\otimes H_\theta ^T + \frac{t\gamma}{2}\left( 2 S_x\otimes S_x^T - S_x^2\otimes\1 -\1\otimes \left(S_x^2\right)^T \right)$. Then the previous equation would look like
\begin{equation}
    \vecc{\psi_\theta(t)} = e^{-\ii t H_\theta \otimes \1 + \ii t \1\otimes H_\theta ^T + \frac{t\gamma}{2}\left( 2 S_x\otimes S_x^T - S_x^2\otimes\1 -\1\otimes \left(S_x^2\right)^T \right)}\vecc{y}\,.
\end{equation}

Indeed, in this way, it is easier to compute the derivative with respect to $\theta$ of this evolved state. To do so we can use Eq. \eqref{eq:derivative_matrixexp} (in Appendix \ref{app:derivative_exp}). Then we see that this accounts for having
\begin{align}
    \partial_\theta\vecc{\psi_\theta(t)} = -\ii t \int_{0}^1 ds &e^{-\ii ts H_\theta \otimes \1 + \ii ts \1\otimes H_\theta ^T + \frac{ts\gamma}{2}\left( 2 S_x\otimes S_x^T - S_x^2\otimes\1 -\1\otimes \left(S_x^2\right)^T \right)} \left( 
 S_z\otimes\1 - \1\otimes S_z^T\right)\\
 \cdot &e^{-\ii t(1-s) H_\theta \otimes \1 + \ii t(1-s) \1\otimes H_\theta ^T + \frac{t(1-s)\gamma}{2}\left( 2 S_x\otimes S_x^T - S_x^2\otimes\1 -\1\otimes \left(S_x^2\right)^T \right)}\vecc{y}\,,
\end{align}
where $\cdot$ refers to the usual product~\footnote{We emphasise the product due to the line-break.}.

If we write this in the eigenbasis of $S_x$, we can remove the transpose $T$ in all the terms in the exponent as $S_z^T = S_z, \ S_x = S_x^T$. With this, we can rewrite the derivative as follows:
\begin{align}
    \partial_\theta\vecc{\psi_\theta(t)} = -\ii t \int_{0}^1 ds &e^{-\ii ts H_\theta \otimes \1 + \ii ts \1\otimes H_\theta  + \frac{ts\gamma}{2}\left( 2 S_x\otimes S_x - S_x^2\otimes\1 -\1\otimes S_x^2 \right)} \left( 
 S_z\otimes\1 -\1\otimes S_z\right)\\
 \cdot &e^{-\ii t(1-s) H_\theta \otimes \1 + \ii t(1-s) \1\otimes H_\theta ^T + \frac{t(1-s)\gamma}{2}\left( 2 S_x\otimes S_x - S_x^2\otimes\1 -\1\otimes S_x^2 \right)}\vecc{y}\,.
\end{align}

If we recall the expression of the CFI
\begin{equation}
        \mathcal{I} = \sum_x\frac{\Tr[\Pi_x\partial_\theta \psi_\theta(t)]^2}{\Tr[\Pi_x\psi_\theta(t)]} \, ,
        \label{eq:classical_fisher_app2}
\end{equation}
we see that we have to compute the following traces, that under the vectorization formalism, they have the following form $\Tr[\Pi_x \partial_\theta\psi_\theta(t)] = \langle\!\langle \Pi_x|\partial_\theta\vecc{\psi_\theta(t)}$. 

In the case of  $\theta = 0$ both $[H_{\theta=0},S_x] = 0$. And it is in this case in which we compute the CFI. We compute the terms for $\Pi_x = \ketbra{S,m_x = \pm 1/2}$. Under the assumption that $\theta = 0$, then using that we are working in the basis of $S_x$, notation $\bra{S,m_x = \pm 1/2}\bra{S,m_x = \pm 1/2^*} = \langle\!\langle \Pi_{\pm}|$
\begin{align}
    \Tr\left[ \Pi_{\pm} \partial_\theta\psi_\theta(t)\right]  =& \langle\!\langle \Pi_{\pm}|\partial_\theta\vecc{\psi_\theta(t)}\\ 
=&-\ii t \bigg\langle\!\!
\!\bigg\langle \Pi_{\pm}\bigg|\int_{0}^1 ds e^{-\ii tsa( S_x^2\otimes\1 - \1\otimes S_x^2) + \frac{ts\gamma}{2}\left( 2 S_x\otimes S_x - S_x^2\otimes\1 -\1\otimes S_x^2 \right)} \left( 
 S_z\otimes\1 -\1\otimes S_z\right)\\
 &\cdot e^{-\ii t(1-s)a(S_x^2 \otimes \1 +  \1\otimes S_x^2) + \frac{t(1-s)\gamma}{2}\left( 2 S_x\otimes S_x - S_x^2\otimes\1 -\1\otimes S_x^2 \right)}\bigg|y\bigg\rangle\!\!
\!\bigg\rangle\,.
\end{align}

Now using the fact that $\vecc{\Pi_{\pm}}$ is an eigenvector of the exponent with eigenvalue $0$, i.e.
\begin{equation}\label{eq:trace_presz}
    \left[-\ii tsa( S_x^2\otimes\1 - \1\otimes S_x^2) + \frac{ts\gamma}{2}\left( 2 S_x\otimes S_x - S_x^2\otimes\1 -\1\otimes S_x^2 \right)\right]\bigg|\Pi_{\pm}\bigg\rangle\!\!
\!\bigg\rangle = 0
\end{equation}
the previous expression can be simplified. Indeed, then we obtain
\begin{align}
    \Tr\left[ \Pi_{\pm} \partial_\theta \psi_\theta(t)\right] = -\ii t \bigg\langle\!\!
\!\bigg\langle \Pi_{\pm}\bigg|\int_{0}^1 ds  \left( 
 S_z\otimes\1 -\1\otimes S_z\right) e^{-\ii t(1-s)a(S_x^2 \otimes \1 -  \1\otimes S_x^2) + \frac{t(1-s)\gamma}{2}\left( 2 S_x\otimes S_x - S_x^2\otimes\1 -\1\otimes S_x^2 \right)}\bigg|y\bigg\rangle\!\!
\!\bigg\rangle
\end{align}

Next, we apply $\langle\!\langle \Pi_{\pm}|$ to $(S_z\otimes\1 + \1\otimes S_z)$, where to do so, we recall that $S_z = \frac{1}{2}(S_+ + S_-)$ i.e. the ladder operators for the eigenvectors of $S_x$. Thus we obtain for the projector $\Pi_+$
\begin{align}
    \langle\!\langle \Pi_{+}| \left( 
 S_z\otimes\1 -\1\otimes S_z\right) =&\bra{S,m_x = +1/2}\bra{S,m_x = +1/2} \left( 
 S_z\otimes\1 -\1\otimes S_z\right)\\
 =&\frac{1}{2}\bigg(\sqrt{\frac{N}{2}\left(\frac{N}{2}+1\right)+\frac{1}{4}}\bra{S,m_x = -1/2} \\
 &+  \sqrt{\frac{N}{2}\left(\frac{N}{2}+1\right)-\frac{3}{4}}\bra{S,m_x = 3/2}\bigg)\bra{S,m_x = +1/2}\nonumber\\
 &-\frac{1}{2}\bra{S,m_x = +1/2}\bigg(\sqrt{\frac{N}{2}\left(\frac{N}{2}+1\right)+\frac{1}{4}}\bra{S,m_x = -1/2} \nonumber\\
 &+  \sqrt{\frac{N}{2}\left(\frac{N}{2}+1\right)-\frac{3}{4}}\bra{S,m_x = 3/2}\bigg)\nonumber\,.
\end{align}

For the projector $\Pi_-$ we find a very similar expression. Indeed
\begin{align}
    \langle\!\langle \Pi_{-}| \left( 
 S_z\otimes\1 -\1\otimes S_z\right) =&\bra{S,m_x = -1/2}\bra{S,m_x = -1/2} \left( 
 S_z\otimes\1 -\1\otimes S_z\right)\\
 =&\frac{1}{2}\left(\sqrt{\frac{N}{2}\left(\frac{N}{2}+1\right)+\frac{1}{4}}\bra{S,m_x = 1/2} +  \sqrt{\frac{N}{2}\left(\frac{N}{2}+1\right)-\frac{3}{4}}\bra{S,m_x = -3/2}\right)\bra{S,m_x = -1/2}\nonumber \\
 &-\frac{1}{2}\bra{S,m_x = -1/2}\bigg(\sqrt{\frac{N}{2}\left(\frac{N}{2}+1\right)+\frac{1}{4}}\bra{S,m_x = +1/2} \\
 &+  \sqrt{\frac{N}{2}\left(\frac{N}{2}+1\right)-\frac{3}{4}}\bra{S,m_x = -3/2}\bigg)\nonumber\,.
\end{align}

Out of these results, we focus for now on the terms that are $\bra{S,m_x = \pm 1/2}\bra{S,m_x = \mp 1/2}$. By doing a similar analysis as before, we can see that these states are eigenvectors of the matrix in the exponent. Indeed
\begin{align}\label{eq:pm1/2}
    &\bra{S,m_x = \pm 1/2}\bra{S,m_x = \mp 1/2}\bigg[-\ii t(1-s)a(S_x^2 \otimes \1 -  \1\otimes S_x^2) \\
    &+ \frac{t(1-s)\gamma}{2}\left( 2 S_x\otimes S_x - S_x^2\otimes\1 -\1\otimes S_x^2 \right)\bigg] = -\frac{t(1-s)\gamma}{2}\,.
\end{align}

We can check that the other terms will banish as follows. First, we check that these are eigenvectors of the same exponential. Indeed, 
\begin{align}
    &\bra{S,m_x = \pm 1/2}\bra{S, m_x = \pm{3}/{2}}\bigg[-\ii t(1-s)a(S_x^2 \otimes \1 -  \1\otimes S_x^2) \\
    &+ \frac{t(1-s)\gamma}{2}\left( 2 S_x\otimes S_x - S_x^2\otimes\1 -\1\otimes S_x^2 \right)\bigg] = t(1-s)(a \ii -2\gamma)
\end{align}
and similarly
\begin{align}
    &\bra{S, \pm 3/2}\bra{S,m_x = \pm 1/2}\bigg[-\ii t(1-s)a(S_x^2 \otimes \1 -  \1\otimes S_x^2) \\
    &+ \frac{t(1-s)\gamma}{2}\left( 2 S_x\otimes S_x - S_x^2\otimes\1 -\1\otimes S_x^2 \right)\bigg] = -t(1-s)(a \ii+2\gamma)\,.
\end{align}

We can put this back in Eq. \eqref{eq:trace_presz}. Then we can see the following
\begin{align}
    &-\ii t\sqrt{\frac{N}{2}\left(\frac{N}{2}+1\right)-\frac{3}{4}}\int_0^1 ds\bigg( \bra{S,m_x = \pm 1/2}\bra{S,m_x = \pm 3/2} e^{t(1-s)(a \ii -2\gamma)} \\
    &- \bra{S,m_x = \pm 3/2}\bra{S,m_x = \pm 1/2}e^{-t(1-s)(a \ii+2\gamma)}\bigg)\bigg|y\bigg\rangle\!\!
\!\bigg\rangle = \order{\frac{1}{a}} 
\end{align}
thus if the interaction is strong enough we can discard this term.

Now let us analise the term proportional to $\bra{S,m_x = +1/2}\bra{S,m_x = -1/2}$. If we put this result and Eq. \eqref{eq:pm1/2} back to Eq. \eqref{eq:trace_presz}, we get
\begin{align}
    \Tr[\Pi_{\pm} \partial_\theta\psi_\theta(t)]  =&-\ii t\frac{1}{2}\sqrt{\frac{N}{2}\left(\frac{N}{2}+1\right)-\frac{3}{4}}( \bra{S,m_x = +1/2}\bra{S,m_x = -1/2} \\
    &-\bra{S,m_x = -1/2}\bra{S,m_x = +1/2}  )|y\rangle\!\rangle \int_{0}^1 ds e^{-\frac{t(1-s)\gamma}{2}} + \order{\frac{1}{a}}\\
=&-\ii \dfrac{1-e^{-\frac{t{\gamma}}{2}}}{{\gamma}}\sqrt{\frac{N}{2}\left(\frac{N}{2}+1\right)-\frac{3}{4}}( \bra{S,m_x = +1/2}\bra{S,m_x = -1/2} \\
&-\bra{S,m_x = -1/2}\bra{S,m_x = +1/2}  )|y\rangle\!\rangle +\order{\frac{1}{a}}\\
=&-\ii \dfrac{1-e^{-\frac{t{\gamma}}{2}}}{{\gamma}}\sqrt{\frac{N}{2}\left(\frac{N}{2}+1\right)-\frac{3}{4}} \left( \braket{S,m_x = +1/2}{y^{\otimes N}} -  \braket{y^{\otimes N}}{S,m_x = -1/2}\right)\,.
\end{align}
This result is the same as Eq. \eqref{eq:trace_upFI} with a different prefactor in time dependence.  Similarly, because $\ket{S,m_x = \pm 1/2}\ket{S,m_x = \pm 1/2}$ is an eigenstate of the exponential dynamics with eigenvalue 0, as shown in Eq. \ref{eq:trace_presz}, then the terms in the CFI $\Tr[\Pi_x\psi_\theta(t)]$ are the same as \ref{eq:trace_downFI}. Therefore the CFI will be the same, just with a different prefactor in time. Indeed, we can find that
\begin{equation}
    \mathcal{I}_{\rm control} = \f \frac{1-e^{-\frac{t{\gamma}}{2}}}{\gamma t^2}  \approx  4\dfrac{(1-\mathrm{e}^{-t\gamma/2})^2}{\gamma^2}\frac{N^{3/2}}{\sqrt{2\pi}}\,.
\end{equation}

\noindent\textbf{\underline{Quantum Fisher with no control, i.e. $\f_{\rm no \,\, cotrol}$.}} We will prove this by computing the Quantum Fisher Information for this setting. That is we assume a Hamiltonian for the coherent dynamics of the form 
\begin{equation}
    H_\theta = \theta S_z \,.
\end{equation}

Let us start with $\ket{\psi} = \ket{+}^{\otimes N}$, where $\ket{+}$ is the eigenstate of $\sigma_x$, as our initial state. This is the state that will dissipate the least under the condition $\theta = 0$ applied before. Then the dynamics are governed by the following differential equation. Then we assume we have a collective noise that is $S_x$. We model this with a Limbladian equation. That is 
    \begin{equation}\label{eq:dynamics_globalnoise_nocontrol}
        \dot{\psi}_{\theta}(t) = -\ii [H_\theta,{\psi}_{\theta}(t)] + \gamma\Big(S_x {\psi}_{\theta}(t) S_x -\frac{1}{2}\{S_x^2, {\psi}_{\theta}(t) 
        \}\Big) \ ,
    \end{equation}
where $\gamma$ is the strength of the noise. By solving this differential equation one obtains 
\begin{equation}
     \psi_\theta(t) = e^{-t\mathcal{L}_\theta }\left( {\psi} \right)\,.
\end{equation}

To compute the Quantum Fisher Information we need to compute the derivative $\partial_\theta \psi_\theta(t)$. To do this we use the vectorization formalism (for more information look at Appendix \ref{app:vec_formalism}). The vectorized form of $\psi_\theta(t)$ is denoted as $\vecc{\psi_\theta(t)}$, with this
\begin{equation}
    \vecc{\psi_\theta(t)} = e^{-\ii t H_\theta \otimes \1 + \ii t \1\otimes H_\theta ^T + \frac{t\gamma}{2}\left( 2 S_x\otimes S_x^T - S_x^2\otimes\1 -\1\otimes \left(S_x^2\right)^T \right)}\left(\ket{+}\ket{+^*}\right)^{\otimes N}\,,
\end{equation}
where $\ket{+^*}$ indicates the conjugate version of $\ket{+}$ that appears when vectorising $\ketbra{+}$. From now on, and to ease the notation, we will denote $\vecc{+} = \left(\ket{+}\ket{+^*}\right)^{\otimes N}$. We also used that the vectorised form of the limbladian $\mathcal{L}_\theta$ is $\hat{\mathcal{L}}_\theta = -\ii t H_\theta \otimes \1 + \ii t \1\otimes H_\theta ^T + \frac{t\gamma}{2}\left( 2 S_x\otimes S_x^T - S_x^2\otimes\1 -\1\otimes \left(S_x^2\right)^T \right)$. Then the previous equation would look like
\begin{equation}
    \vecc{\psi_\theta(t)} = e^{-\ii t H_\theta \otimes \1 + \ii t \1\otimes H_\theta ^T + \frac{t\gamma}{2}\left( 2 S_x\otimes S_x^T - S_x^2\otimes\1 -\1\otimes \left(S_x^2\right)^T \right)}\vecc{+}\,.
\end{equation}

Indeed, in this way, it is easier to compute the derivative with respect to $\theta$ of this evolved state. To do so we can use Eq. \eqref{eq:derivative_matrixexp} (in Appendix \ref{app:derivative_exp}). Then we see that this accounts for having
\begin{align}
    \partial_\theta\vecc{\psi_\theta(t)} = -\ii t \int_{0}^1 ds &e^{-\ii ts H_\theta \otimes \1 + \ii ts \1\otimes H_\theta ^T + \frac{ts\gamma}{2}\left( 2 S_x\otimes S_x^T - S_x^2\otimes\1 -\1\otimes \left(S_x^2\right)^T \right)} \left( 
 S_z\otimes\1 - \1\otimes S_z^T\right)\\
 \cdot &e^{-\ii t(1-s) H_\theta \otimes \1 + \ii t(1-s) \1\otimes H_\theta ^T + \frac{t(1-s)\gamma}{2}\left( 2 S_x\otimes S_x^T - S_x^2\otimes\1 -\1\otimes \left(S_x^2\right)^T \right)}\vecc{+}
\end{align}
where $\cdot$ refers to the usual product~\footnote{We emphasise the product due to the line-break.}.

If we write this in the eigenbasis of $S_x$, we can remove the transpose $T$ in all the terms in the exponent as $S_z^T = S_z, \ S_x = S_x^T$. With this, we can rewrite the derivative as follows:
\begin{align}
    \partial_\theta\vecc{\psi_\theta(t)} = -\ii t \int_{0}^1 ds &e^{-\ii ts H_\theta \otimes \1 + \ii ts \1\otimes H_\theta  + \frac{ts\gamma}{2}\left( 2 S_x\otimes S_x - S_x^2\otimes\1 -\1\otimes S_x^2 \right)} \left( 
 S_z\otimes\1 -\1\otimes S_z\right)\\
 \cdot &e^{-\ii t(1-s) H_\theta \otimes \1 + \ii t(1-s) \1\otimes H_\theta ^T + \frac{t(1-s)\gamma}{2}\left( 2 S_x\otimes S_x - S_x^2\otimes\1 -\1\otimes S_x^2 \right)}\vecc{+}\,.
\end{align}

Furthermore, in the same way we did before, we set $\theta = 0$, and thus $H_\theta = 0$, therefore, the only dynamics come from the noise term $S_x$. Then we can rewrite the previous equation as
\begin{align}
    \partial_\theta\vecc{\psi_\theta(t)} = -\ii t \int_{0}^1 ds &e^{\frac{ts\gamma}{2}\left( 2 S_x\otimes S_x - S_x^2\otimes\1 -\1\otimes S_x^2 \right)} \left( 
 S_z\otimes\1 -\1\otimes S_z\right)\\
 \cdot &e^{\frac{t(1-s)\gamma}{2}\left( 2 S_x\otimes S_x - S_x^2\otimes\1 -\1\otimes S_x^2 \right)}\vecc{+}\,.
\end{align}

Now it is easy to see that $\vecc{+}$ is an eigenstate of the matrix $2 S_x\otimes S_x - S_x^2\otimes\1 -\1\otimes S_x^2$ with eigenvalue $0$. Indeed $\vecc{+} = \ket{+}\ket{+}$. Therefore, we can simplify the previous expression further:
\begin{align}
    \partial_\theta\vecc{\psi_\theta(t)} = -\ii t \int_{0}^1 ds &e^{\frac{ts\gamma}{2}\left( 2 S_x\otimes S_x - S_x^2\otimes\1 -\1\otimes S_x^2 \right)} \left( 
 S_z\otimes\1 -\1\otimes S_z\right)\vecc{+}
\end{align}
we further use that $S_z = \frac{1}{2}(S_+^{x} + S_-^{x})$ and the fact that $\ket{+}^{\otimes N} = \ket{S = N/2,m_x = S}$ where we are following the notation in Eq. \eqref{eq:basis_sx_app}. Therefore, we can compute the following
\begin{align}
     \partial_\theta\vecc{\psi_\theta(t)} =& -\ii t \int_{0}^1 ds e^{\frac{ts\gamma}{2}\left( 2 S_x\otimes S_x - S_x^2\otimes\1 -\1\otimes S_x^2 \right)} \left( S_z\otimes\1 -\1\otimes S_z\right)\vecc{+}\\
 =&-\ii t \frac{\sqrt{N}}{2}\int_{0}^1 ds e^{\frac{ts\gamma}{2}\left( 2 S_x\otimes S_x - S_x^2\otimes\1 -\1\otimes S_x^2 \right)} (\ket{S = N/2,m_x = S-1}\ket{S = N/2,m_x = S} \\
 &- \ket{S = N/2,m_x = S}\ket{S = N/2,m_x = S-1})
\end{align}
where we applied the ladder operators onto the states following $S_{\pm}^{(x)}\ket{S, m_x} = \sqrt{(S+1)S - m_x(m_x\pm1)}\ket{S,m_x\pm 1}$, as well as using the fact that $S_+\ket{S,m_x = S} = 0$. Therefore because the states obtained are still eigenstates of the operator in the exponential we can trivially solve this and undo the vectorization to find that 
\begin{align}
    \partial_\theta\psi_\theta(t) = -\ii \sqrt{N} \frac{1-e^{-t\gamma/2}}{\gamma} (\ketbra{S = N/2,m_x = S-1}{S = N/2,m_x = S} - \ketbra{S = N/2,m_x = S}{S = N/2,m_x = S-1})\,.
\end{align}

Finally, we can use that the QFI can be computed as follows~\cite{qmqi}
\begin{equation}
     \f = 2\sum_{i,j}\text{Re}\frac{\bra{j}\partial_\theta\psi_\theta(t)\ket{i}\bra{i}\partial_\theta\psi_\theta(t)\ket{j}}{a_i+a_j}
\end{equation}
where $\{\ket{i}\}$ are the eigenbasis of $\psi_\theta$ and $\lambda_i$ the eigenvectors. We recall that under the dinamics from Eq. \eqref{eq:dynamics_globalnoise_nocontrol} and with $\theta = 0$, the initial state is an steady state $\ket{\psi_\theta(t)} = \ket{+}^{\otimes N}$. And therefore we find that 
\begin{equation}
    \f = 4 N \frac{(1-e^{-t\gamma/2})^2}{\gamma^2}\,.
\end{equation}

\end{proof}

\end{document}